\documentclass[12pt,preprint]{aastex}
\usepackage{rotating}

\shorttitle{Mixing in Supernova Explosion}

\shortauthors{Mao et al.}

\begin{document}



\title{Matter Mixing in Core-collapse Supernova Ejecta: Large Density Perturbations
in the Progenitor Star?}


\author{
Jirong Mao\altaffilmark{1,2,5,6},
Masaomi Ono\altaffilmark{2},
Shigehiro Nagataki\altaffilmark{1},
Masa-aki Hashimoto\altaffilmark{2},
Hirotaka Ito\altaffilmark{1},
Jin Matsumoto\altaffilmark{1},
Maria G. Dainotti\altaffilmark{1,3},
Shiu-Hang Lee\altaffilmark{1,4}
}
\altaffiltext{1}{Astrophysical Big Bang Lab, RIKEN, Wako, Saitama 351-0198, Japan}
\altaffiltext{2}{Department of Physics, Kyushu University, Fukuoka 812-8581, Japan}
\altaffiltext{3}{Astronomical Observatory, Jagiellonian University, ul. Orla 171, Krakow 31-501, Poland}
\altaffiltext{4}{Institute of Space and Astronautical Science (ISAS), Japan Aerospace Exploration Agency (JAXA), Kanagawa 252-5210, Japan}
\altaffiltext{5}{Yunnan Observatory, Chinese Academy of Sciences, Kunming, Yunnan Province, 650011, China}
\altaffiltext{6}{Key Laboratory for the Structure and Evolution of Celestial Objects, Chinese Academy of Sciences, Kunming, China}

\email{jirong.mao@riken.jp}

\begin{abstract}
Matter mixing is one important topic in the study of core-collapse supernova (CCSN) explosions. In this paper,
we perform two-dimensional hydrodynamic simulations to reproduce 
the high velocity $^{56}$Ni clumps observed in SN 1987A. 
This is the first time that large density perturbation is proposed in the CCSN progenitor to generate Rayleigh-Taylor (RT) instability
and make the effective matter mixing. In the case of a spherical explosion,
RT instability is efficient at both C+O/He and He/H interfaces of the SN progenitor. Radial coherent structures shown in perturbation patterns are important for obtaining high velocity $^{56}$Ni clumps. 
We can also obtain matter mixing features and high velocity $^{56}$Ni clumps  
in some cases of aspherical explosion.
We find that one of the most favorable models in our work has a combination of bipolar and equatorially asymmetric explosions in which at least 25\% of density perturbation is introduced at different composition interfaces of the CCSN progenitor. These simulation results are comparable to the observational findings of SN 1987A.
\end{abstract}




\section{Introduction}
Heavy elements in the universe can be released by core-collapse supernova (CCSN) explosions. It has been found in the observations of SN 1987A that some fast-moving $^{56}$Ni clumps ejected outside have similar velocities of helium and hydrogen (e.g., Hanuschik et al. 1988).
This phenomenon is called matter mixing in this paper. The radioactive decay of $^{56}$Ni$\rightarrow$ $^{56}$Co$\rightarrow$ $^{56}$Fe (the half-lives of $^{56}$Ni and $^{56}$Co are 5.9 and 77.1 days, respectively) and the related energy heating of SN 1987A provide excellent opportunities to study matter mixing problems and explosion morphology.
Early detections of hard X-ray emission \citep{dotani87,sunyaev87} and gamma-ray lines from $^{56}$Co \citep{matz88,leising90} in SN 1987A
indicate a high velocity component of radioactive $^{56}$Ni from the CCSN explosive nucleosynthesis.
The explosion energy of SN 1987A is given as $E/M_{\rm{env}}=(1.1 - 1.5)\times 10^{50}~\rm{erg~{\it M}_\odot^{-1}}$ where $M_{\rm{env}}$ is the mass of the hydrogen-rich envelope in the supernova (SN) progenitor and the total mass of synthesized $^{56}$Ni in SN 1987A is about 0.07$M_{\odot}$ (Shigeyama et al. 1988; Shigeyama \& Nomoto 1990; see also Handy et al. 2014 for the summary of SN 1987A explosion energy).
The mass of high velocity
$^{56}$Ni clumps was determined to be about a few times of $10^{-3}~M_{\odot}$ \citep{utrobin95} from the measurement of H$\alpha$ fine structure \citep{hanuschik88}.
\citet{haas90} and \citet{spyromilio90} observed
doppler-broadened line profiles of [Fe II] in SN 1987A.
\citet{haas90} found that the velocity can reach
about 4,000 $\rm{km~s^{-1}}$, and $4-17\%$ of the total $^{56}$Ni has a high velocity larger than 3,000 $\rm{km~s^{-1}}$.
Furthermore, the line profile is asymmetric, and its peak is located in the redshifted side of about 1,000 $\rm{km~s^{-1}}$
\citep{haas90,spyromilio90}.
Another interesting radioactive element produced by the explosive nucleosynthesis is $^{44}$Ti. Models of the spectrum and lightcurve of SN 1987A powered by $^{44}$Ti have been presented \citep{fransson93,jerkstrand11}. Although the half-life of $^{44}$Ti$\rightarrow$$^{44}$Sc$\rightarrow$$^{44}$Ca is about 59 years (the half-lives of $^{44}$Ti and $^{44}$Sc are $58.9\pm 0.3$ years and 3.97 hr, respectively, see Ahmad et al. 2006), hard X-ray emission lines from the decay
of $^{44}$Ti in SN 1987A were also detected and the estimated mass of $^{44}$Ti is $(3.1\pm 0.8)\times 10^{-4}$ $M_\odot$ \citep{grebenev12}. The inner ejecta of SN 1987A is heated by the radioactivity of $^{44}$Ti \citep{kjar10}. Moreover, three-dimensional (3D) structures reconstructed by observations show that the inner ejecta of SN 1987A may lie in the plane of the equatorial inner ring \citep{kjar10,larsson13,sinnott13}. As the collision between this inner ejecta and the edge of inner ring may happen, this morphology can naturally explain the observed inner ejecta rebrightening (e.g., Gr\"{o}ningsson et al. 2008, Dwek et al. 2010).

The physical mechanisms of CCSN explosions are under debate. In general, there are at least two interesting aspects
of CCSN explosion mechanisms (e.g., Janka 2012; Kotake et al. 2012; Burrows 2013). In the framework of a neutrino heating
mechanism, standing accretion shock instability \citep{blondin03} and neutrino-driven convection \citep{herant94,burrows95} were proposed to generate a globally anisotropic structure and enhance the effective heating.
Many hydrodynamic simulations have been performed during recent years (e.g., Nordhaus et al. 2010; Suwa et al. 2011; Kuroda et al. 2012; M\"{u}ller et al. 2012; Bruenn et al. 2013; Couch \& Ott 2013;
Hanke et al. 2013; Murphy et al. 2013; Ott et al. 2013, Wongwathanarat et al. 2013; Takiwaki et al. 2014). On the other hand,
MHD effects can trigger a jet-like CCSN explosion, as shown in some numerical simulations \citep{kotake04,sawai05,sawai13,burrows07,takiwaki09}.

So far, some tests to solve the matter mixing problems of CCSN explosions have been carried out under the hydrodynamic framework, and the explosive nucleosynthesis has been adopted in particular. Rayleigh-Taylor (RT) instability (or Richtmyer-Meshkov instability in the condition of entropy impulsive acceleration) can happen naturally in CCSN explosion models, and it plays a vital role on shock propagation and matter mixing.
For example, it was found in early works that He/H and C+O/He layers in an SN progenitor can be unstable against strong RT
instability \citep{ebi89,benz90}. RT fingers first dominate at the H/He interface, and at late times are prominently shown at the He/C+O interface \citep{muller91}.

Perturbation in the progenitor star has been considered to be an origin of RT instability and this mechanism may be adopted to solve matter mixing problems.
In principle, perturbations are related to the physics of convection and turbulence in the stellar interior. The integrated convection turnover time across the whole convective region is comparable to the nuclear timescale. This indicates that the composition in the CCSN progenitor is not homogeneous, thus the progenitor system has strong fluctuations \citep{bazan98}.
Stellar convection is thought to make significant perturbations based on the dynamical simulations of progenitor stars (e.g., Bazan \& Arnett 1994; Bazan \& Arnett 1998; Meakin \& Arnett 2007b; Arnett \& Meakin 2011a).
\citet{bazan94} suggested progenitor perturbations of density, temperature, pressure, and electron fraction in the order of 5\%, 1\%, 3\%, and 0.08\%, respectively.
\citet{arnett94} expected 2.5\% of density perturbation and 0.8\% of velocity perturbation in the oxygen shell of progenitor star.
CCSN progenitor features were also explored by an extended work of density perturbation at the edges of metal convective zones \citep{bazan98}.
Furthermore, some violent dynamic eruptions in the convective zones with simultaneous carbon and oxygen shell burning inside the progenitor star were found in 3D simulations, and about 10\% of density perturbation was detected in those convective zones \citep{meakin07b,arnett11a}.

The position that RT instability occurs in the CCSN progenitor has also been discussed. \citet{arnett89} noticed the possibility that RT instability begins at the edge of the dense mass shell, where heavy elements mix with hydrogen through the He envelope. \citet{muller91} mentioned that RT fingers form at both He/C+O and H/He interfaces. \citet{nagataki98} introduced velocity perturbation at the layer between He and H.
\citet{ellinger12} found that this instability at He/H interface is caused by reverse shock. \citet{hammer10} and \citet{kif03,kif06} found an instability at the Si/O interface, which was not shown in the results of \citet{ellinger12}.
Although all of instabilities at Si/O, He/C+O, and He/H interfaces were described by \citet{kif03}, the instability at the C+O/He boundary may be the most prominent one \citep{ellinger12}. This is consistent with the conclusion given by \citet{kane00}.

It is expected that perturbations in the CCSN progenitor may generate or enhance the matter mixing.
Simulations with introducing an initial density perturbation of 5\% amplitude in the progenitor model failed to reproduce the high velocity ($>3500~\rm{km~s^{-1}}$) $^{56}$Ni clumps \citep{herant91,herant92}.
On the other hand, the possibility of large perturbations in the CCSN progenitor has been investigated.
The matter mixing can be reproduced by simulation models if 30\% of the initial velocity perturbation and 30\% of the initial energy perturbation are introduced in the CCSN progenitor \citep{yamada91}.
\citet{nagataki98} introduced 30\% of velocity perturbation at the H/He boundary layer in the CCSN progenitor
and obtained high velocity ($>3000~\rm{km~s^{-1}}$) $^{56}$Ni clumps. However, the spacial resolution of these simulations is rather low.
Three-dimensional simulations with different metallicities in the CCSN progenitor were also performed to develop RT instability \citep{jogg09,jogg10a,jogg10b}.
It is interesting that
large asymmetric structures during late times of a CCSN explosion may be originated by the significant perturbation that arises from an initial explosion itself, but not from merging RT structures.

The asymmetric effect of a CCSN explosion is also involved as an important point for the matter mixing.
\citet{nagataki98} and \citet{nagataki00} found that asymmetric or jet-like explosions can effectively change the velocity distribution of $^{56}$Ni. \citet{hung03,hung05} performed 3D smoothed particle hydrodynamics (SPH) simulations and compared the difference between symmetric and asymmetric CCSN explosions. They confirmed that an asymmetric explosion makes the matter mixing in supernova ejecta, although high velocity $^{56}$Ni clumps could not be reproduced.
\citet{kif03} comprehensively discussed RT instability and explosion evolution in one-dimensional (1D) and two-dimensional (2D) cases. Introducing low-mode perturbation in a CCSN explosion progenitor may reproduce high velocity ($>3300~\rm{km~s^{-1}}$) $^{56}$Ni clumps \citep{kif06}, although robust conclusions of 2D simulations were further discussed by \citet{gaw10}.
Three-dimensional simulations of CCSN explosions were also performed by \citet{hammer10}, and high velocity $^{56}$Ni could be reproduced.
However, their models neglected the effect of matter fallback into the compact remnant.
As recently suggested by \citet{ono13}, not only perturbations in the SN progenitor but also asymmetric explosions (clumpy morphologies in explosions) may be necessary to reproduce the observed high velocity $^{56}$Ni clumps of SN 1987A.

In this paper we focus on RT instability as one possible solution of matter mixing problems in SN 1987A. We have already performed 2D hydrodynamic simulations to examine the matter mixing of CCSN (see Paper I).
If a large radial velocity perturbation of 30\% amplitude is introduced just before the shock wave
reaches the hydrogen envelope of the progenitor star, high velocity $^{56}$Ni clumps of 3,000 $\rm{km~s^{-1}}$ may be reproduced, in the case of bipolar explosion.
This indicates that RT instability originating from
large perturbations in the progenitor may be one reason for the matter mixing in a CCSN explosion.
Here, we further extend the work in Paper I.
In particular, this is the first time that large density perturbation is systematically considered in a CCSN progenitor to make
strong RT instability, and consequently dig out a part of inner heavy elements to outer layers during CCSN explosion.
Here, we mention some possible motivations of the large density perturbations introduced in our CCSN progenitor. The global turbulent feature of the CCSN progenitor interior was explored \citep{meakin07a, arnett09, arnett11a, arnett11b, viallet13}, and
it was realized that a realistic progenitor model is crucial for CCSN explosion \citep{smith14}.
In this paper,
we adopt the CCSN progenitor that has been successfully used to produce the lightcurve, $^{56}$Ni mass, and explosion energy of
SN 1987A \citep{shigeyama87,shigeyama88,hashimoto89,shigeyama90}. Then, we introduce large density perturbations to this progenitor.
We suggest that this progenitor with a large density perturbation may be physically
equivalent/corresponding to a realistic model with a global turbulent feature and strong dynamic
instabilities.
Furthermore, Figure 2 shows that
the radial coherent features at large scales exist in some of our perturbation patterns.
These radial coherent features are important for reproducing
the matter mixing of SN 1987A. The coherent features shown in our perturbation
patterns may capture the similar behavior driven by the turbulent
convection in the stellar interior \citep{meakin07a,arnett09, arnett11a, arnett11b, viallet13}.
Moreover, \citet{smith14} argued two important
effects of the hydrodynamic instability and turbulence in the progenitor model: (1) finite amplitude fluctuations in velocity and
temperature, and (2) nonlinear interaction with nuclear burning during the latest stages of the stellar evolution.
These effects are not included in current stellar evolution models.
\citet{smith07} claimed that SN 1987A is related to the luminous blue variable (LBV), which shows violent dynamical
eruption or enhanced mass loss prior to the core-collapse. This invokes large density perturbation in the hydrogen
envelope of the CCSN progenitor. We also propose Betalgeuse as an observational example: The clumpy structure in the inner envelope may be related to
the giant convection cells of the outer atmosphere \citep{decin12}. The convection
pattern can be also unveiled \citep{montarges14}. These observational evidences also suggest large density perturbation in the hydrogen envelope of the CCSN progenitor.

We suggest some physical factors$-$such as C+O layer convection, non-radial pulsations in LBV stars,
cool hydrogen envelopes, and convection-driven asymmetries in the deep interior$-$ are affecting the
development of RT instability and the matter mixing of SN 1987A. Although we introduce large
density perturbations in the progenitor model as a major cause to develop RT instability and
produce the matter mixing in this paper, there are many pathways to the final asymmetries after the shock breakout. We focus on perturbations in the progenitor star, and its growth due to RT instability during the shock wave propagation. As mentioned, perturbations could be introduced by combinations of different sources, such as convection in very deep, cool hydrogen envelopes; convective shell burnings in the deep interior; and nonlinear, non-radial pulsations in an LBV star. Aside from the origins of the perturbations, the growth of RT instability is affected by the progenitor structure. For giant RT fingers, a large hydrogen envelope, as in normal red supergiant (RSG) stars, is required (e.g., Wongwathanarat et al. 2015). On the other hand, RT instabilities in Wolf-Rayet progenitors may be significantly reduced by the lack of an H/He boundary with a strong density gradient. However, deeper asymmetries stemming from the RT instabilities at inner composition boundaries and/or an aspherical explosion could still produce structure. Because the progenitor of SN 1987A is known to be a blue supergiant (BSG), which had a compact hydrogen envelope, it is difficult to produce prominent RT fingers compared to those of RSG stars. In addition, to dig out the innermost ejecta to outer layers, other factors-such as the compactness of the C+O shell, the density structure of He shell, and the density gradient at He/H boundary-may also be important (Wongwathanarat et al. 2015).

We propose a large density perturbation in the progenitor model as follows:
We test three cases of perturbations for spherical explosion. For all cases, the amplitude of perturbations are randomly assigned, but the length scale of the perturbations are different.
First, perturbations are introduced at the interfaces of different composition layers, with an exponential decay of its amplitude in the radial direction.
At each boundary layer, the length scale of the perturbation in the angular direction is controlled by a fixed wave number. In this case, perturbations have the radial coherent structure.
Second, the perturbation length scale in the radial direction is proportional to the density scale height determined by the progenitor model. The perturbation length scale in angular direction is controlled by a fixed wave number.
Third, we constrain the perturbation length scale in both radial direction and angular direction, as it is proportional to the density scale hight.
All of these perturbation modes are comprehensively described in Section 2.3 and illustrated in Figures 1 and 2. We also test three cases for aspherical explosion under the condition that perturbations are introduced at interfaces of different composition layers.
First, we take the case that explosion along the polar direction is stronger than that along the equatorial direction. This corresponds to a bipolar or jet-like SN explosion.
Second, we propose the case of an SN explosion that is asymmetric to the equatorial plane.
Third, we combine the two cases to take into account both bipolar and equatorially asymmetric effects of the CCSN explosion. The observed [Fe II] line in SN 1987A shows an asymmetric feature in the velocity profile \citep{haas90, spyromilio90}.
The observation of SN 1987A in 3D structure presents an elongated morphology \citep{kjar10}. These observational evidences invoke us to make a combination of mildly bipolar and equatorially asymmetric explosions.
In our work,
we concern the radial velocity profile and line of sight velocity profile of $^{56}$Ni, the total ejected mass of $^{56}$Ni, and the mass of high velocity ($>3000~\rm{km~s^{-1}}$) $^{56}$Ni. We also discuss the total ejected mass of $^{44}$Ti and the related explosion energy. In Section 2, we review the procedures of the numerical simulation presented in Paper I. Numerical simulation methods and the progenitor star are briefly described and many density perturbation modes are given in detail. In Section 3, we comprehensively show the matter mixing results of $^{56}$Ni in our simulations. Some important issues associated with our results are discussed in Section 4. Finally, we summarize our achievements in Section 5.

\section{Numerical Simulations}
We now briefly describe numerical methods, initial conditions, and progenitor model, which are presented in Paper I. We list several key points for our simulation models in Section 2.1 and 2.2. We
extend the work of Paper I to explore the possibility of reproducing high velocity $^{56}$Ni clumps by using the progenitor with large density perturbation, and comprehensive perturbation modes are presented in Section 2.3.

\subsection{Numerical Arrangements and Initial Conditions}
We perform an adaptive mesh refinement (AMR) hydrodynamic code, FLASH \citep{fryxell00}, and a  2D spherical coordinate ({\it r}, $\theta$) is adopted. The initial computation domain covers the region with a radius of $1.4\times 10^8~\rm{cm}<{\it r}<3.0\times 10^9~\rm{cm}$ and an angle of $0<\theta<\pi$. The outer boundary of the progenitor model is limited at $r=3.4\times 10^{12}~\rm{cm}$, which corresponds to the surface of the hydrogen-rich envelope. A wind component is added beyond this radius.
The base-level grid points are $48(r)\times 12(\theta)$ and the maximum refinement level is set to be 7. Thus, the effective maximum numbers of grid points are $3072(r)\times 768(\theta)$. The minimum effective cell sizes are about
10 km and 0.23 deg in radial direction and angular direction, respectively. As the SN shock propagates and reaches close to the radial outer boundary, the computation domain is extended via remappings. When the physical values of the extended regions
are set to the values of the progenitor model, 44 remappings are required. We use the monotonic cubic interpolation scheme
to interpolate physical values \citep{steffen90}. The computational cost is about 20 thousand CPU hr for each simulation model.

The remapping and boundary conditions applied in this work were presented in Paper I.
If the forward shock and/or RT fingers reach close to the radial outer boundary (approximately 20\% of the radial computational domain from the radial outer boundary), the radial size of the
computational region is extended by a factor of 1.2.
At the radial inner boundary, a ``reflection" boundary condition is employed at the start of simulation.
After the forward shock has reached the composition interface of C+O/He
(corresponding to the radius of $6\times 10^9$ cm), it is switched to a ``diode" boundary condition that allows matter
to flow out of the computational domain, but inhibits it from entering the computational domain through the
inner boundary in order to include the effect of matter fallback.
Although changing the switch timing can somewhat affect the degree of the innermost matter fallback, we fix the switch timing to make sure that the mass of $^{56}$Ni that remains in the computational domain does not become too
small compared with that for SN 1987A.
At the radial outer boundary, we adopt the "diode" boundary condition over the simulation.
In the angular direction, we take the ``reflection" boundary condition.
Meanwhile, the mass inside the inner computation domain is regarded as a point source, and both the point source gravity and spherically averaged self-gravity are considered.
The Helmholtz equation of state (EOS) is adopted for our simulations, whereas a modified ideal gas and radiation EOS is used for a very low density region, and we blend the two kinds of EOS at the transition density region (see Paper I for the details).

Explosive nucleosynthesis is applied and we obtain the following elements from
FLASH code: n, p, $^1$H, $^3$He, $^4$He, $^{12}$C, $^{14}$N, $^{16}$O, $^{20}$Ne, $^{24}$Mg, $^{28}$Si,
$^{32}$S, $^{36}$Ar, $^{40}$Ca, $^{44}$Ti, $^{48}$Cr, $^{52}$Fe, $^{54}$Fe, and $^{56}$Ni. Advection equations are solved to trace
the distribution of these elements. Energy depositions due to the radioactive decay of $^{56}$Ni to $^{56}$Fe are included, as described in Paper I. However, some isotopes of iron group elements (e.g., $^{57}$Ni and $^{58}$Ni) are not included in the original FLASH code.
Therefore,
the masses of $^{56}$Ni and $^{44}$Ti obtained by FLASH code may be overestimated.
In order to
clarify the effect of this limited nuclear reaction network on the mass of heavy elements,
an explosive nucleosynthesis calculation with a large nuclear network is performed as a post-process.
We stress this issue in Section 4. The process is presented in the Appendix.

\subsection{Progenitor Model}
The progenitor model of CCSN explosion adopted in this paper is a 16.3 $M_\odot$ BSG star with 6 $M_\odot$ helium core and 10.3 $M_\odot$ hydrogen envelope. This progenitor model has a radius set as $3.4\times 10^{12}$ cm, and has been successfully adopted for reproducing the lightcurve, explosion energy, and total $^{56}$Ni mass (0.07 $M_\odot$) of SN 1987A \citep{shigeyama87, shigeyama88,nomoto88,shigeyama90}.
A wind component with the density profile of $\rho\propto r^{-2}$ is added and
the uniform temperature of $T=10^4$ K
is adopted. The inner density of the wind component is $3.0\times 10^{-10}~\rm{g~cm^{-3}}$. The wind component
is extended to the radius of $4.5\times 10^{12}$ cm and simulations are carried out until just before the shock
waves reach the radius. The density of the wind component is smoothly connected to that of the
stellar surface. If we assume that the wind velocity is 15 $\rm{km~s^{-1}}$, the mass loss rate is $1.8\times 10^{-3}~M_\odot~\rm{yr^{-1}}$;
this value is larger than $1.0\times 10^{-5}~M_\odot~\rm{yr^{-1}}$ used
in Gawryszczak et al. (2010). We restrict our simulation to just after the shock breakout, and the selected
wind profile does not affect the matter mixing.

We set the total input energy to be $2.5\times 10^{51}$ erg in most simulation models and divide this input energy into two equal parts: kinetic energy and thermal energy. The final explosion energy is related to the input energy. Modification of total input energy value is also attempted in the two simulation models.

Si, C+O, He, and H layers are settled as an onion-like structure in this progenitor. The outer boundary of the Si-rich layer is at
the radius of $3\times 10^8$ cm, the interface between C+O and He core is at $6\times 10^9$ cm, and the
hydrogen envelope begins at $5\times 10^{10}$ cm. It is also noted that oxygen shell burning is limited at $3\times 10^9$ cm \citep{arnett94,bazan98}.
Therefore, we propose that these numbers are four critical radii in the progenitor star.

\subsection{Explosion and Density Perturbation Modes}
\subsubsection{Spherical Explosion of SC1, SC2, and SC3}
In this subsection, we introduce three density perturbation modes that affect matter mixing in a spherical CCSN explosion. From the dynamical simulations of oxygen shell burning \citep{bazan98,meakin07b}, density perturbations of $\delta \rho/\rho \sim$ 10\% could be introduced at the boundaries of the convective shell burning. We assume that density perturbations are prominently introduced at the boundaries of shell burnings, and further assume perturbations with large amplitude (up to 50\%) to obtain successful matter mixing. Then,
we consider density perturbations at four critical radii ($3\times 10^8$ cm at Si/C+O interface, $3\times 10^9$ cm at oxygen shell burning position, $6\times 10^9$ cm at C+O/He interface, and $5\times 10^{10}$ cm at He/H interface), which roughly correspond to the boundaries of shell burnings in the progenitor.
Density perturbation at each critical radius with the shape of $1+\epsilon[2~\rm{rand}(m\theta/\pi)-1]$ in the angular direction is given, where rand is the random number within the range between 0 and 1, and is a function of $\theta$.
We take $m+1$ random numbers at sample points, $\theta=0$, $\pi/m$, $2\pi/m,$..., $(m-1)\pi/m$, $\pi$. The random number for an arbitrary $\theta$ is obtained by an interpolation with the values of adjacent sample points, and $\epsilon$ is the amplitude of the perturbation.
The amplitude has an exponential decay in the radial direction and can be presented as $\epsilon=\epsilon_0~\rm{exp}(-|{\it r}-{\it r}_0|/{\it r}_0)$, where $\epsilon_0$ is the amplitude at each boundary layer and $r_0$ is the radius of the certain boundary.
We call this case SC1. Thus, two parameters, $\epsilon_0$ and $m$, are involved in SC1. In order to understand the perturbation effect at each boundary on the matter mixing, we perform four more one-by-one boundary perturbation models, in which perturbations at the Si/C+O interface, the C+O/He interface, the He/H interface, and both C+O/He and He/H interfaces are considered, respectively.
The RT instabilities at the H/He and He/C+O boundaries can be generated by the passage of
the shock and/or reverse shock, and the condition of RT instability ($\nabla \rho \cdot \nabla p<0$) was given
by \citet{chevalier76}. The growth rate of RT
instabilities at H/He and He/C+O interfaces has been investigated (e.g., Ebisuzaki et al. 1989; Paper I) and such interfaces are unstable against RT instabilities after the passage of the shock waves. The four one-by-one boundary perturbation models are used to see which boundaries are critical for the matter mixing.
In particular, we see the radial coherent structure in the perturbation pattern (Figure 2) that was obtained from large density perturbations imposed on those interfaces of the progenitor, which is important to the matter mixing of SN 1987A.

Besides the possibility of large density perturbations at interfaces of different composition layers, we propose that perturbations may also be introduced at any position in the progenitor. Although this global perturbation is not motivated by a known physical process, if the progenitor of SN 1987A was an LBV, such a perturbation might have existed. We propose the cases of SC2 and SC3, which are two different perturbation modes, to search the parameter space and explore the potential impacts.
In the case of SC2, the density perturbation may be related to the density distribution of the progenitor star. We derive the density scale height $H(r)$, which can be presented as $H(r)=|dr/d~\rm{ln}\rho|$.
We assume that the radial length scale of the density perturbation $l_r(r)$ is proportional to the density scale height $H(r)$. For simplicity, we introduce a scaling factor $f$ of $l_r(r)=fH(r)$. If we adopt $f=5$ in a simulation model, we can obtain the radial coherent structure shown in Figure 2. A simulation model of $f=1$ is also applied to compare with the corresponding model of $f=5$. For a given radius, the angular length scale of the density perturbation is $l_\theta(r)=r\pi/m$, and we fix the number of $m$.
We take random numbers at 2D sample points at intervals of $l_r(r)$ and $l_\theta(r)$. The perturbation at arbitrary ($r$, $\theta$) is given by $1+\epsilon_0[2~\rm{rand}({\it r},\theta)-1]$, where the random number is obtained by an interpolation with the values of neighboring sample points.
In the case of SC2, the perturbations are modulated not only in the angular direction but also in the  radial direction.
Thus, three parameters, $\epsilon_0$, $m$ and $f$, are involved in the case of SC2.

Third, we assume that the length scales of the perturbation in both the radial and angular direction, $l_r(r)$ and $l_\theta(r)$, respectively, are proportional to the density scale hight $H(r)$. In other words, we assume $l_r(r)=fH(r)$ and $l_\theta(r)=fH(r)$. We take random numbers at 2D sample points at intervals of $l_r(r)$ and $l_\theta(r)$. The perturbation at arbitrary ($r$, $\theta$) is given by $1+\epsilon_0[2~\rm{rand}({\it r},\theta)-1]$, where the random number is obtained by an interpolation with the values of the neighboring sample points. We call this case SC3.
For SC3, two parameters, $\epsilon_0$, and $f$, are involved\footnote{Before we perform simulations of SC1, SC2, and SC3, one simple case, ``Basicmodel", is considered. In this case, we introduce density perturbations at the four critical radii in the progenitor star as mentioned
in SC1. However, we set the amplitude of the density perturbation at each critical radius by the way of
$\epsilon=\epsilon_0{\rm{exp}}(-|r-r_0|/h)$, where $\epsilon_0=0.5$, and $h$ corresponds to the density scale height at each critical radius.
From the function $H(r)$, we have $h_1=2.2\times 10^8$ cm, $h_2=1.2\times 10^9$ cm, $h_3=2.0\times 10^9$ cm, and $h_4=2.0\times 10^{10}$ cm at each critical radius, respectively.
Density perturbation at each critical radius in angular direction is as same as that given in SC1. In this simple
case, we cannot reproduce high velocity $^{56}$Ni clumps. Then, we turn to complicated cases of SC1, SC2, and
SC3 in the spherical explosion to reproduce the matter mixing.}.

The density scale hight, $H(r)$, of the progenitor and the ratio between $r$ and $H(r)$ as a function of radius is illustrated in the upper panel of Figure 1.
We can see that $H(r)$ is roughly proportional to $r$ and the ratio $r/H(r)$ ranges between 2 and 8. The exception is the case for the regions around the stellar surface. In addition, $H(r)$ tends to be small around the composition interfaces of C+O/He and He/H. Moreover, we show the radial perturbations involved with the cases of SC2 and SC3 in the lower panel of Figure 1.
We take the perturbation mode of SC3 ($\epsilon_0=50\%$ and $f=1$) as an example, and see that the amplitude of the density perturbation in the radial direction is randomly distributed. Distributions of the 2D density perturbation clumps with amplitude $\epsilon_0=50\%$ for SC1, SC2, and SC3 are shown in Figure 2. For the case of SC1, perturbations are introduced around the critical
radii, and angular perturbation is adjusted by the number of {\it m}.
Perturbations extend from the chosen radii in both
upward and downward directions.
For SC2, perturbations are
adjusted in angular direction and radial direction by {\it m} and {\it f}, respectively. For SC3, perturbations in both angular direction and radial direction are adjusted by the same number of {\it f}.
Thus, from these perturbation patterns,
we see that the coherent structure in the radial direction shown in some panels of Figure 2 plays a key role to reproduce high velocity $^{56}$Ni clumps. In particular, we notice that the density perturbations driven by the turbulent
convection in stellar interiors show the coherent feature on large scales, and wave numbers have an
order of the circumference at the outer boundary divided by the depth of the convection zone \citep{meakin07a, arnett09,arnett11a,arnett11b,
viallet13}. The coherence shown in our perturbation patterns with $m=8$ can
capture a similar feature.

\subsubsection{Aspherical Explosion}
The density perturbation modes mentioned in Section 2.3.1 are under the condition of spherical explosion.
Because the observations of SN 1987A indicate an aspherical explosion, we can mimic some aspherical explosion cases that were adopted in Paper I.
All of the following aspherical explosion models are under the case of SC1 density perturbation mode.

We take the initial radial velocity $v_r\propto r[1+\alpha \rm{cos}(2\theta)]/(1+\alpha)$ that was written as Equation (7) of Paper I.
Through this function form, we can consider a bipolar explosion. In order to represent the bipolar
explosion proposed by Nagataki et al. (1997, 1998), Nagataki (2000), and Paper I, we obtain the ratio of the radial velocity along the polar
direction to that along the equatorial plane as $v_{\rm{pol}}/v_{\rm{eq}}=(1+\alpha)/(1-\alpha)$, where $\alpha$ is the parameter to determine the degree of asymmetry ($\alpha=0$ corresponds to a spherical explosion). In this paper, $\alpha$ is set to be 3/5 , i.e. $v_{\rm{pol}}/v_{\rm{eq}}=4$.

We also propose the asymmetric feature of the CCSN explosion to the equatorial plane as mentioned in Paper I (e.g., due to neutrino-driven explosion aided by SASI; Suwa et al. 2010).
An equatorially asymmetric explosion is mimicked by changing the normalization of the radial velocity across $\theta=\pi/2$. We define the initial radial velocity at $\theta=0$ and that at $\theta=\pi$ as $v_{\rm{up}}$ and $v_{\rm{down}}$, respectively. We adjust the normalization so that $v_{\rm{up}}/v_{\rm{down}}=1.8$. If we take $v_{\rm{up}}/v_{\rm{down}}=2$, as given by Paper I, the estimated neutron star kick velocity becomes larger than that typical of young pulsars. Therefore, we adopt a smaller value for $v_{\rm{up}}/v_{\rm{down}}$ in this paper.

We can further combine the cases of bipolar explosion and equatorially asymmetric explosion together as a case of global asymmetric explosion.
The combined global asymmetric explosion model may be used to examine SN 1987A explosion, as indicated by the asymmetric feature of [Fe II] line \citep{haas90,spyromilio90} and the elongated morphology \citep{kjar10} in SN 1987A.

\section{Results}
Our 2D hydrodynamic simulation results on the matter mixing of CCSN explosions are presented in this section. We list those parameters adopted in our simulation models in Table 1. Some major simulation results are listed in Table 2.

\subsection{Spherical Explosion}
We first present our simulation results for SC1. The results of the case with the perturbation amplitude $\epsilon_0=50\%$ and $m=8$ are shown in Figure 3, whereas the results of the case with the perturbation amplitude $\epsilon_0=50\%$ and $m=20$ are shown in Figure 5. We show density distribution, distribution of the $^{56}$Ni mass fraction, radial velocity of several elements, and line of sight velocity profiles of $^{56}$Ni, respectively\footnote{In this paper, for simplicity, the mass distributions of elements as a function of radial velocity at the ends of simulation time are called radial velocity profiles, and mass distributions of elements as a function of line of sight velocity at the ends of simulation time are called as line of sight velocity profiles.}. We clearly see that prominent RT fingers are formed and distributions of $^{56}$Ni have anisotropic structures. Although most $^{56}$Ni is within the inner part of the contour, we see that the RT fingers of $^{56}$Ni are beyond the radial velocity of 3,000 $\rm{km~s^{-1}}$.
The matter mixing happens, as we can see in radial velocity profiles. The line of sight velocity of $^{56}$Ni is extended beyond 3,000 $\rm{km~s^{-1}}$, and high velocity $^{56}$Ni clumps are reproduced. The perturbation modes of SC1 show a prominent coherent structure in radial direction (Figure 2), which we believe is important to reproduce high velocity $^{56}$Ni clumps. A different set of random numbers may induce different results. To test this, we use the model of SC1p50m8 with a different set of random numbers and run the simulation again. We obtain the following results: total $^{56}$Ni mass is 0.15 $M_\odot$; high velocity ($>3000~\rm{km~s^{-1}}$) $^{56}$Ni is 21.7\%; total $^{44}$Ti mass is $2.1\times 10^{-4}~M_\odot$. Thus, our
simulation results using this different set of random numbers are consistent with those of the original one. We plot the final density distribution, final distribution of $^{56}$Ni mass fraction, radial velocity profile, and line of sight velocity profile of $^{56}$Ni with different observational view angles in
Figure 4.

The perturbation mode in the angular direction is dominated by the parameter $m$. Figures 3 and 5 show that different values of parameter $m$ provide different RT fingers: More RT fingers in the angular direction are developed if we take a larger value of $m$. However, as high velocity $^{56}$Ni clumps are mainly determined by the strong coherent structure in the radial direction, a larger percentage number of high velocity $^{56}$Ni can be obtained by the simulation models with a smaller number of $m$.

In order to investigate the effect of the perturbation amplitude on the matter mixing of SN 1987A, simulations with 10\% of density perturbation are also performed in the case of SC1. We show the results using parameter $m=20$ as an example in Figure 6. Density distribution, distribution of $^{56}$Ni mass fraction, radial velocity profiles of several elements, and line of sight velocity profiles of $^{56}$Ni are given, respectively. Although RT instability is developed at the last stage, $^{56}$Ni has roughly spherical distribution. We see that line of sight velocity of $^{56}$Ni is limited within 2,000 $\rm{km~s^{-1}}$, and the mixing of the innermost elements (e.g., $^{56}$Ni and $^{44}$Ti) into high velocity ($>3,000~\rm{km~s^{-1}}$) regions has failed. We conclude that a density perturbation as small as 10\% cannot reproduce the high velocity $^{56}$Ni clumps of SN 1987A.

It is interesting to examine which boundary layer where density perturbation occurs is important for the matter mixing.
We perform three simulation models in which perturbations are introduced at a single composition (Si/C+O, C+O/He, He/H) interface. One simulation model with perturbations at both C+O/He and He/H interfaces is also performed.
The parameters of $m=20$ and $\epsilon_0=50\%$ are set for each simulation model. We do not obtain high velocity $^{56}$Ni when the density perturbation is introduced at the boundary layer between silicon and oxygen. If density perturbation is introduced at the C+O/He interface or He/H interface, high velocity $^{56}$Ni clumps appear. However, the mass of high velocity $^{56}$Ni is only about 0.7\% of the total $^{56}$Ni produced by the CCSN explosion, which is less than the observational value ($>4\%$, Haas et al. 1990). If we take perturbations at both C+O/He and He/H interfaces, high velocity clumps of $^{56}$Ni are successfully obtained. These results indicate that most effective density perturbations for the matter mixing occur at both C+O/He and He/H interfaces, which is consistent with the examination shown in Figure 4 of Paper I$-$the growth factors of seed perturbation around the interfaces of C+O/He and He/H in the progenitor model are prominent.
Figure 7 shows results obtained from the model in which the perturbations at C+O/He and He/H interfaces are introduced.

We then present our simulation results for SC2. The results of the case with perturbation amplitude $\epsilon_0=50\%$ and $m=8$ are shown in Figure 8, while the results of the case with perturbation amplitude $\epsilon_0=50\%$ and $m=20$ are shown in Figure 9. For both cases, we take parameter $f=5$. We see that RT fingers are developed and the matter mixing works: We obtain high velocity $^{56}$Ni clumps. Cases with a small perturbation amplitude $\epsilon_0=10\%$ are also presented, and the results obtained from $m=8$ and $m=20$ are given. We show the results from the case of $\epsilon_0=50\%$, $m=20$, and $f=5$ as an example in Figure 10. Although we see some developed RT fingers, we confirm that the SC2 model with such small density perturbation cannot reproduce the matter mixing.

We further investigate the effect of radial coherent structure on the matter mixing for SC2. This effect is dominated by the parameter $f$. A small number of $f$ indicates a small length scale of density perturbation. Figure 11 shows the results of the simulation of $\epsilon_0=50\%$, $m=20$, and $f=1$ as an example. We cannot reproduce the matter mixing by this simulation model. As we can see from the density perturbation patterns in Figure 2, the perturbation mode of $f=5$ provides the prominent coherent structure in the radial direction, and the matter mixing is obtained. The perturbation mode of $f=1$ provides a far less coherent structure in the radial direction, so the matter mixing cannot be obtained.

We present our simulation results for SC3 at last. The density perturbation with 50\% amplitude is considered in two simulation models, and $f=1$ and $f=5$ are taken, respectively, in each one. However, we do not find high velocity $^{56}$Ni clumps in these simulation results. Figure 12 shows the results from the simulation model with $\epsilon_0=50\%$ and $f=1$.
As we see from the density perturbation patterns in Figure 2, the perturbation modes of SC3 do not provide any clear coherent structure in the radial direction. Thus, the matter mixing cannot be obtained.

We summarize our main results of the spherical explosion models: (1) The density perturbation amplitude is the most important parameter and RT instability dominates the matter mixing. RT fingers can be developed and the matter mixing can be reproduced by some simulation models with 50\% of density perturbation. While 10\% of density perturbation introduced in simulation models can generate RT fingers, but it is not enough to reproduce the matter mixing.
(2) The synthesized $^{56}$Ni and $^{44}$Ti in the stellar interior can be effectively dug out because the radial coherent structures shown in perturbation patterns make the successful matter mixing in the CCSN explosion.
(3) The matter mixing cannot be reproduced by the simulation model with the density perturbation introduced at the Si/C+O interface, but density perturbations at both the C+O/He and He/H interfaces can effectively achieve high velocity $^{56}$Ni clumps. In other words, prominent density perturbations at large scales are   successful for the matter mixing of a CCSN explosion. (4) The distribution of $^{56}$Ni is roughly spherical in the simulation models where the matter mixing cannot be reproduced. If the simulation models are successful for the matter mixing, the distribution of $^{56}$Ni is expected to be strongly anisotropic with prominent RT fingers. The properties of the heavy elements distribution shown in our simulation results indicate that SN 1987A may not favor spherical explosion models. (5) The distribution of $^{44}$Ti follows that of $^{56}$Ni. (6) Although we initially assume a spherical explosion in our simulation models, large density perturbations introduced in the progenitor may break down this spherical property. The set of initial conditions is important because it affects the evolution of the CCSN explosion.

We focus on
$^{56}$Ni velocity profiles in these spherical explosion models, and obtain high velocity $^{56}$Ni clumps. We think that the matter mixing of SN 1987A can be qualitatively reproduced by these models. We also further discuss two issues.
First, the total mass of synthesized $^{56}$Ni in our spherical simulation results is about two to three times of that of observational determinations \citep{shigeyama88}. Second, the explosion energy produced by our spherical simulations is about two times that of the observational fittings by \citet{shigeyama88} and \citet{shigeyama90}, although the explosion energy has a range of $(1 - 2)\times 10^{51}$ erg \citep{handy14}.
We note that \citet{kif06} produced the matter mixing in the CCSN with an explosion energy of $2\times 10^{51}$ erg, and \citet{hammer10} provided an explosion energy of $1.8\times 10^{51}$ erg in their 2D models.
In order to examine the matter mixing of CCSN spherical explosions affected by the input energy, we reduce the input energy from $2.5\times 10^{51}$ to $1.7\times 10^{51}$ erg and re-perform one simulation model for SC1. Those results are shown in Figure 13.
The explosion energy of $1.08\times 10^{51}$ erg calculated from this simulation model is consistent with the observational one given by \citet{shigeyama88}. But the total synthesized $^{56}$Ni mass is still larger than the observational one, and we do not obtain enough high velocity $^{56}$Ni clumps. Therefore, it is necessary to perform simulation models of aspherical CCSN explosion to study matter mixing.

\subsection{Aspherical Explosion}
We present the results of aspherical explosion models in this subsection. First, we show the results of a bipolar model.
Density distribution, distribution of the $^{56}$Ni mass fraction, radial velocity of several elements, and line of sight velocity profiles of $^{56}$Ni are shown in Figure 14.
Due to this jet-like CCSN explosion, newly synthesized heavy elements in the stellar interior are pushed outside, mainly along the polar direction.
Distribution of $^{56}$Ni is roughly along the polar regions, and we see from the radial velocity profiles that the matter mixing works. We also successfully obtain high velocity $^{56}$Ni clumps.

The results of the equatorially asymmetric explosion model are shown in Figure 15. Due to the asymmetric explosion across the equatorial plane, the density distribution and the distribution of $^{56}$Ni are both equatorially asymmetric. The matter mixing is effective and high velocity $^{56}$Ni clumps are obtained.

We also obtain the results of the combination model from the bipolar explosion and equatorial asymmetric explosion. In this global asymmetric case, the matter mixing is successfully produced. However, we overestimate masses of high velocity $^{56}$Ni and $^{44}$Ti when we introduce 50\% of the density perturbation in the progenitor. In order to reproduce the proper high velocity $^{56}$Ni clumps, we reduce the density perturbation amplitude from 50\% to 25\%. The resulting mass and velocity profiles of $^{56}$Ni from the simulation model with the reduced perturbation amplitude are consistent with the observational ones of SN 1987A.
We also perform one simulation model of a global asymmetric case with 10\% density perturbation amplitude. We confirm that the obtained high velocity $^{56}$Ni clumps are not enough if only 10\% of the density perturbation is introduced in the progenitor.

We summarize our main results for the matter mixing in an aspherical CCSN explosion: (1) Either bipolar explosion or equatorially asymmetric explosion can effectively reproduce high velocity $^{56}$Ni clumps if we introduce 50\% of the density perturbation in the progenitor. (2) We cannot obtain enough high velocity $^{56}$Ni clumps from the simulation model with 10\% of the density perturbation; at least 25\% of the density perturbation is necessary to reproduce the observed matter mixing and high velocity $^{56}$Ni clumps. (3) The distribution of $^{44}$Ti follows that of $^{56}$Ni. (4) The total $^{56}$Ni mass and the mass of high velocity $^{56}$Ni produced in aspherical explosion models roughly fit the observational values. However, the explosion energy is still overestimated compared to the number given by \citet{shigeyama88} and \citet{shigeyama90}.

In order to examine the matter mixing of the CCSN aspherical explosion affected by the input energy, we can reduce the input energy
and re-perform simulation models. For example, we obtain our simulation results if the input energy is $2.0\times 10^{51}$ erg: the total $^{56}$Ni mass is $0.07M_\odot$, high velocity ($>3,000~\rm{km~s^{-1}}$) $^{56}$Ni mass is $6.5\%$ of the total $^{56}$Ni mass, and explosion energy is $1.43\times 10^{51}$ erg. The density distribution, distribution of $^{56}$Ni mass fraction, radial velocity profiles of several elements, and line of sight velocity profiles of $^{56}$Ni are shown in Figure 16. These results are consistent with the observational ones of SN 1987A. We regard this model (Asph+SC1p25m20$\ast\ast$ listed in Tables 1 and 2) as one of the most favorable simulation models in our work to explain the observational phenomena of the matter mixing in SN 1987A. The bipolar feature inferred from rotational CCSN explosion has been also explored by theoretical models \citep{nakamura14}.

\section{Discussion}
We introduce large density perturbations in the CCSN progenitor to investigate the matter mixing of the CCSN explosion. The consequent question is: What is the mechanism for the large density perturbation that occurs in stellar interiors? MLT is usually considered in the convection zones of the stellar interior \citep{kipp90}. However, MLT is a local and spherically symmetric treatment in stellar models. Non-local and anisotropic convective mixing theories have been developed \citep{xiong77,grossman93,xiong97,canuto98,deng06,li07,hotta14}, but those treatments focus on the evolution of main-sequences stars. In addition, it seems that the related overshooting in solar models and post-main-sequence star models \citep{tian09,li12,zhang12a,zhang12b} cannot strongly affect the matter mixing of CCSN explosion because the overshooting model is applied for the globally stable state of stars. However, the large density perturbation proposed in this paper is one requirement for the successful matter mixing of the CCSN explosion.
\citet{meakin06} noticed that the spherically symmetric features in stellar interior can be broken as density perturbation is induced by turbulence within stellar convection zones. This kind of perturbation is related to carbon and oxygen shell burning. Analytic explorations and 3D simulations of turbulent convection have been performed, where turbulent entrainment and turbulent kinetic energy have been considered \citep{meakin07a,arnett09}. About 10\% of density perturbation is obtained in convection zones as carbon and oxygen shell burning \citep{meakin07b}.
Recently, \citet{arnett11a,arnett11b} found that violent dynamic eruptions occur in convective zones with simultaneous C, O, and Si shell burning of a CCSN progenitor star via 3D simulations.
A realistic stellar model should provide aspherical properties from hydrodynamic instabilities and turbulence for CCSN explosion \citep{smith14}.
Moreover, it has been found recently that the evolution of matter mixing is different between an BSG progenitor model and an RSG progenitor model \citep{wong14}.
In our 2D hydrodynamic simulation models, although 10\% of density perturbation in the progenitor can produce RT fingers, 50\% the of density perturbation is required in order to reproduce high velocity $^{56}$Ni clumps. Our results provide additional clues to further analyze the turbulent convection of the CCSN explosion.

We have performed 2D hydrodynamic simulations to research of CCSN matter mixing in this work. \citet{kane00} pointed out that RT instability grows faster in 3D simulation models, and \citet{hammer10} reached a similar conclusion after comparing 2D and
 3Dsimulation results. As \citet{hammer10} explained, heavy element clumps have toroidal topology in 2D simulations and bubble-sphere topology in 3D simulations, so these heavy element clumps experience less drag force in 3D simulations than in 2D ones. Thus, heavy element clumps can retain higher velocities in 3D models. If these physical processes mentioned above can really happen in CCSN explosions, we expect that less density perturbation should be introduced in 3D simulation models to reproduce high velocity $^{56}$Ni clumps. However, it seems that \citet{jogg10b} did not find qualitative differences of matter mixing efficiency between 2D and 3D simulations. One possible reason is that the interaction of the small RT fingers may reduce the final growth of RT instability, even if RT instability grows faster at the beginning of a CCSN explosion in 3D simulations. We can use 3D simulations with different density perturbation amplitudes to further investigate this CCSN matter mixing issue in the future.

We notice that RT fingers do not show a prominent feature at the early stage of CCSN explosion.
\citet{couch13} proposed velocity perturbations in an angular direction at the Si/O interface of a progenitor. Their 3D simulations showed that this kind of perturbation can affect CCSN explosion.
If we consider the neutrino-driven convection, the CCSN shock will be
asymmetric around Fe/Si or Si/O boundaries. 
In fact, \citet{kif06} claimed that the initial global deformation of the shock can trigger
strong instability at the He/H boundary, which makes the global anisotropy of the inner ejecta.
We find that large density perturbations placed at the Si/O interface may only provide and/or enhance aspherical features at the beginning of the CCSN explosion, but this does not give any direct effects on the matter mixing at the later time of the CCSN explosion.
This discrepancy may be due to the application of different progenitors in different hydrodynamic simulation models.
In our simulations, RT fingers are clearly identified after the shock propagates though the C+O/He interface. Giant RT fingers are developed quickly after the forward shock passes the He/H interface, and finally dominate at the late stage of CCSN explosion. During this later explosion process, we see a few small RT fingers merging into a large one, which was also mentioned by other previous works \citep{arnett89,kif03,hammer10,ellinger12}.
Moreover, when we successfully reproduce high velocity $^{56}$Ni clumps, we clearly see the huge and prominent
RT fingers with the length scale of about $2 \times 10^{12}$ cm in these simulation models. However, the RT fingers with a length scale smaller than $1\times 10^{12}$ cm shown in some simulation models are
short and weak, and we cannot reproduce high velocity $^{56}$Ni clumps by these models.

The obtained total $^{56}$Ni mass from spherical explosion models in this paper tends to be overestimated compared with that of SN 1987A.
There are two possible reasons: First, if we reduce the input energy in spherical explosion models, the total $^{56}$Ni mass can be reduced; second, a limited nuclear reaction network is included in the FLASH code and some isotopes of $^{56}$Ni are neglected. Although a detailed discussion on the production of isotopes goes far beyond the scope of this paper, we attempt to examine the effect of neglecting isotopes on the production of $^{56}$Ni by using a large nuclear reaction network. We perform a 1D hydrodynamic simulation of CCSN explosion without any perturbation using the same progenitor model mentioned in Section 2.
The input energy is set to be $2.5\times10^{51}$ erg. Then, explosive nucleosynthesis is calculated as a post-process
using the large nuclear reaction network, including 464 nuclei. As a result, the following values
are obtained: $M(^{56}\rm{Ni})=0.14~{\it M_\odot}$, $M(^{57}\rm{Ni})=0.01~{\it M_\odot}$, and $M(^{58}\rm{Ni})=0.05~{\it M_\odot}$.
On the other hand, $M(^{56}\rm{Ni})$ obtained by 1D FLASH code is 0.21 $M_\odot$.
This indicates that the nuclear reaction network in the FLASH code may overestimate the mass of $^{56}$Ni in a CCSN explosion. 
We present the detailed calculation process using the large nuclear reaction network in the Appendix.

We also notice that the masses of high velocity $^{56}$Ni and $^{44}$Ti clumps could be affected by the dimensionality in this paper. It is declared that 2D axisymmetric hydrodynamic simulations have numerical problems due to the computational grid-geometry. \citet{kif06} and \citet{gaw10} mentioned that RT fingers may grow faster along the polar direction due to the discretization errors and the non-penetrating feature around the polar axis. Thus, the artificial axial jets are shown in axisymmetric simulation results. We have also observed these kinds of features in our simulation contours. We speculate that some extra high velocity clumps of $^{56}$Ni and $^{44}$Ti along the polar direction may be artificially produced by these axial jets.
The productions of total $^{56}$Ni and $^{44}$Ti may be also affected by this geometric effect. In order to examine the axial jet effect, we perform a simple test. We consider the computational domain with angular direction from 15 degree to 165 degree and calculate total $^{56}$Ni mass and high velocity $^{56}$Ni clumps. We call this case ``exclude15". For comparison, we also consider the computational domain with angular direction from 30 degree to 150 degree and calculate the same quantities again. We call this case ``exclude30". Two simulation models-the spherical explosion model ``SC1p50m20" and the bipolar explosion model ``Bipo+C1p50m20"-are chosen. Line of sight velocity profiles and radial velocity profiles are shown in Figure 17. We see that the axial jet effect may produce artificial high velocity $^{56}$Ni clumps. Bipolar aspherical explosion models may suffer from this effect more seriously than spherical explosion models. The related quantities are as follows: for the case of ``exclude15" and ``SC1p50m20", total $^{56}$Ni mass is 0.16 $M_\odot$ and high velocity ($>3,000~\rm{km~s^{-1}}$) $^{56}$Ni clumps have 5.5\% of the total $^{56}$Ni mass; for the case of ``exclude30" and ``SC1p50m20", total $^{56}$Ni mass is 0.15 $M_\odot$ and high velocity ($>3,000~\rm{km~s^{-1}}$) $^{56}$Ni clumps have 5.1\% of the total $^{56}$Ni mass; for the case of ``exclude15" and ``Bipo+C1p50m20", total $^{56}$Ni mass is 0.08 $M_\odot$ and high velocity ($>3,000~\rm{km~s^{-1}}$) $^{56}$Ni clumps have 8.5\% of the total $^{56}$Ni mass; for the case of ``exclude30" and ``Bipo+C1p50m20", total $^{56}$Ni mass is 0.06 $M_\odot$ and high velocity ($>3,000~\rm{km~s^{-1}}$) $^{56}$Ni clumps have 0.9\% of the total $^{56}$Ni mass. As a summary, the axial jet effect may only reduce a few percent of high velocity $^{56}$Ni clumps in a spherical explosion model, but it may reduce at least 20\% of high velocity $^{56}$Ni clumps in a bipolar explosion model. 3D simulations are necessary for further quantitative analysis.

The observed line profiles of [Fe II] in SN 1987A are not symmetric and their peak is redshifted \citep{haas90}.
The inner ring of SN 1987A is inclined at about $45^\circ$ to the sky: the northern part of the ring is close to observers and the southern part of the ring is away from observers \citep{tziamtzis11}. The inner ejecta of SN 1987A is elongated and it is roughly in the same plane of the inner ring \citep{kjar10}.
We calculate the line of sight velocity profiles of $^{56}$Ni to different observational view angles. We define positive velocity and negative velocity in the line of sight velocity profiles as the redshifted velocity and blueshifted velocity, respectively. From our analysis of density contours and line of sight velocity profiles of $^{56}$Ni,
it looks like that the profile with observational view angle of $45^\circ$ or $135^\circ$ is morphologically close to the observational one. In particular, the peak shift is dominated by the ratio of $v_{\rm{up}}$ and $v_{\rm{down}}$ when we consider equatorially asymmetric explosion models. A large value of the ratio leads to a strong peak shift. On the other hand, the density perturbation also affects the peak shift in equatorially asymmetric explosion models. We find that the large density perturbation tends to weaken the peak shift. Some fine tuning in simulation models is required to reproduce exactly the observational redshifted peak of line profile, although this is not the main scope in this paper.
The results of our 2D simulations are helpful to understand the morphology of SN 1987A, but it is obvious that 3D simulations, which we plan to perform in the future, are more suitable to explain the reconstructed 3D features of SN 1987A \citep{kjar10,larsson13}.

Direct detections of emission lines from the decay of $^{44}$Ti in two young supernova remnants have been reported.
The mass of $^{44}$Ti in SN 1987A was derived as $(3.1\pm 0.8)\times 10^{-4}~M_\odot$ \citep{grebenev12} and the mass of $^{44}$Ti in Cassiopeia A is about $1.6^{+0.6}_{-0.3}\times 10^{-4}~M_\odot$ \citep{gre14}. We also obtain the mass
of $^{44}$Ti in our simulation models. Our values obtained from the spherical explosion models are roughly comparable to the observational one, while the values obtained from the aspherical explosion models are larger than the observational one. Moreover, the ratios between the $^{56}$Ni mass and $^{44}$Ti mass in spherical explosion models are larger than those of aspherical models.
If we overestimate the mass of $^{56}$Ni in the spherical explosion models, the process of an $\alpha$-rich freeze-out may enhance the production of $^{44}$Ti \citep{woosley91,nagataki00}. The so-called mass-cut, which arbitrarily determines the location of the ejecta from the compact remnant may affect the mass of $^{44}$Ti. However, it seems in Paper I that the differences of velocity profiles between the results with mass-cut and those without mass-cut are not significant. We also note that the overestimation of both $^{56}$Ni and $^{44}$Ti may be induced by the small nuclear reaction network in the FLASH code. Compared to the productions from spherical explosion models, it seems that those obtained from aspherical explosion models are more consistent with observational results. We confirm the speculation of Paper I that $^{44}$Ti may be an indicator of explosion asphericity. However, we also realize that the production of $^{44}$Ti and $^{56}$Ni in the CCSN explosions is related to not only temperature and density, but also the electron fraction and detailed thermodynamic trajectories \citep{mag10}. Postprocess calculations of the nucleosynthesis with a large nuclear reaction network are required. These complicated physical processes make some additional uncertainties to the ratio between $^{44}$Ti and $^{56}$Ni.

We can calculate radial velocities not only for the elements of H, $^{56}$Ni, and $^{44}$Ti, but also for the elements of $^{4}$He, $^{12}$C, $^{16}$O, and $^{28}$Si. The radial velocities of the elements in each simulation model are summarized in Table 3. If the simulation models of the spherical explosion with large density perturbation (50\% amplitude) can produce high velocity $^{56}$Ni and $^{44}$Ti, the radial velocity profiles of $^{12}$C, $^{16}$O, and $^{28}$Si also extend to a large velocity part. On the other hand, the simulation models of a spherical explosion with small density perturbation (10\% amplitude) cannot produce high velocity $^{56}$Ni and $^{44}$Ti, and it is also difficult to obtain more high velocity heavy elements of $^{12}$C, $^{16}$O, and $^{28}$Si.
In order to examine the effect of large density perturbation on the distributions of $^4$He, $^{12}$C, $^{16}$O, and $^{28}$Si, we plot the line of sight velocity profiles of these elements in Figure 18 (view angle 90 degree) and Figure 19 (view angle 45 degree). Three simulation models (SC1p50m20, SC1p10m20, and Asp+C1p50m20($\ast$)) are considered. We see that $^{12}$C, $^{16}$O, and $^{28}$Si are extended into the high velocity region in the spherical explosion model with 50\% of density perturbation. Aspherical explosion with the view angle of 90 degree takes effects on the distribution of $^{12}$C, $^{16}$O, and $^{28}$Si, so that these elements are concentrated in the center of line of sight velocity profiles. From the observational point of view, \citet{larsson13} have obtained [Si I]+[Fe II] line profiles of SN 1987A.
The [Si I]+[Fe II] emission is close to the plane of the ring and
the profile is north-south asymmetry that is partially due to dust absorption. The emission is concentrated
in the velocity interval of $1500-3000~\rm{km~s^{-1}}$.
\citet{tanaka12} have measured the polarization features of [He I], [O I], [Ca II], and [Fe II] in type Ib/Ic SN 2009jf and SN 2009mi. In principle, this kind of spectral observation provides the opportunity for us to extract the velocity profiles of these elements. We expect that those velocity profiles calculated from our simulation models can be further examined by future observations and consequently provide additional hints to explore clumpy distributions and multi-dimensional geometries of CCSN explosions.

\section{Summary}
We perform 2D hydrodynamic simulations of a CCSN to explain the high velocity $^{56}$Ni clumps of SN 1987A. The density perturbation with 50\% amplitude is introduced in the progenitor model, such that the matter mixing can be reproduced in spherical explosion models. We take this large density perturbation to be one possible way to explain the final aspherical feature of a CCSN explosion indicated by the distributions of $^{56}$Ni and $^{44}$Ti. On the other hand, simulation models with 10\% of density perturbation cannot reproduce high velocity $^{56}$Ni clumps.
The RT instability is efficient at both the C+O/He interface and He/H interface to reproduce high velocity $^{56}$Ni clumps. The radial coherent structures shown in perturbation patterns are important to produce the matter mixing.

When we consider the case of aspherical explosion, at least 25\% of density perturbation introduced in the progenitor is necessary to reproduce the high velocity $^{56}$Ni clumps of SN 1987A. Either bipolar explosion or equatorially asymmetric explosion is effective for the explanation of observational matter mixing properties.

Explosion energy, total $^{56}$Ni mass, total $^{44}$Ti mass, and high velocity ($>3,000~\rm{km~s^{-1}}$) $^{56}$Ni mass have been obtained in each CCSN simulation model. The distribution of $^{44}$Ti follows that of $^{56}$Ni in all simulations models. After the comparison of our simulation results to the observations of SN 1987A, we note that one of the most favorable simulation models in this work is an aspherical one with the combination of bipolar explosion and equatorially asymmetric explosion, in which 25\% of density perturbation is introduced at different composition interfaces of the CCSN progenitor.

As noted in Section 4, we have performed 2D simulations for the research of the matter mixing. However, 3D simulations may result in different matter mixing compared with 2D ones \citep{hammer10}. If we perform 3D simulations using the same progenitor, we expect that less density perturbations will be introduced to make the same matter mixing of SN 1987A. To obtain a more robust conclusion, we will perform 3D simulations for the matter mixing of SN 1987A in the future.

\acknowledgments
We are grateful to Dr. A., Wongwathanarat and T. Takiwaki for their useful discussion. The software used in this work was in part developed by the DOE NNSA-ASC OASCR Flash Center at the University of Chicago. Numerical computation in this work was carried out on SR16000 at the Yukawa Institute Computer Facility. Mao, J. is supported by the Hundred-Talent Program (Chinese Academy of Sciences). This work is also supported by the Ministry of Education, Culture, Sports, Science and Technology (MEXT: No. 23105709 and No. 26105521) and the Japan Society for the Promotion of Science (JSPS: No. 19104006, No. 23340069, No. 2503786, No. 25610056, No. 26287056, No. 26800159, No. 26800141 and No. 24.02022).

\appendix
\section{Production of $^{56}\rm{Ni}$ and $^{44}\rm{Ti}$ with a Large Nuclear Reaction Network}
CCSN explosive nucleosynthesis has been comprehensively studied \citep{woosley73,hashimoto95} and applied for SN 1987A \citep{hashimoto89,woosley91}.
In this paper, we perform 2D hydrodynamic simulations using the FLASH code to study the matter mixing of CCSN explosion. As noted in \citet{nagataki97}, some isotopes of $^{56}$Ni should be included in the explosive nucleosynthesis. In
order to clarify the effect of the small nuclear reaction network in the FLASH code to the production of $^{56}$Ni and $^{44}$Ti, we calculate
the explosive nucleosynthesis using a large nuclear reaction network and a fiducial pure spherical explosion
model.

We first perform a 1D hydrodynamic simulation of CCSN explosion using the same progenitor model and methods. The input
energy of $2.5\times 10^{51}$ erg is adopted for a fiducial case. To calculate explosive nucleosynthesis as a post-process, we adopt the so-called tracer particle method (Nagataki et al. 1997): 2,000 particles are distributed on the computational domain of the hydrodynamic simulation
($1.36\times 10^8~\rm{cm}<{\it r}<6\times 10^9~\rm{cm}$, which covers up to the oxygen-rich layer),
and mass and composition are assigned to each particle to reproduce the mass distribution and
the composition of the progenitor model. Then, by using the results from the time evolution of
the velocity field in Eulerian grids, particles are passively moved to obtain the time
evolution of density and temperature.
Using the initial composition and
the time evolution of density and temperature, explosive nucleosynthesis is calculated
for each ejected particle. We adopt a large nuclear reaction network including 464 nuclei
(up to $^{94}$Kr), which was used in Ono et al. (2009). It should be noted that the innermost
region ($r=1.36\times 10^8$ cm) is slightly neutron-rich (electron fraction $Y_e\sim 0.485$) and the production of neutron-rich isotopes, $^{57}$Ni and $^{58}$Ni, is expected.
As a result, we obtain the following values:
$M(^{56}\rm{Ni})=1.4\times 10^{-1}~{\it M_\odot}$,
$M(^{57}\rm{Ni})=1.0\times 10^{-2}~{\it M_\odot}$,
$M(^{58}\rm{Ni})=5.4\times 10^{-2}~{\it M_\odot}$, and
$M(^{44}\rm{Ti})=1.0\times 10^{-4}~{\it M_\odot}$.
The masses of $^{56}$Ni and $^{44}$Ti estimated
by the FLASH code are $2.1\times 10^{-1}~M_\odot$ and $3.0\times 10^{-4}~M_\odot$
respectively.
Thus, the masses of $^{56}$Ni and $^{44}$Ti calculated by the FLASH code are overestimated by a factor of 0.7 and a factor of 3, respectively, compared with those calculated by the large nuclear reaction network.
Because the amount of $^{44}$Ti is much less than that of $^{56}$Ni, the overestimation of $^{44}$Ti due to
neglecting other elements is more severe than that of $^{56}$Ni.

\clearpage

\begin{figure}
\includegraphics[scale=1.1]{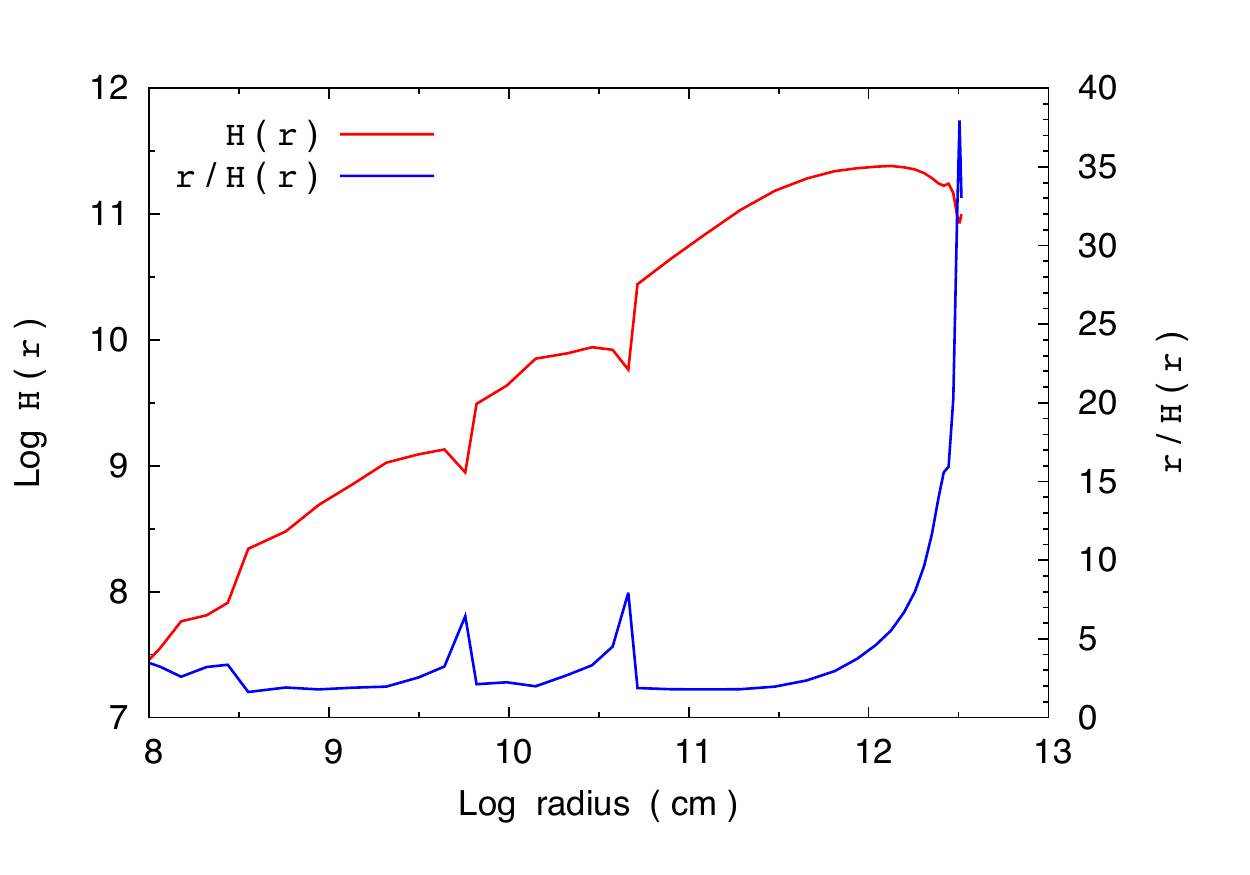}
\includegraphics[scale=1.1]{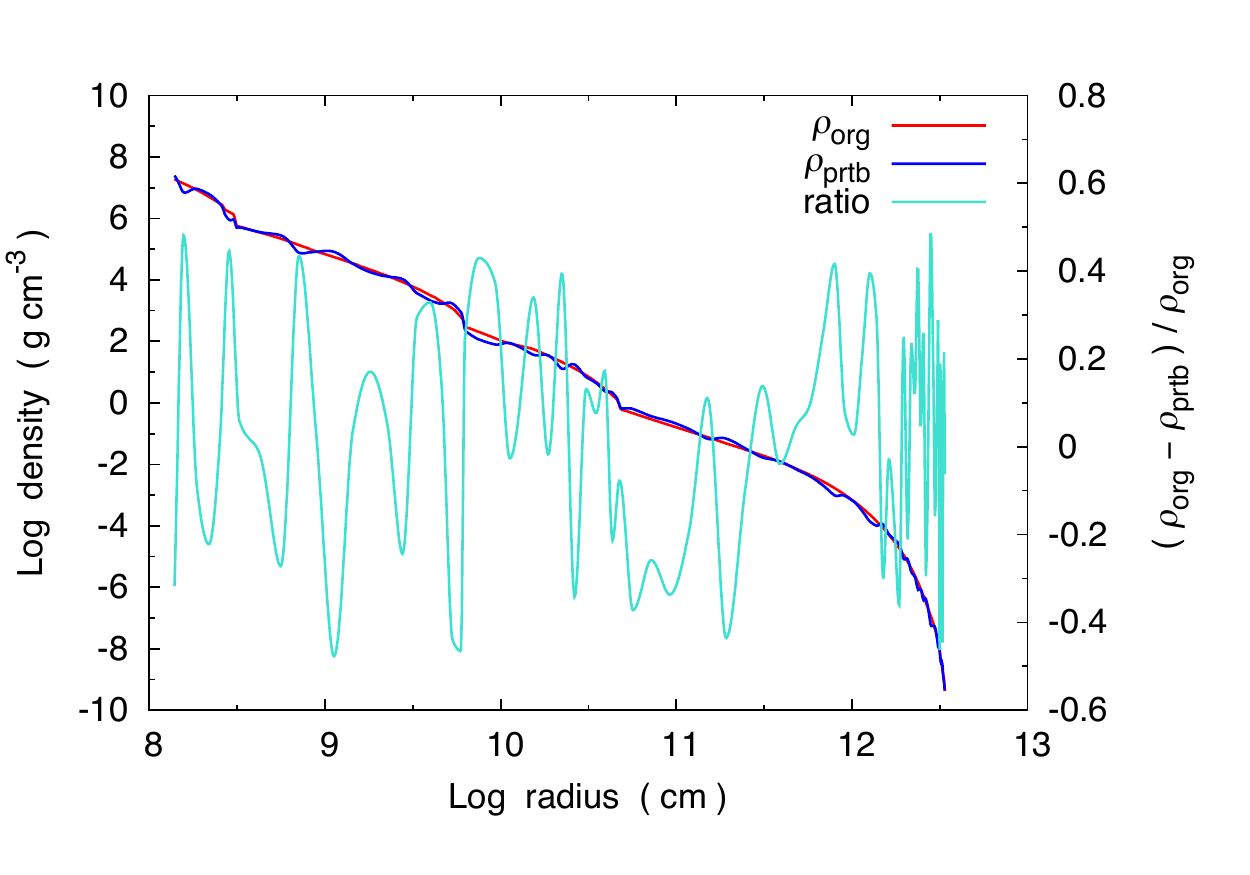}
\caption{\footnotesize{Upper panel: density scale height $H(r)$ and $r/H(r)$ of the progenitor model as a function of radius;
lower panel:
profiles of density and density perturbation in the progenitor model. Distributions of original density, perturbative density, and perturbation amplitude are denoted by red, blue, and cyan lines, respectively. Perturbation amplitude is randomly distributed in the radial direction and the maximum value is 50\%. We take this perturbation mode from the case of SC3 ($\epsilon_0=0.5$ and $f=1$).}
\label{fig1}}
\end{figure}

\begin{figure}
\includegraphics[scale=0.18]{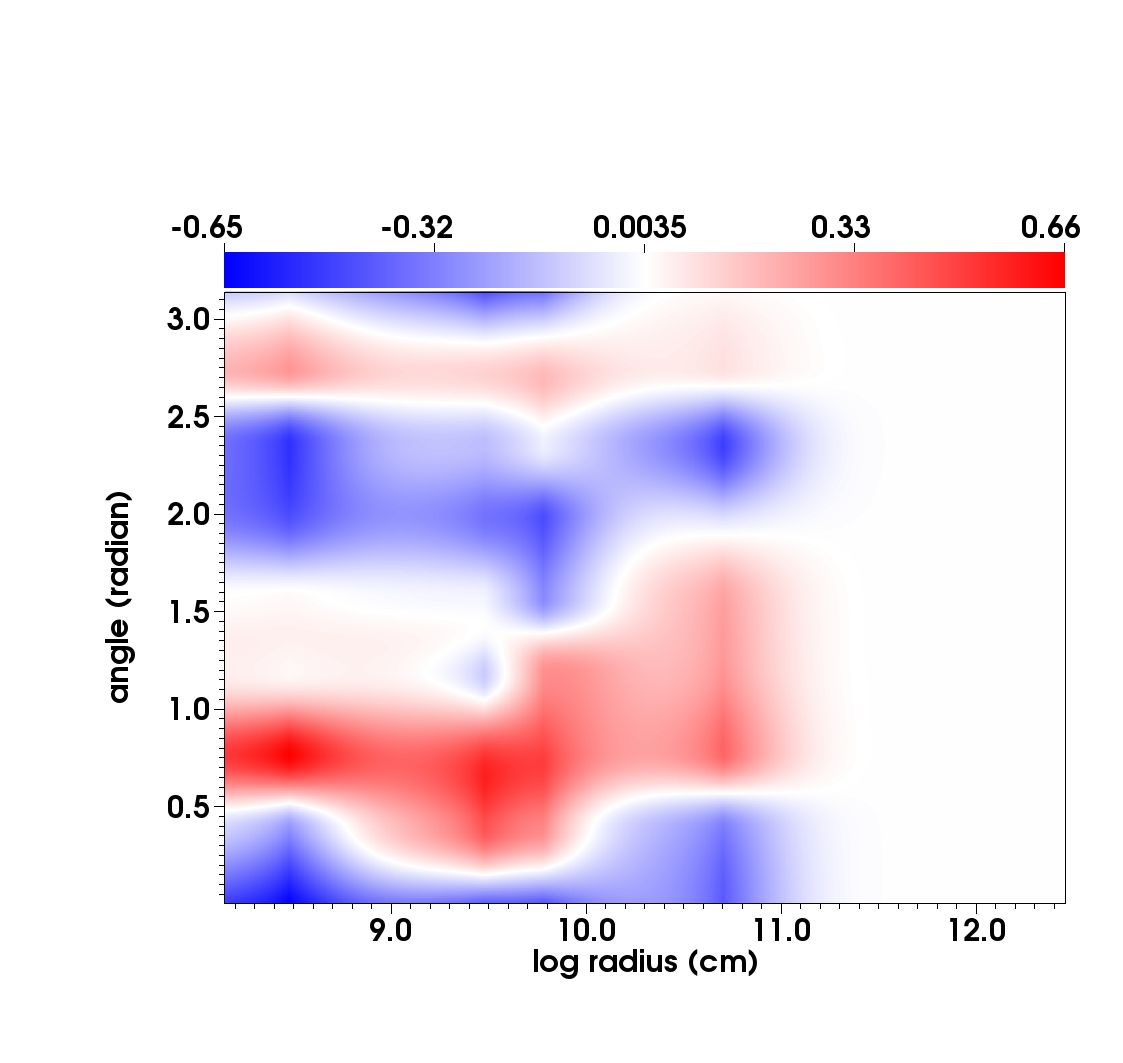}
\includegraphics[scale=0.18]{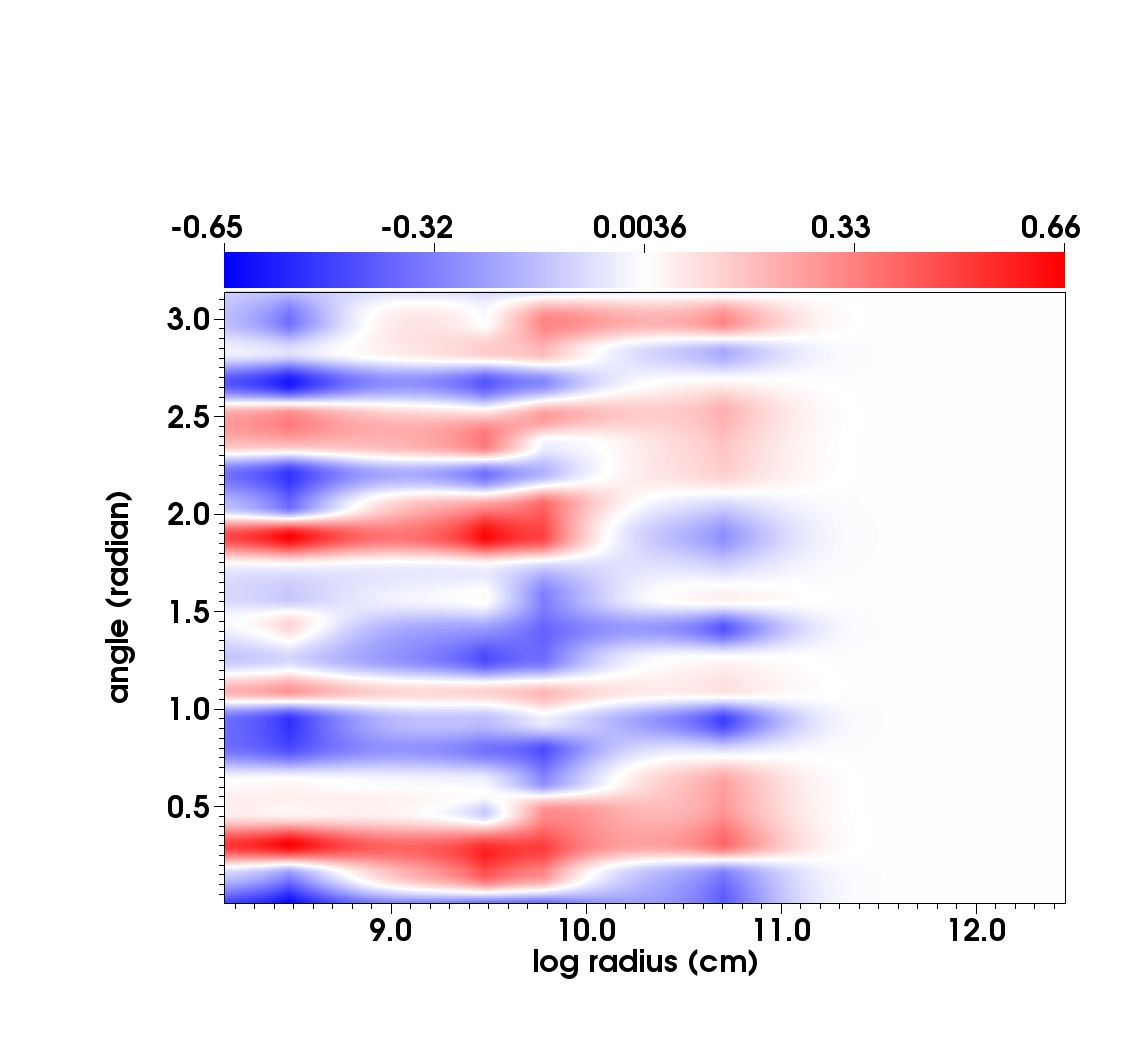}\\
\includegraphics[scale=0.18]{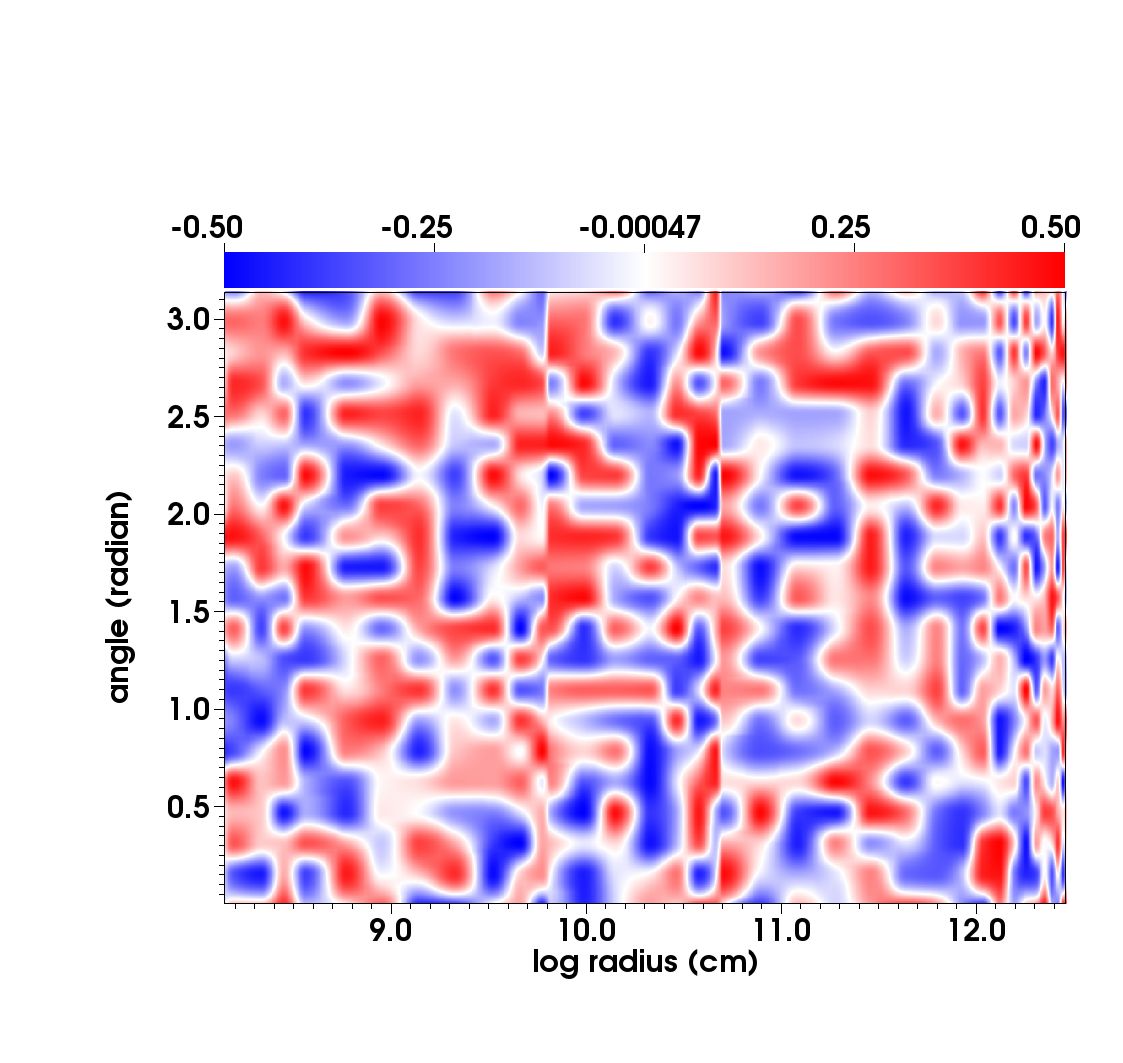}
\includegraphics[scale=0.18]{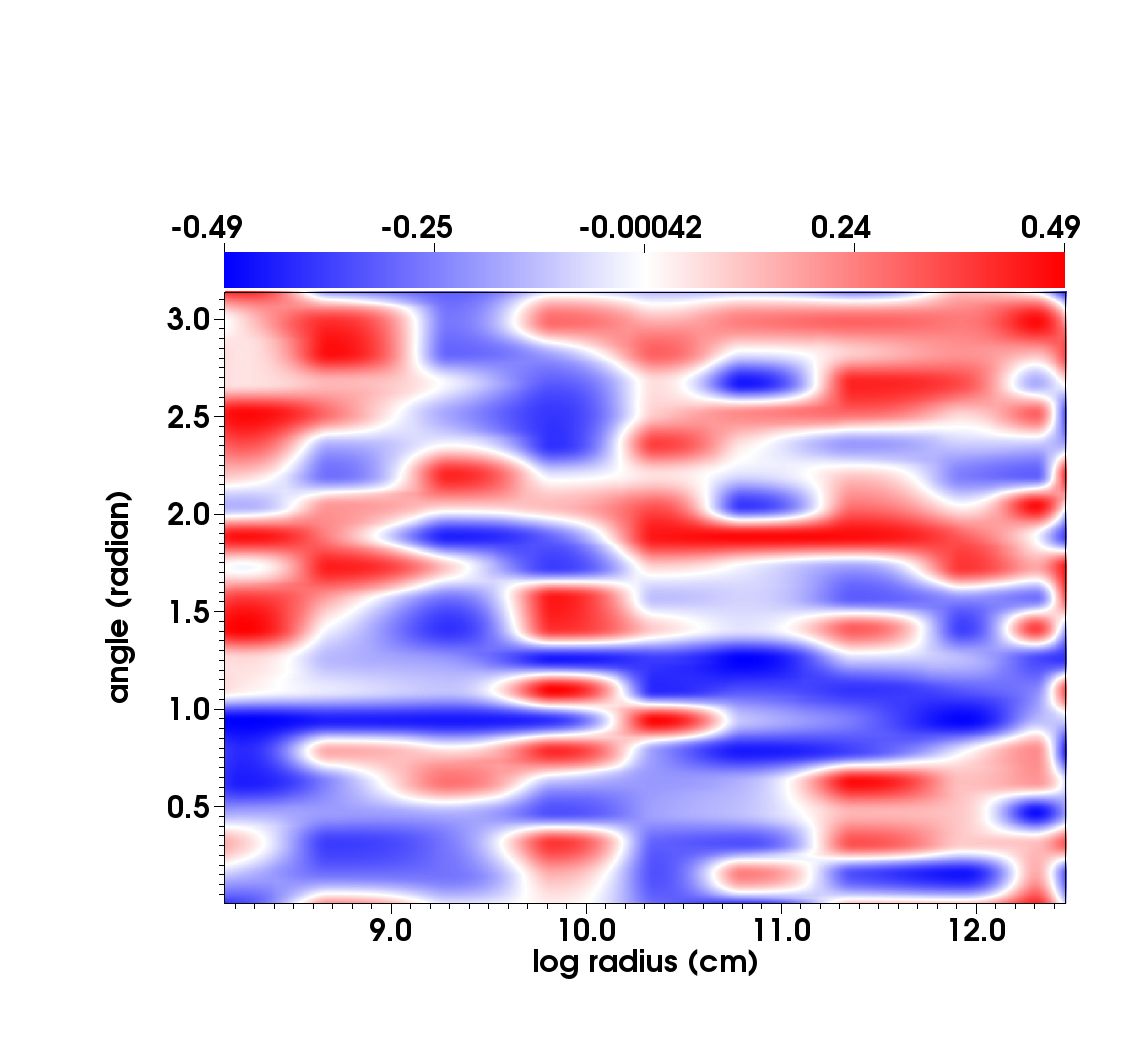}\\
\includegraphics[scale=0.18]{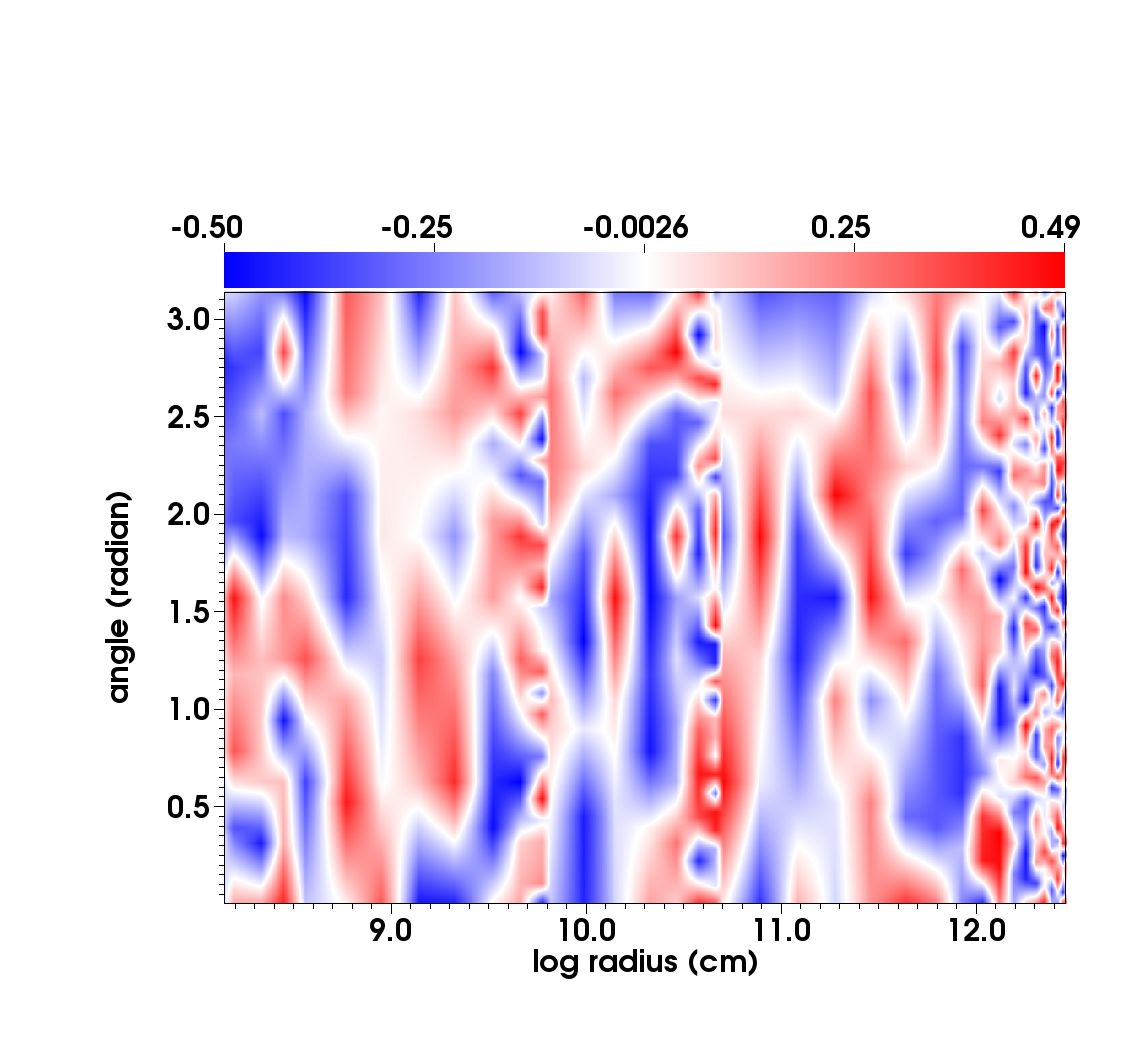}
\includegraphics[scale=0.18]{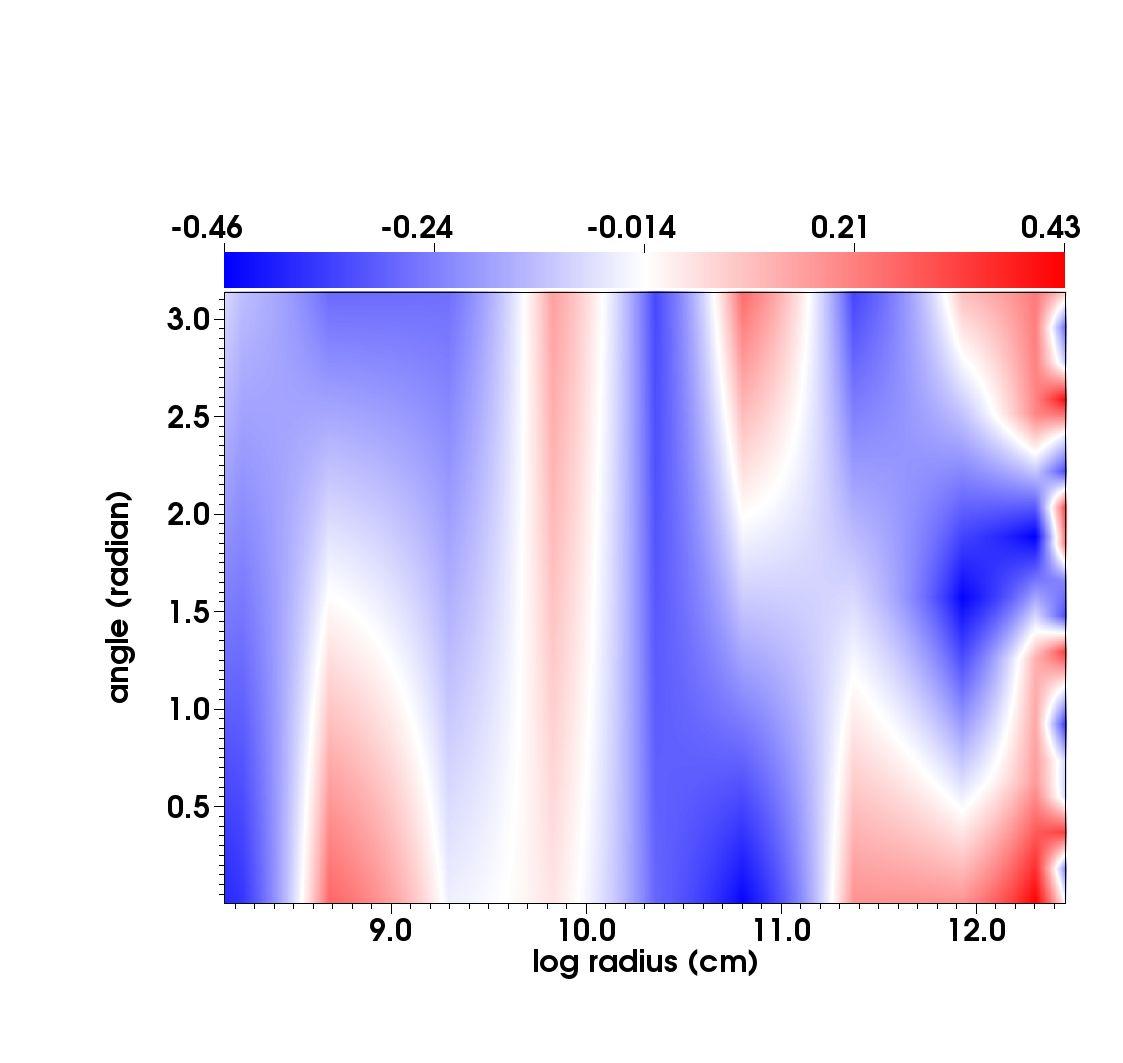}
\caption{\scriptsize{Distributions of density perturbation in the $r$-$\theta$ plane. The ranges of $r$ and $\theta$ are $1.4\times 10^8~\rm{cm}<{\it r}<3.4\times 10^{12}~\rm{cm}$ and $0<\theta<\pi$, respectively, and the color bar in the top of each panel
denotes the amplitude range of perturbation.
upper left: perturbation mode of SC1 (m=8); Upper right: perturbation mode
of SC1 (m=20); middle left: perturbation mode of SC2 ($m=20$ and $f=1$); middle right: perturbation mode of SC2 ($m=20$ and $f=5$);
lower left: perturbation mode of SC3 ($m=20$ and $f=1$); lower right: perturbation mode of SC3 ($m=20$ and $f=5$). We note four critical radii:
$3\times 10^8$ cm at Si/C+O interface, $3\times 10^9$ cm at oxygen shell burning position, $6\times 10^9$ cm at C+O/He interface, and $5\times 10^{10}$ cm at He/H interface. Only the models with coherent-structured perturbation in the radial direction can reproduce high velocity clumps of $^{56}$Ni (see upper left, upper right, and middle right panels).}
\label{fig2}}
\end{figure}

\begin{figure}
\includegraphics[scale=0.25]{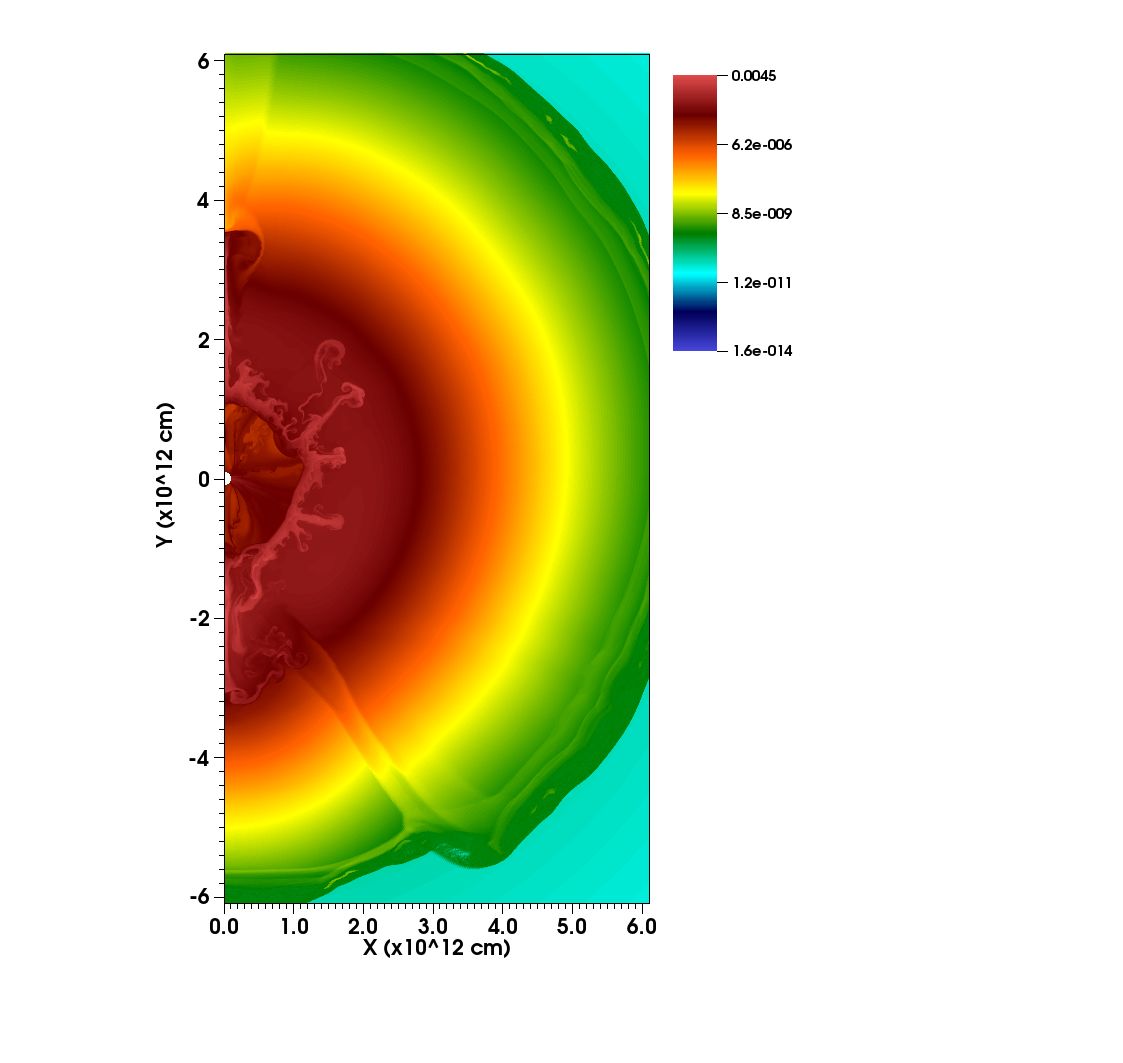}
\includegraphics[scale=0.25]{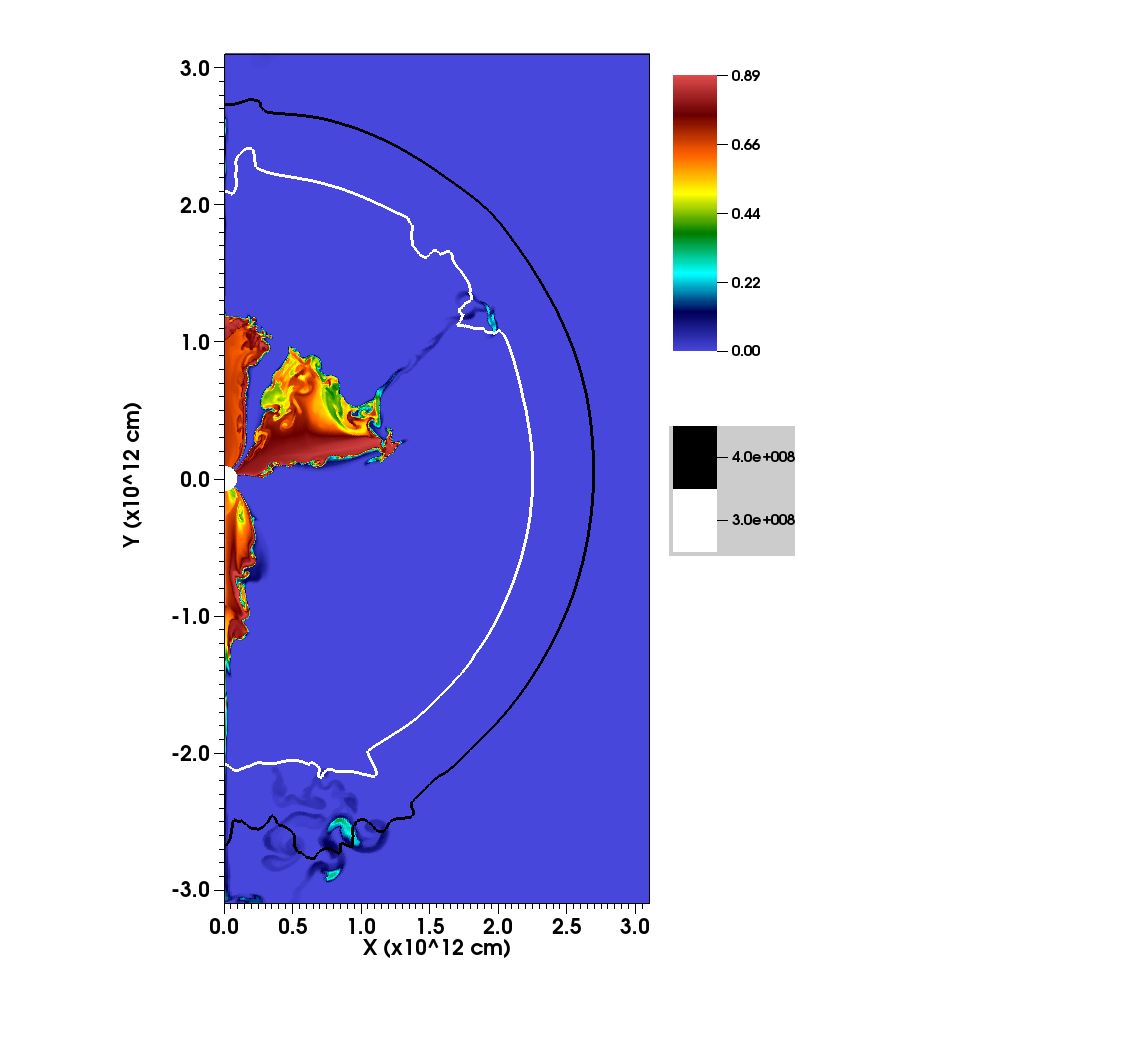}
\includegraphics[scale=0.7]{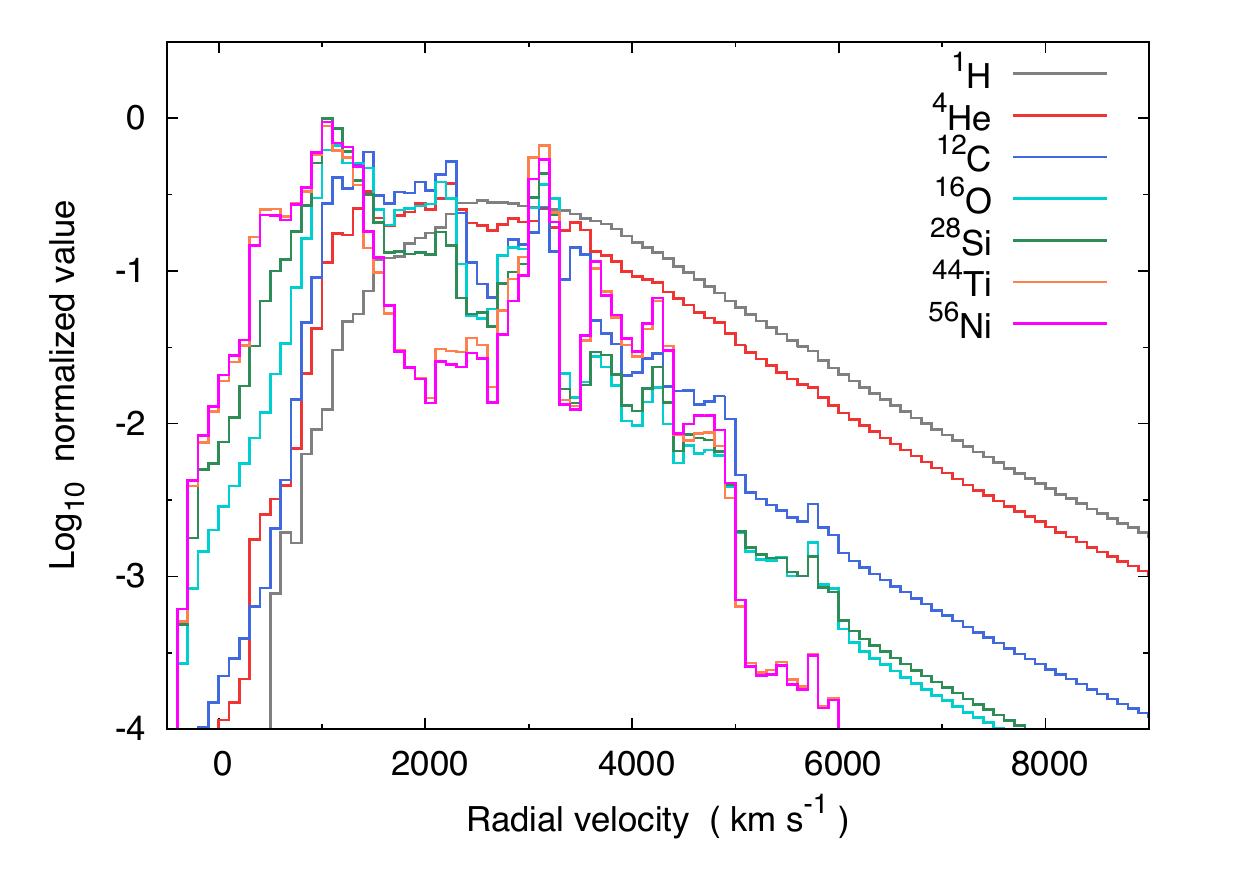}
\includegraphics[scale=0.7]{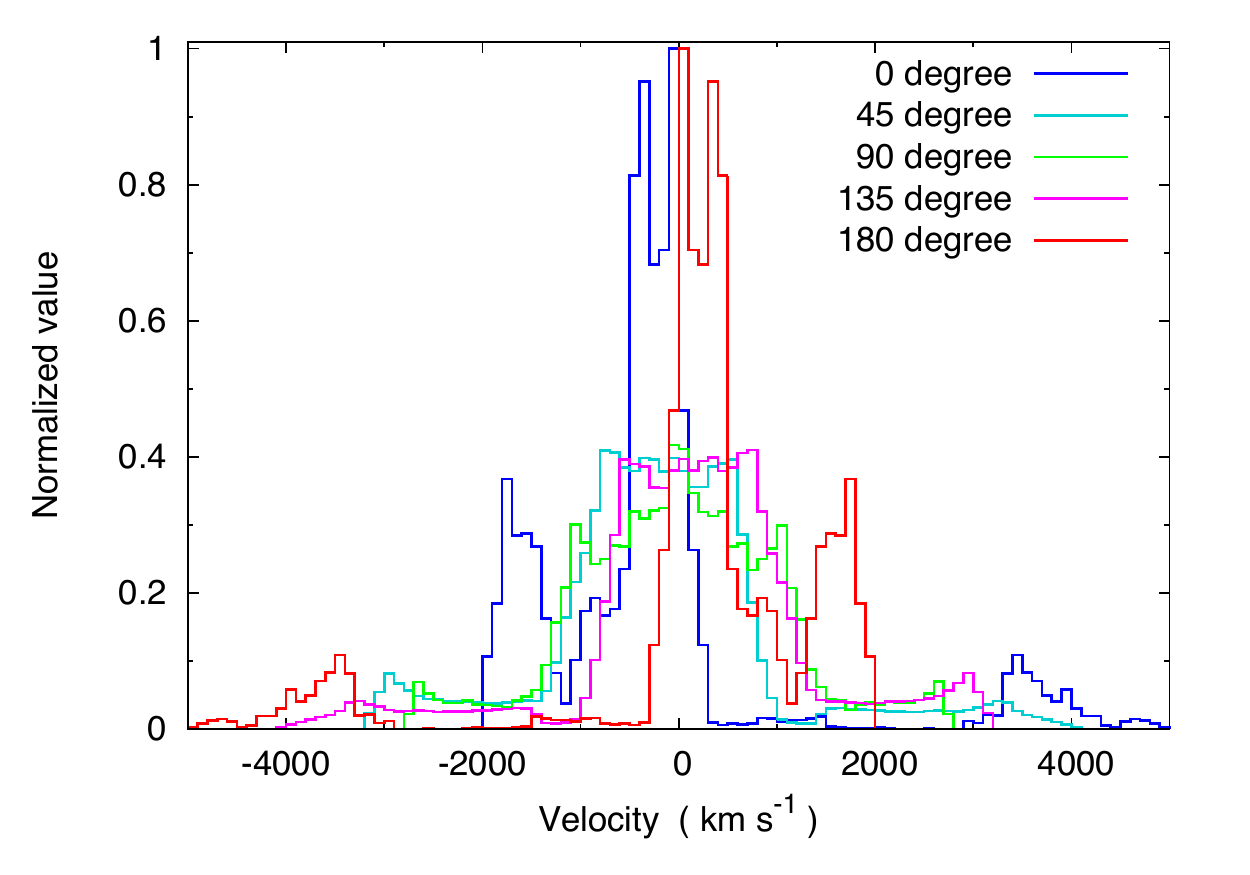}
\caption{Results obtained from the model SC1p50m8. Upper left: final density distribution; upper right: final distribution of $^{56}$Ni mass fraction. The white and black lines indicate the contours of radial velocities of 3,000 $\rm{km~s^{-1}}$ and 4,000 $\rm{km~s^{-1}}$, respectively; Lower left: radial velocity profile; lower right: line of sight velocity profile of $^{56}$Ni with different observational view angles. For each lower panel, the velocity bin is 100 $\rm{km~s^{-1}}$, numbers of $\Delta M/M$ are normalized by their maximum value, where $\Delta M$ is the mass in the velocity range of $v\sim v+\Delta v$; and $M$ is the total mass.
\label{fig3}}
\end{figure}

\begin{figure}
\includegraphics[scale=0.25]{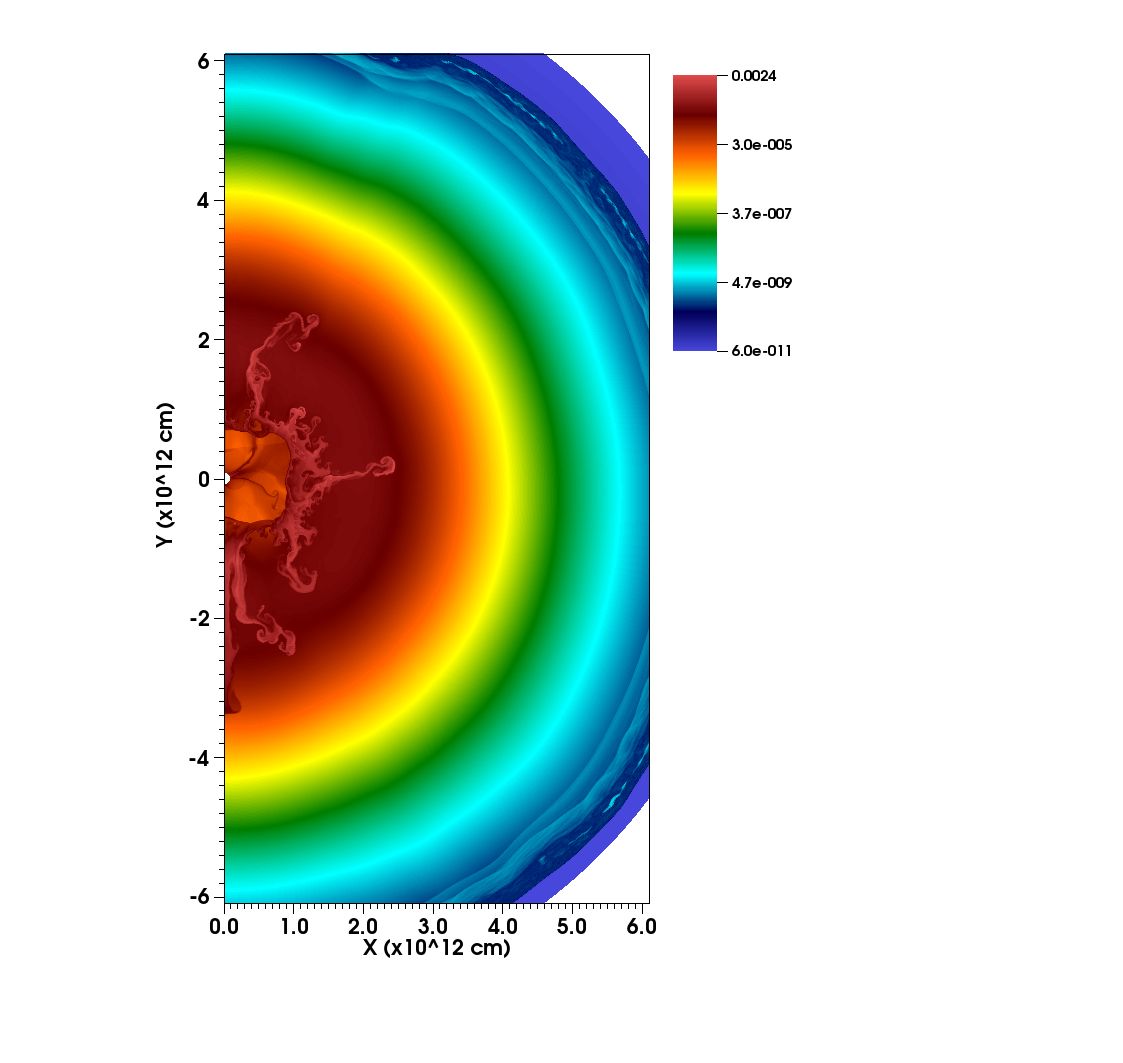}
\includegraphics[scale=0.25]{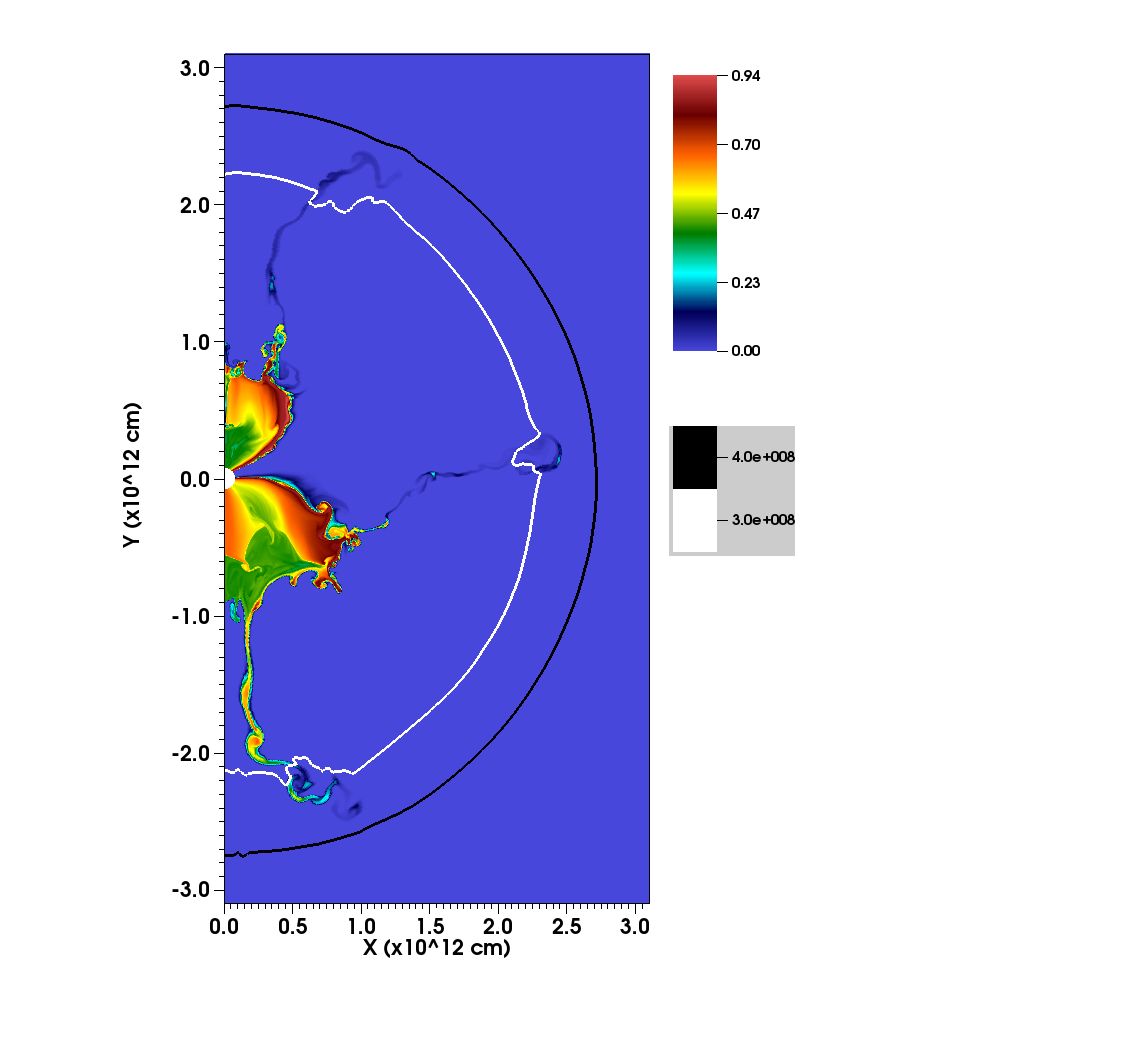}
\includegraphics[scale=0.7]{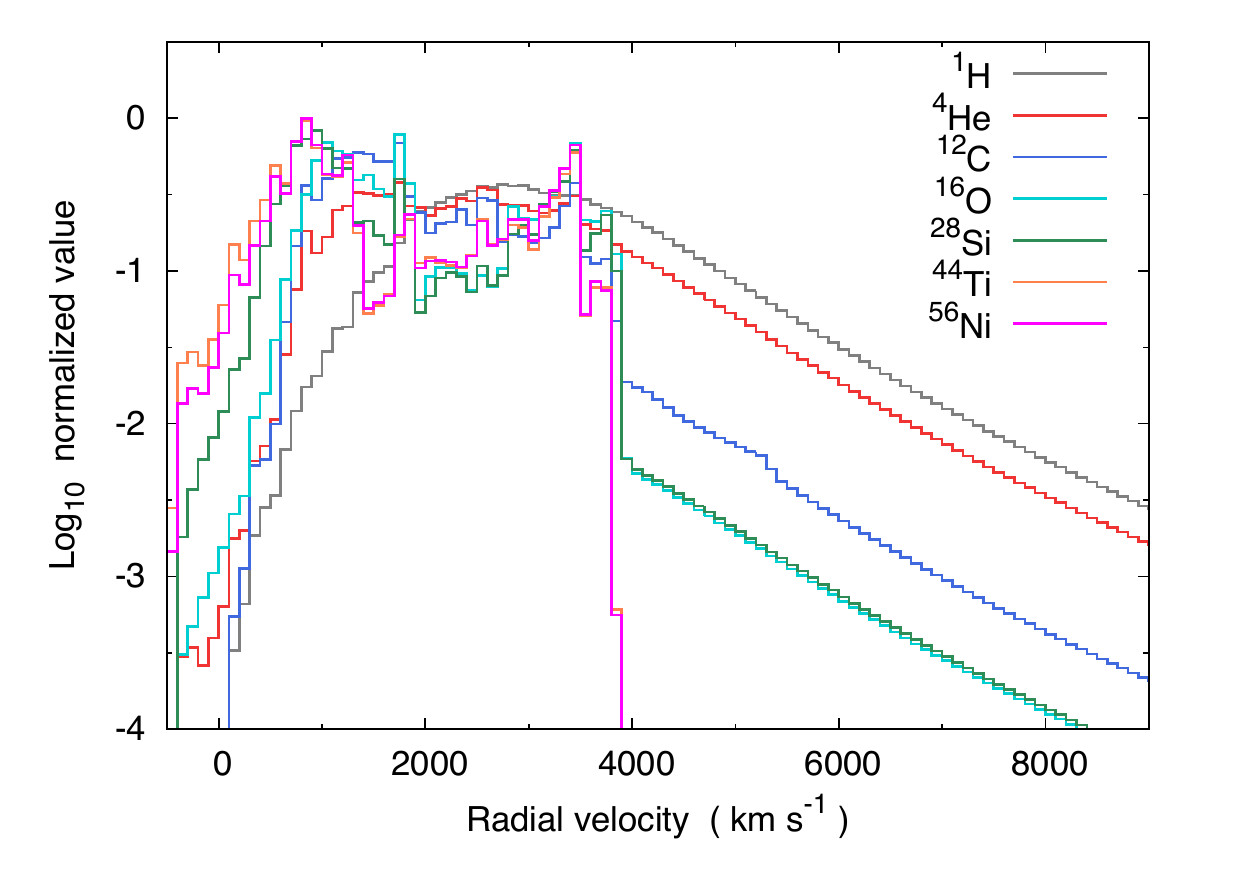}
\includegraphics[scale=0.7]{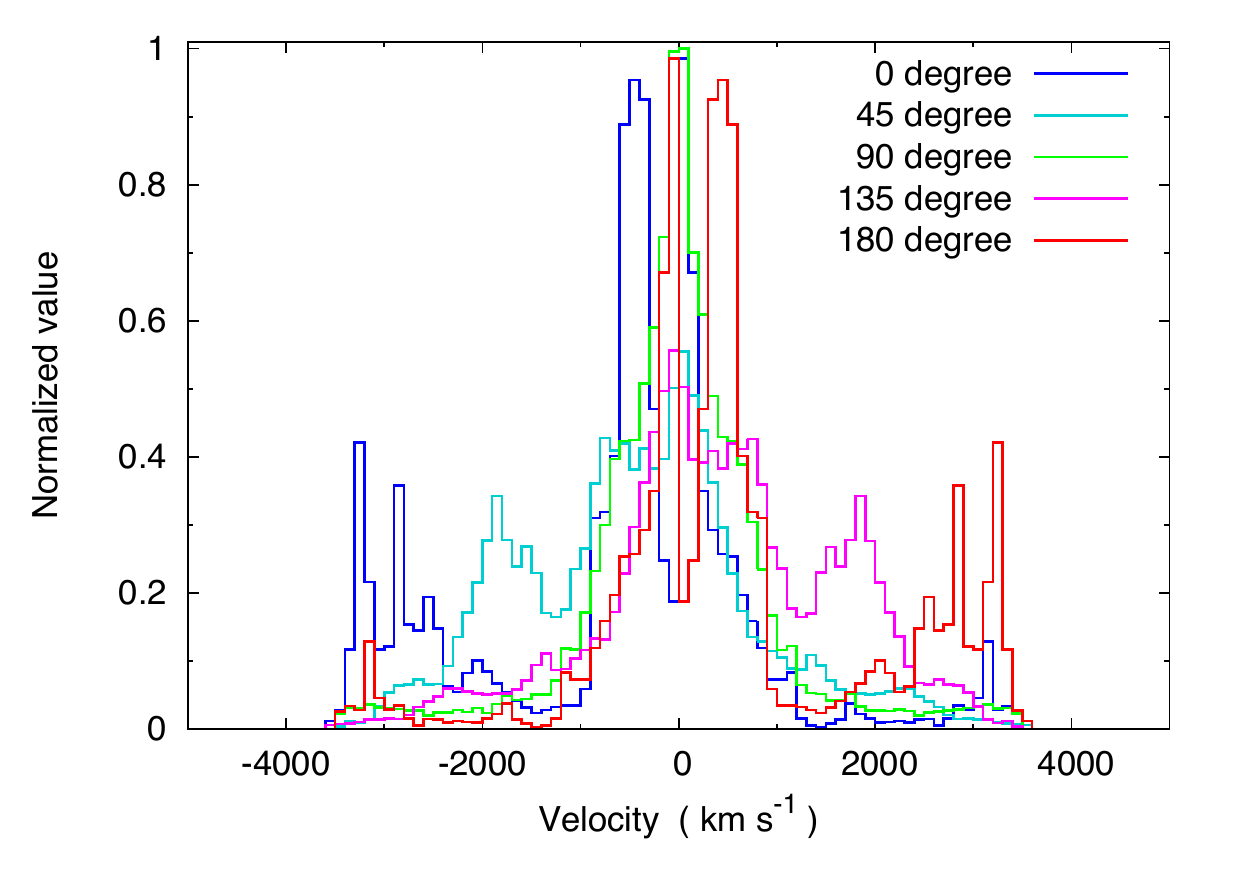}
\caption{Results obtained from the model SC1p50m8 with a different set of random numbers. Panel descriptions are the same as those in Figure 3.
\label{fig4}}
\end{figure}


\begin{figure}
\includegraphics[scale=0.25]{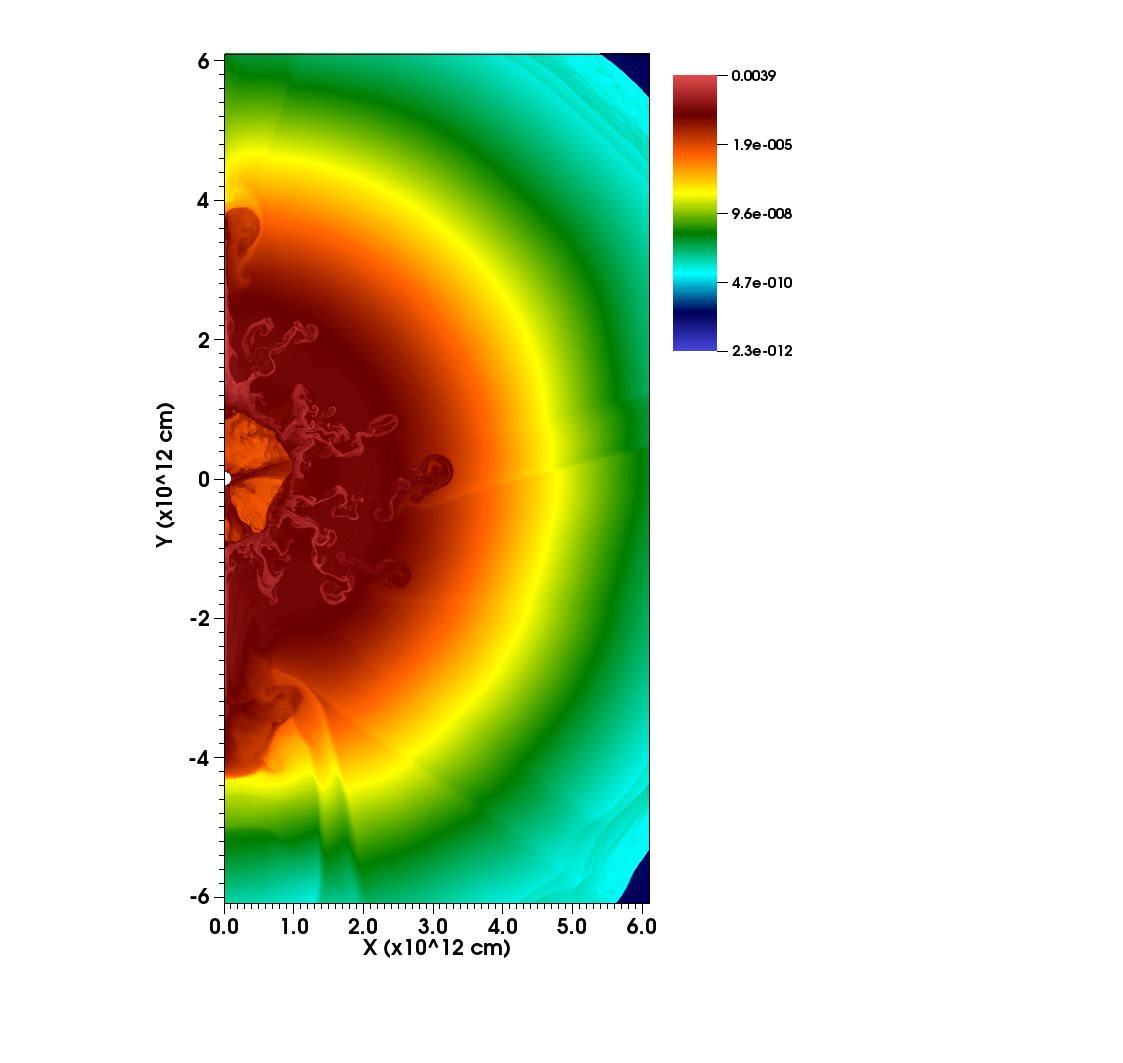}
\includegraphics[scale=0.25]{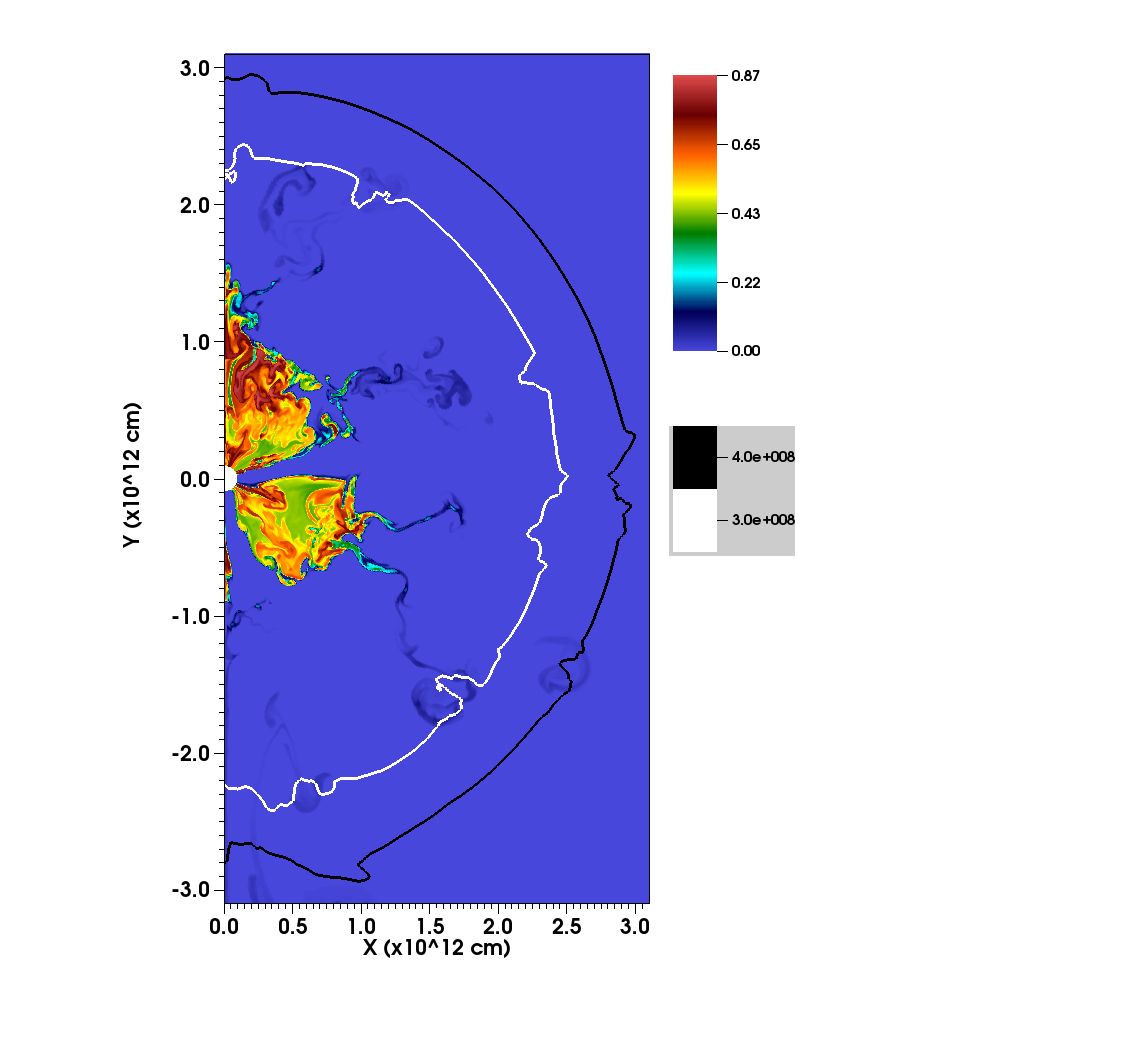}
\includegraphics[scale=0.7]{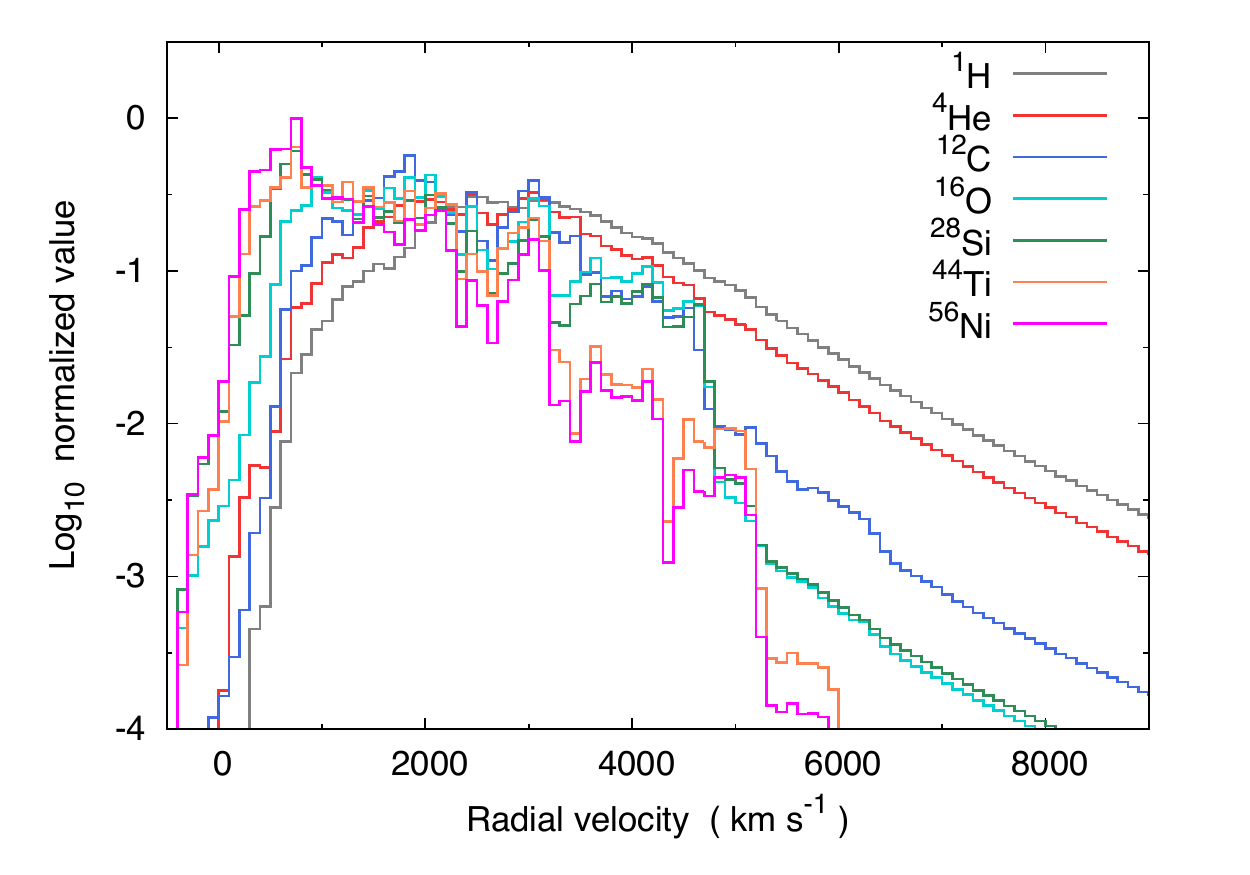}
\includegraphics[scale=0.7]{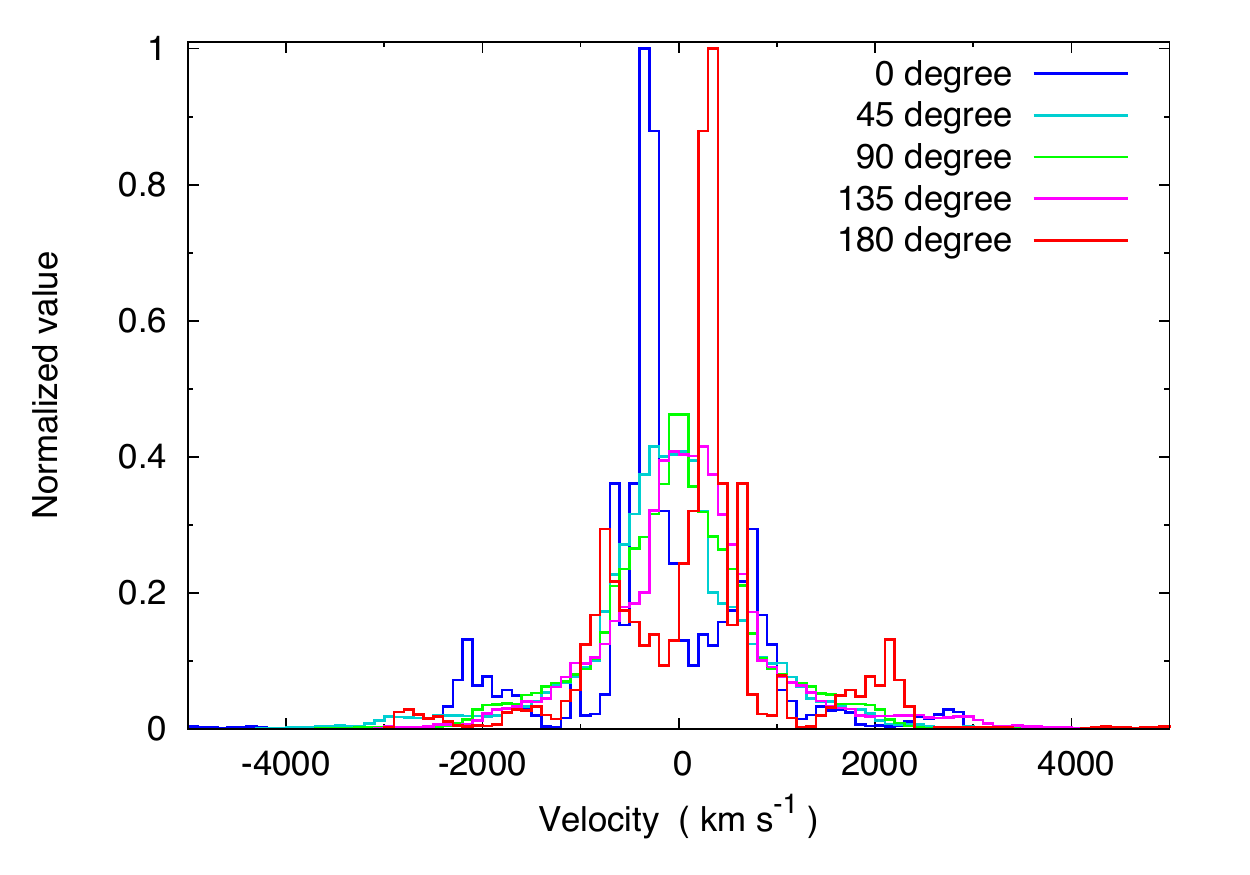}
\caption{Results obtained from the model SC1p50m20. Panel descriptions are the same as those in Figure 3.
\label{fig5}}
\end{figure}


\begin{figure}
\includegraphics[scale=0.25]{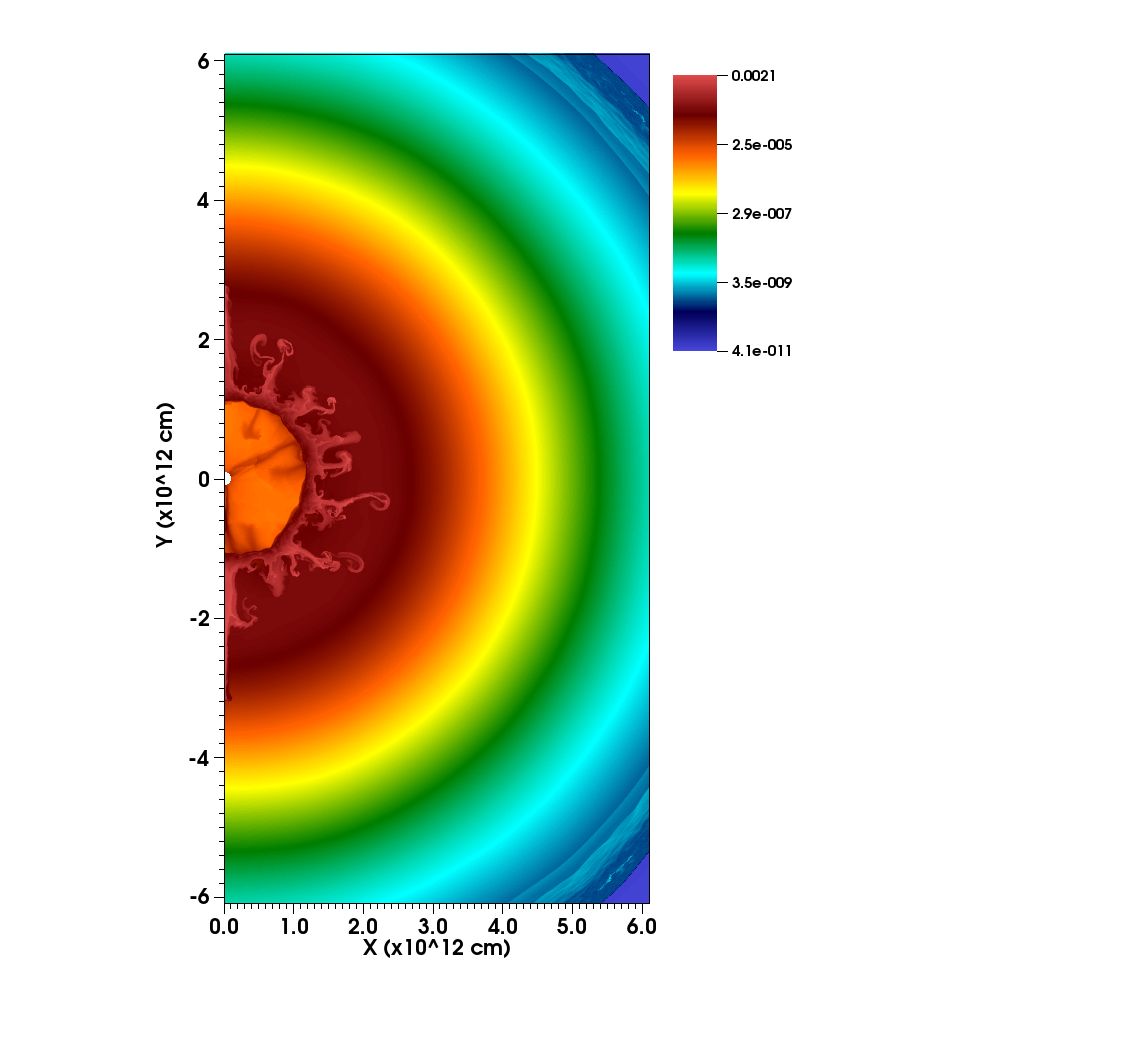}
\includegraphics[scale=0.25]{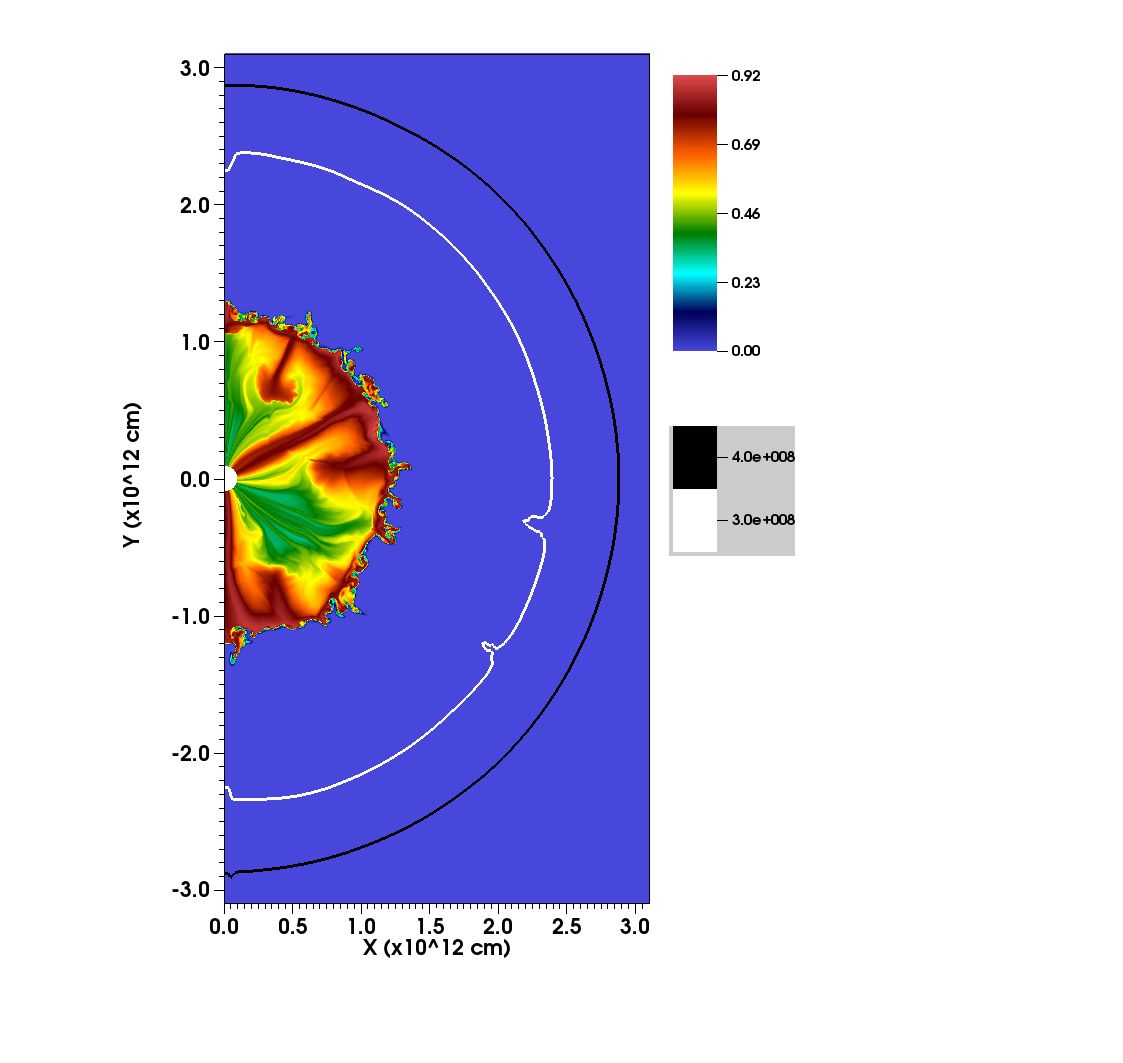}
\includegraphics[scale=0.7]{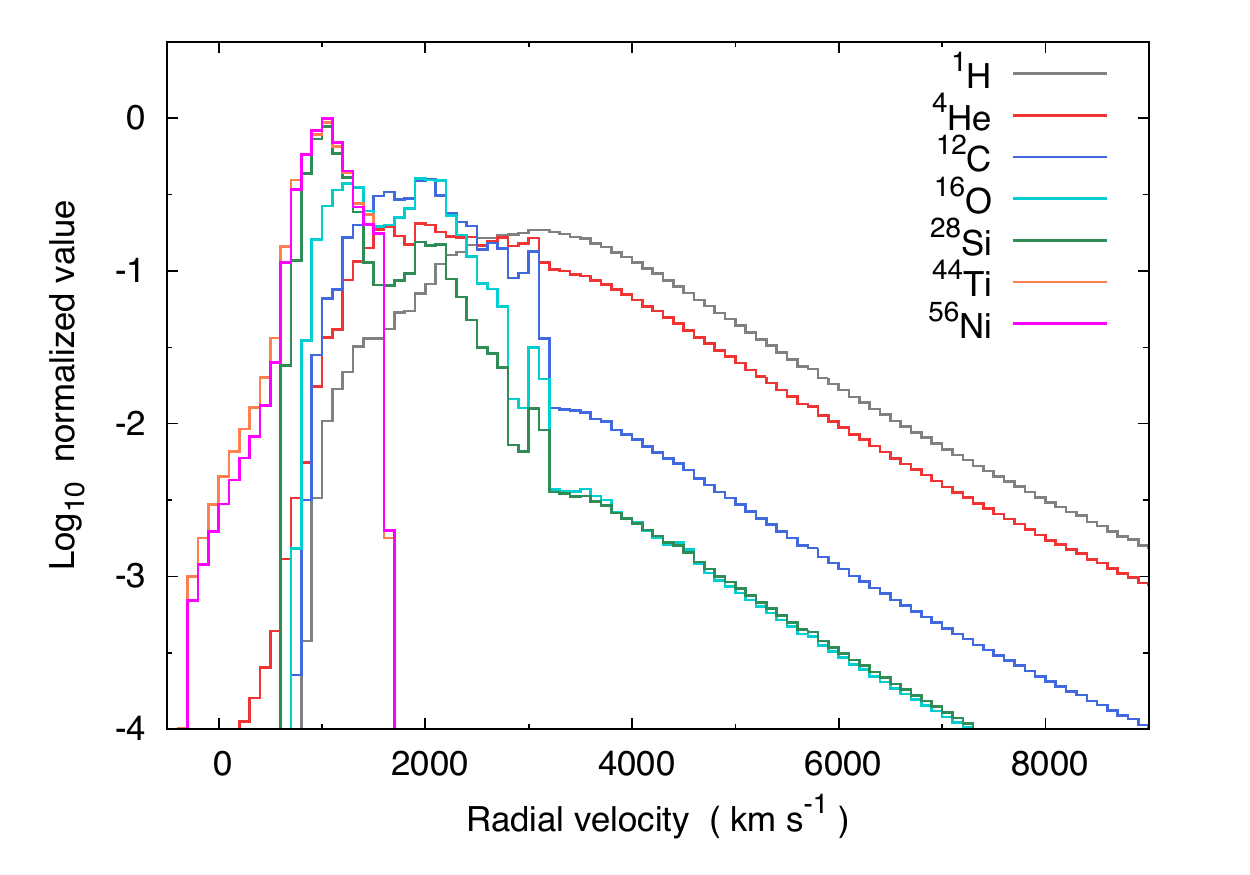}
\includegraphics[scale=0.7]{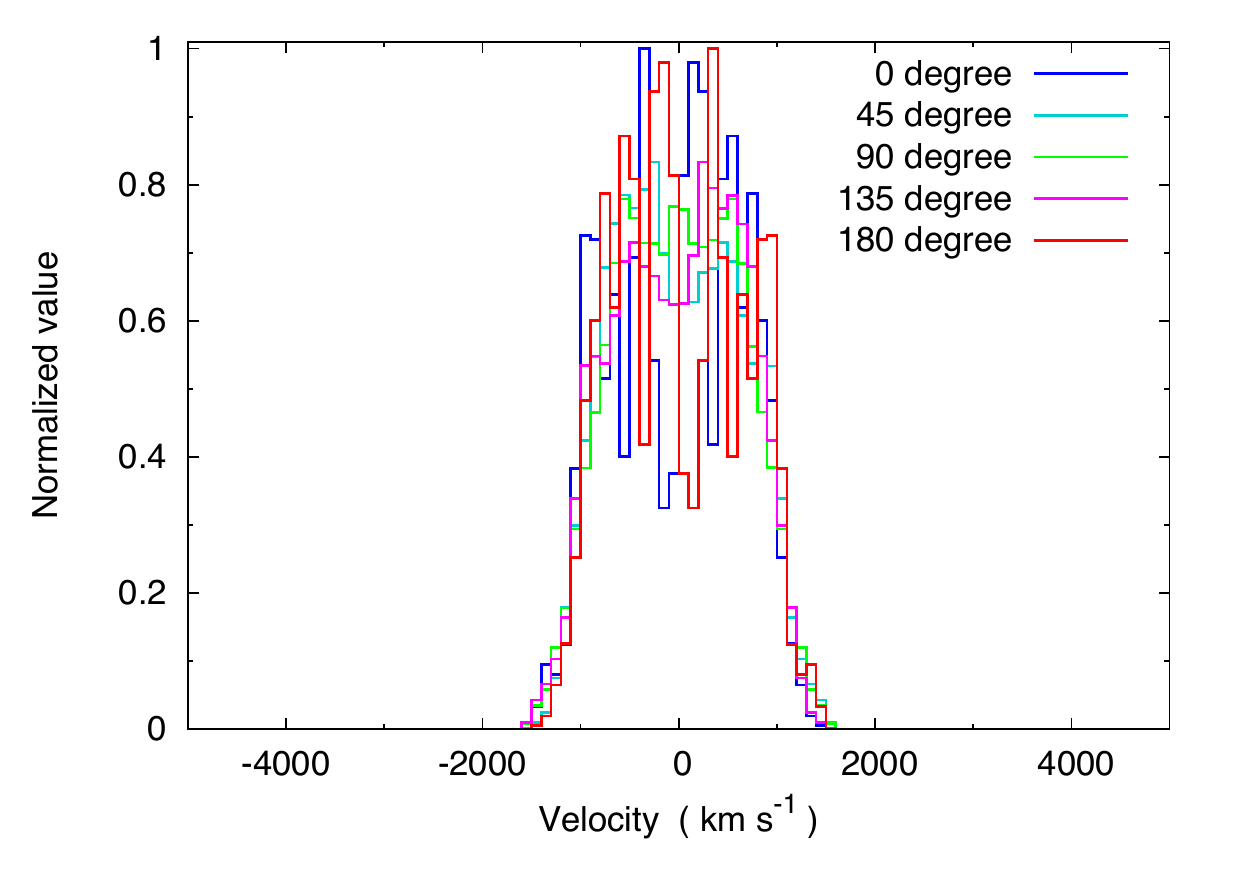}
\caption{Results obtained from the model SC1p10m20. Panel descriptions are the same as those in Figure 3.
\label{fig6}}
\end{figure}


\begin{figure}
\includegraphics[scale=0.25]{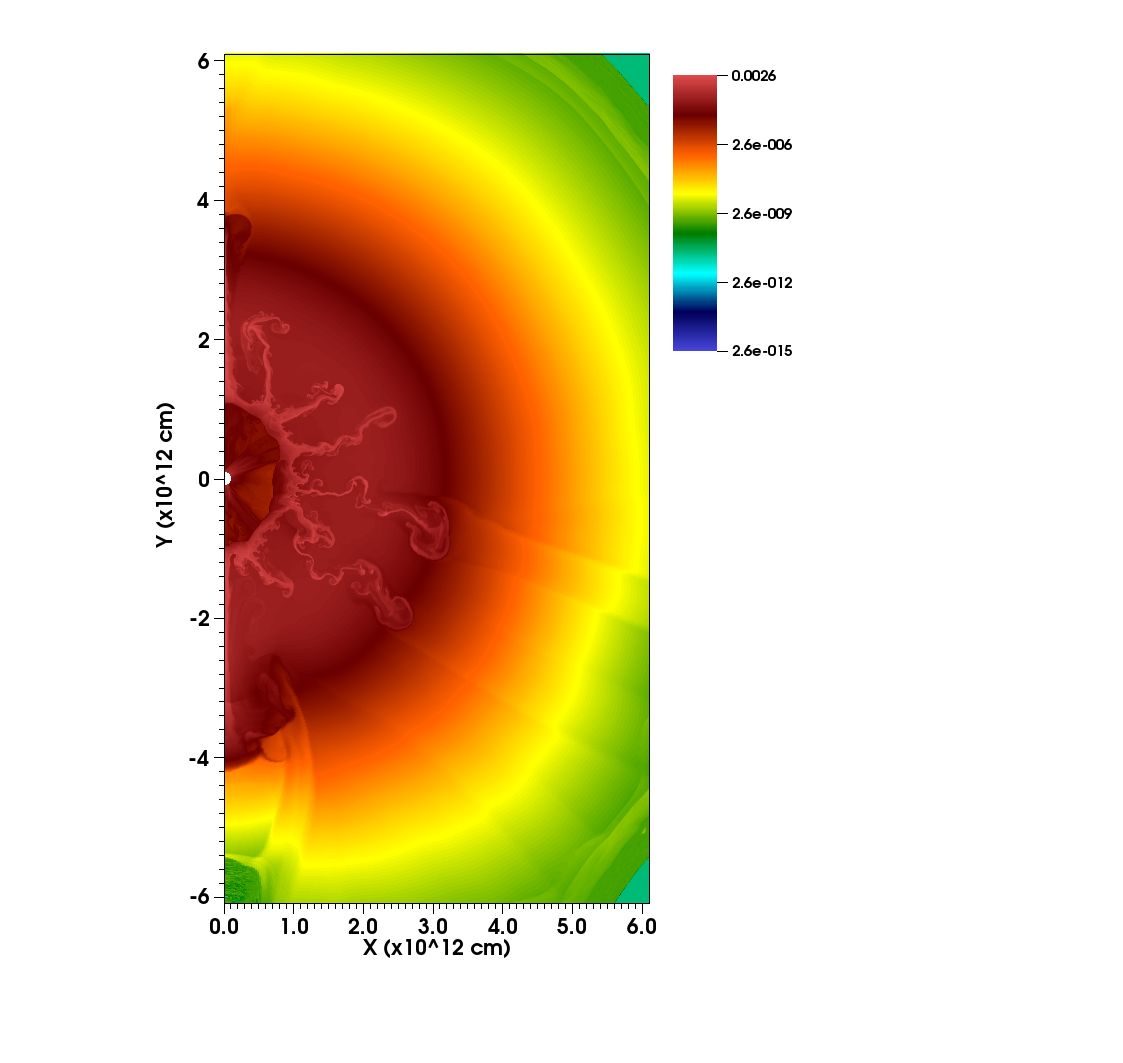}
\includegraphics[scale=0.25]{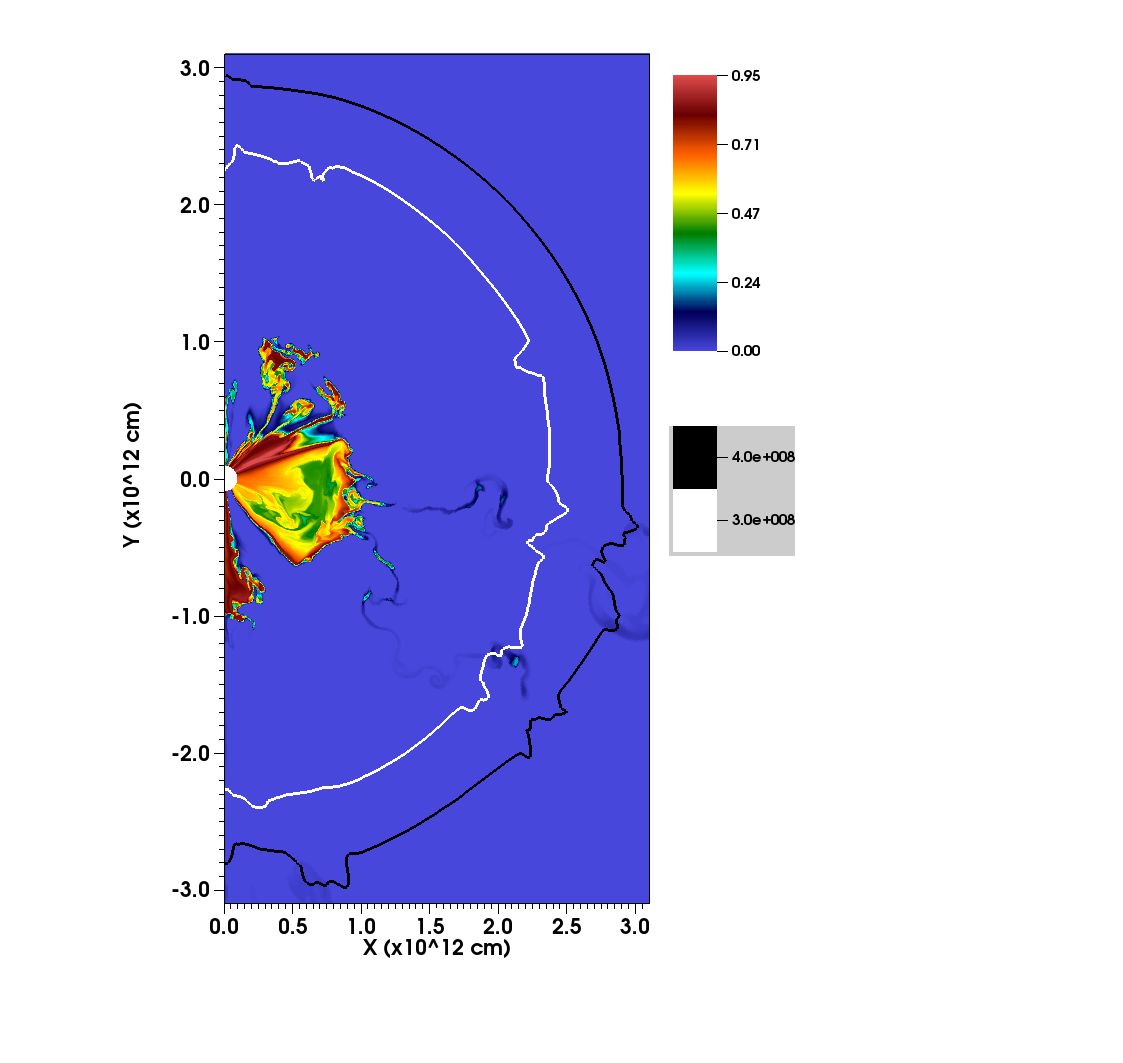}
\includegraphics[scale=0.7]{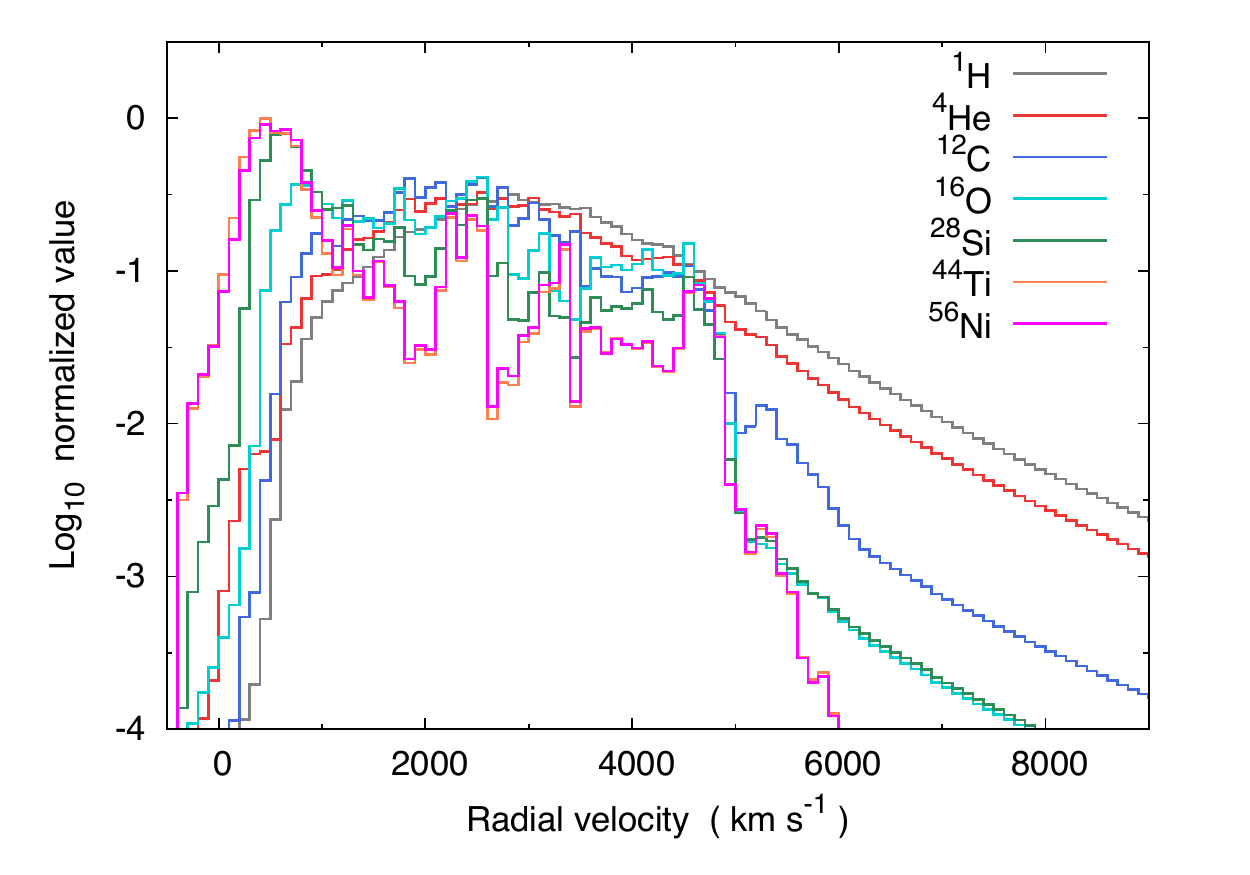}
\includegraphics[scale=0.7]{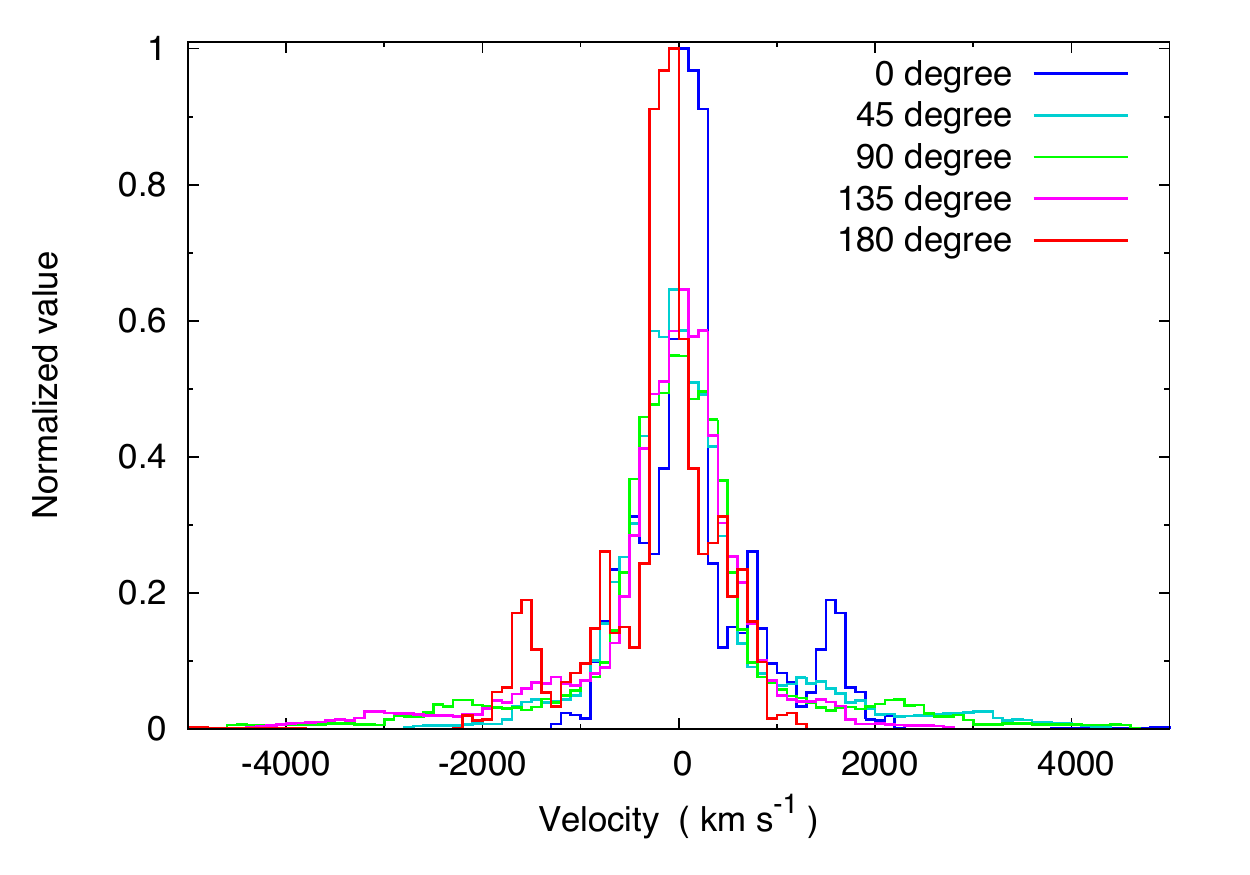}
\caption{Results obtained from the model SC1C+O/He/H. Panel descriptions are the same as those in Figure 3.
\label{fig7}}
\end{figure}


\begin{figure}
\includegraphics[scale=0.25]{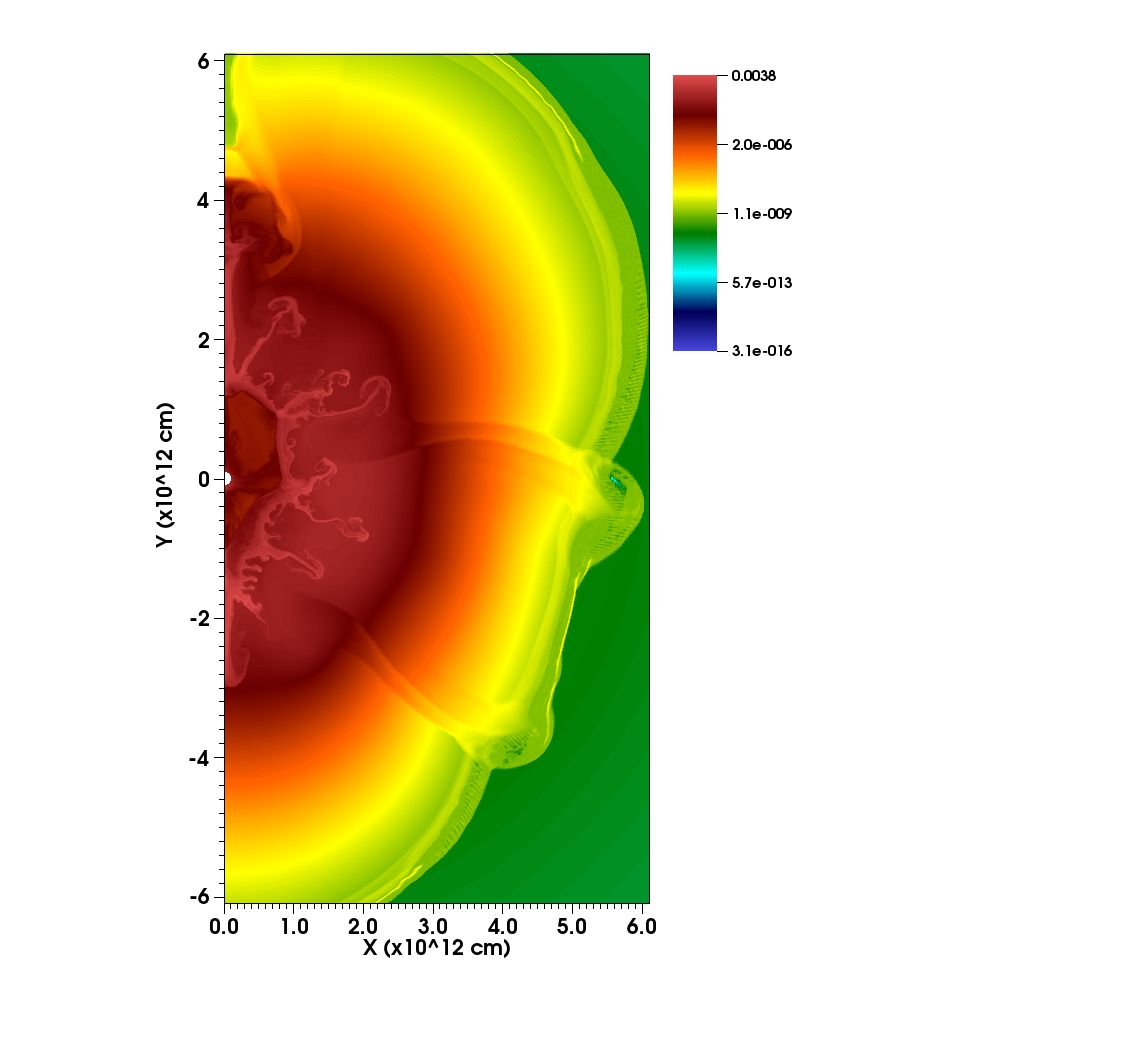}
\includegraphics[scale=0.25]{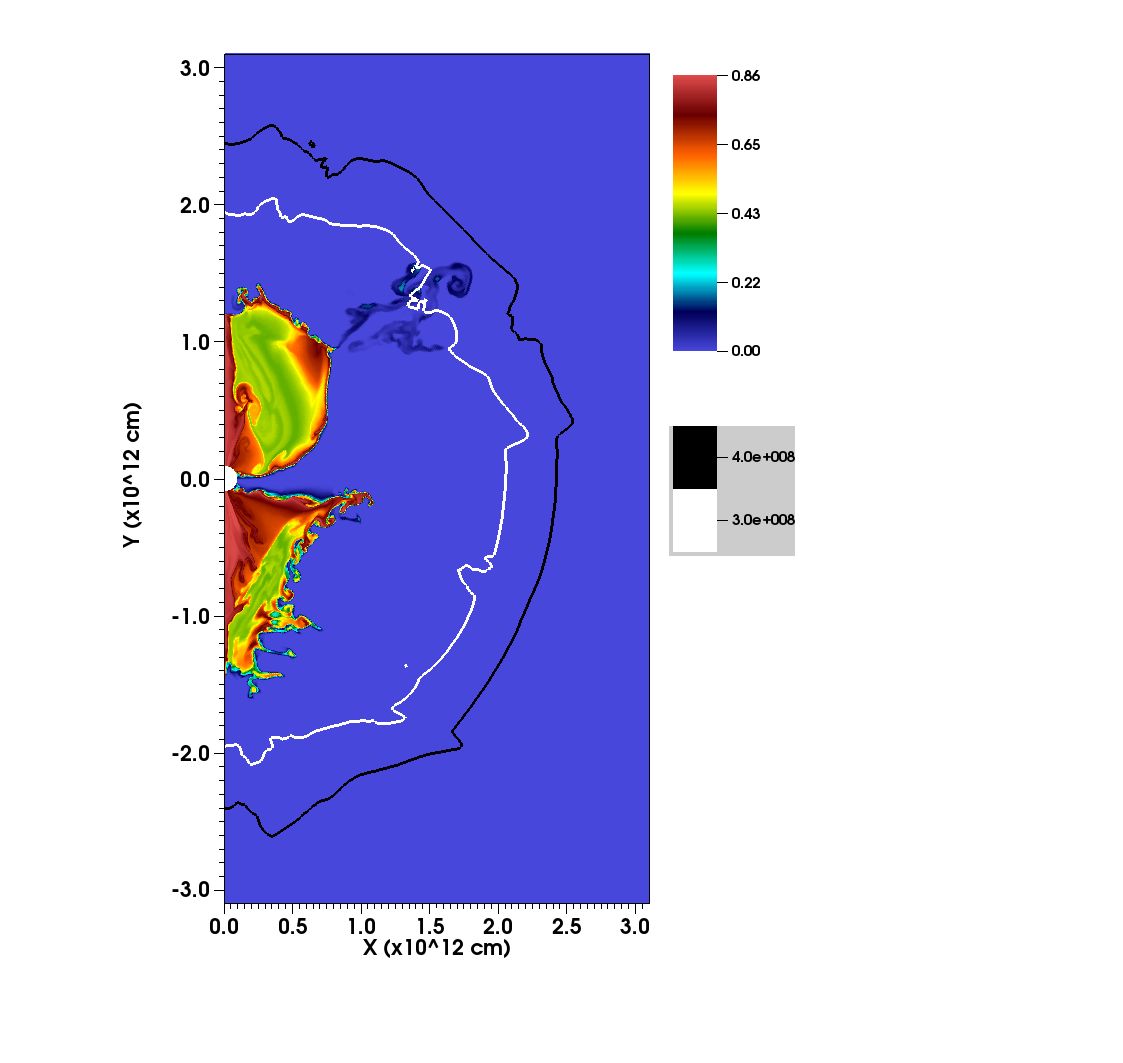}
\includegraphics[scale=0.7]{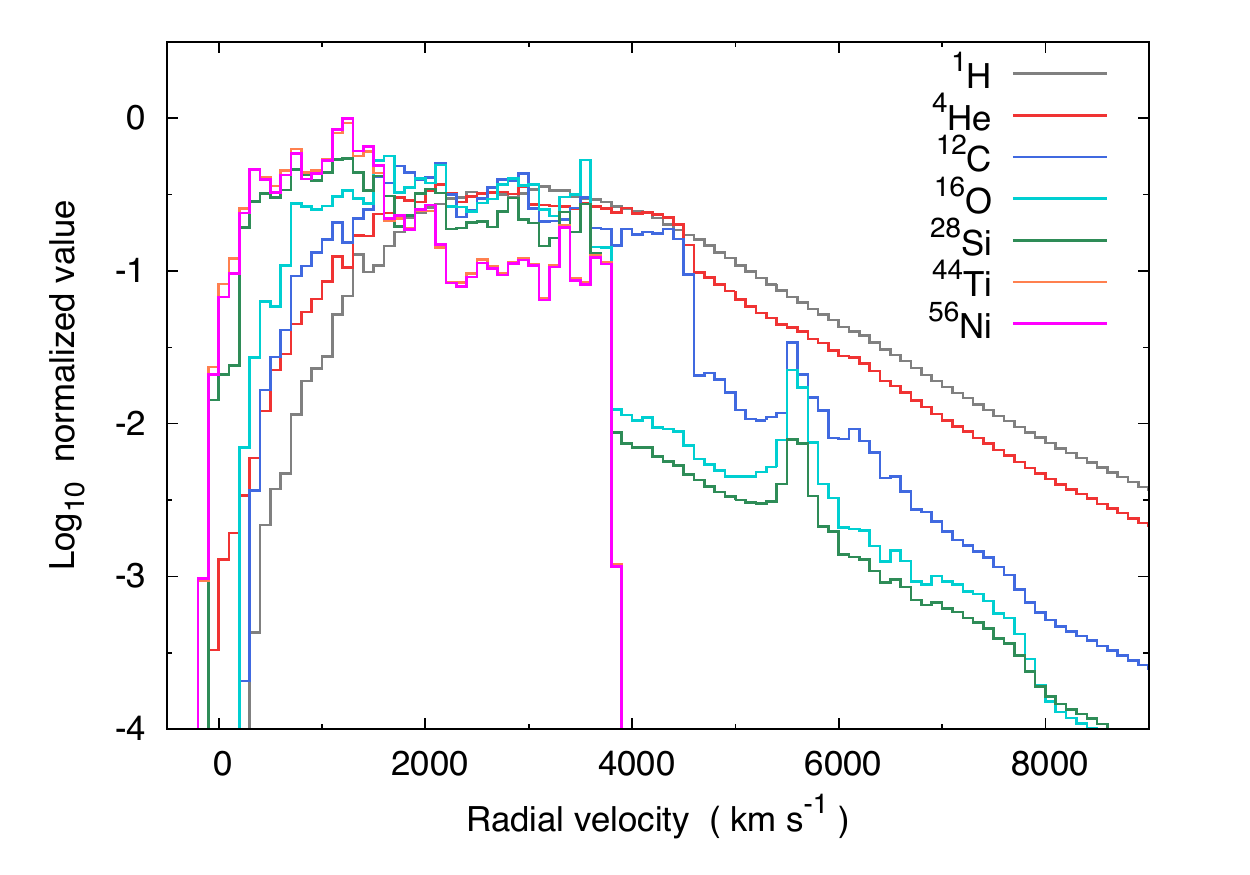}
\includegraphics[scale=0.7]{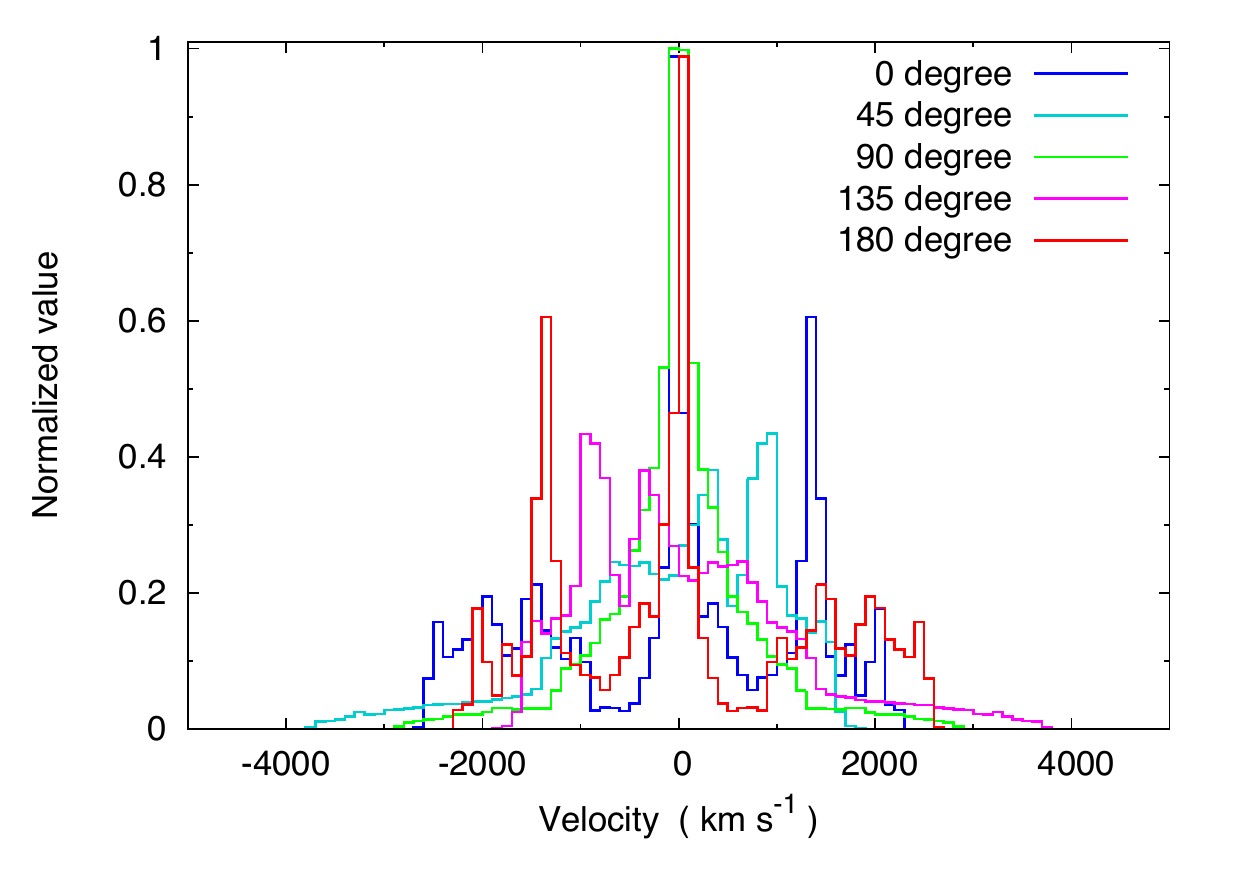}
\caption{Results obtained from the model SC2p50m8f5. Panel descriptions are the same as those in Figure 3.
\label{fig8}}
\end{figure}


\clearpage

\begin{figure}
\includegraphics[scale=0.25]{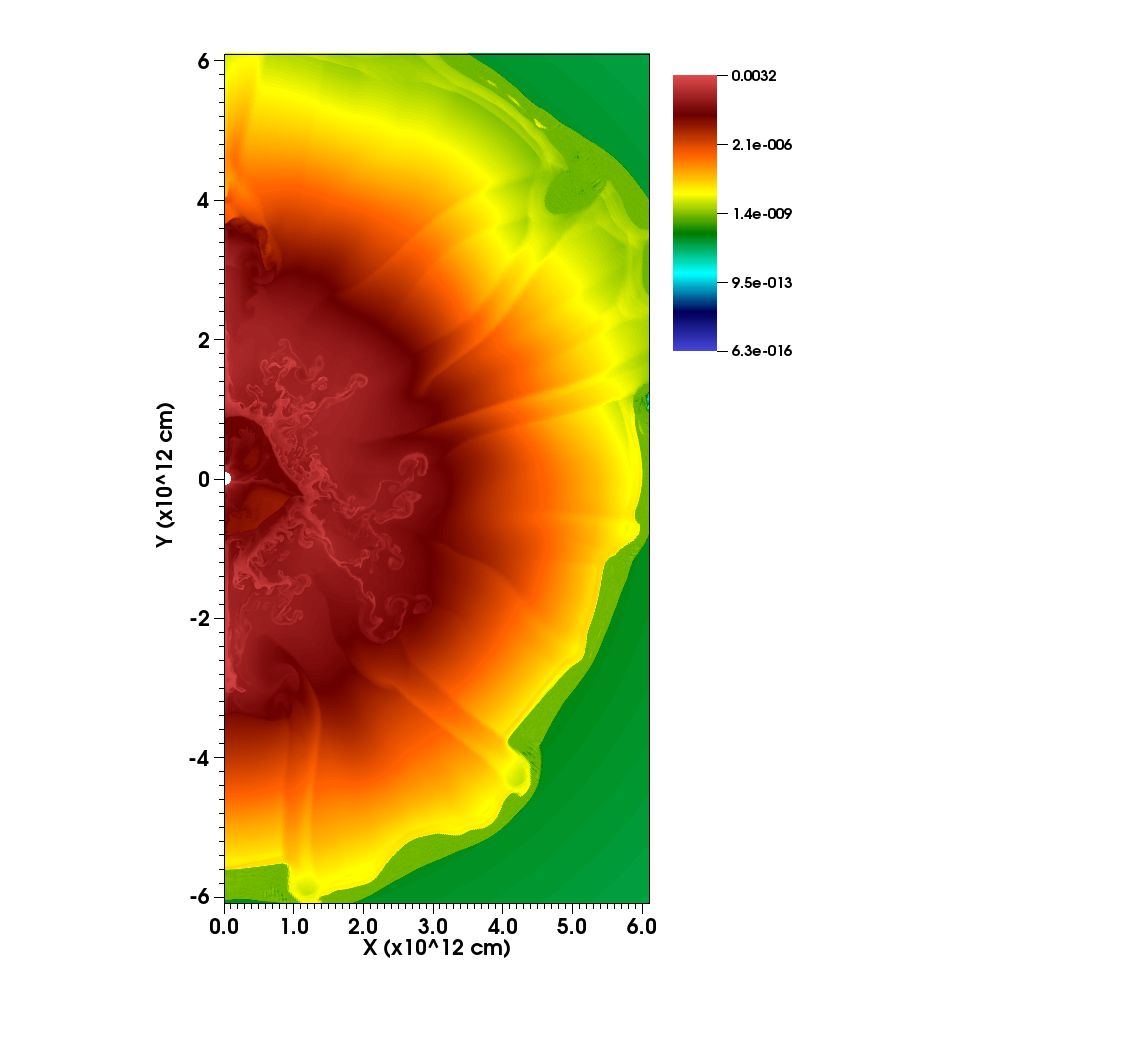}
\includegraphics[scale=0.25]{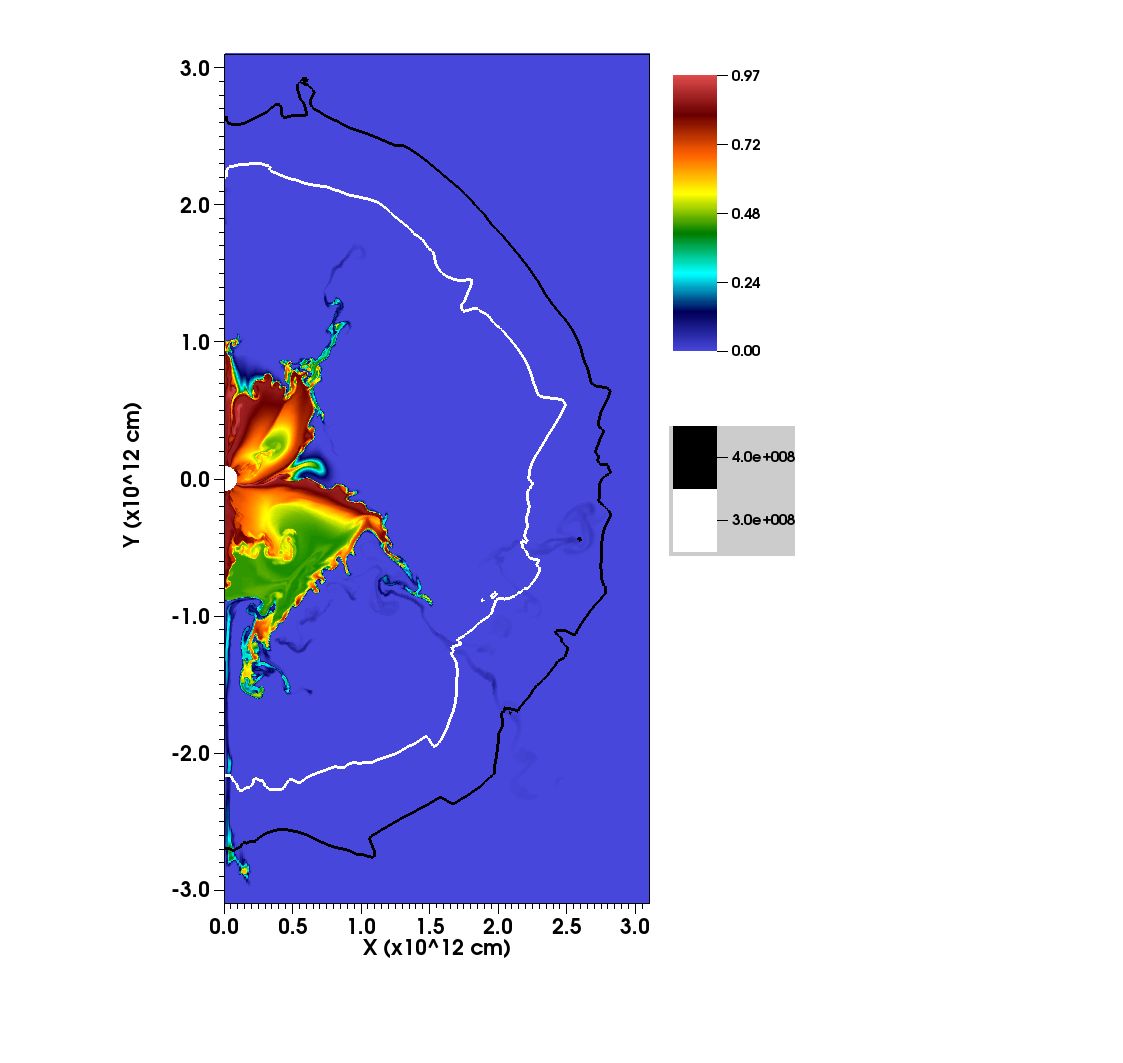}
\includegraphics[scale=0.7]{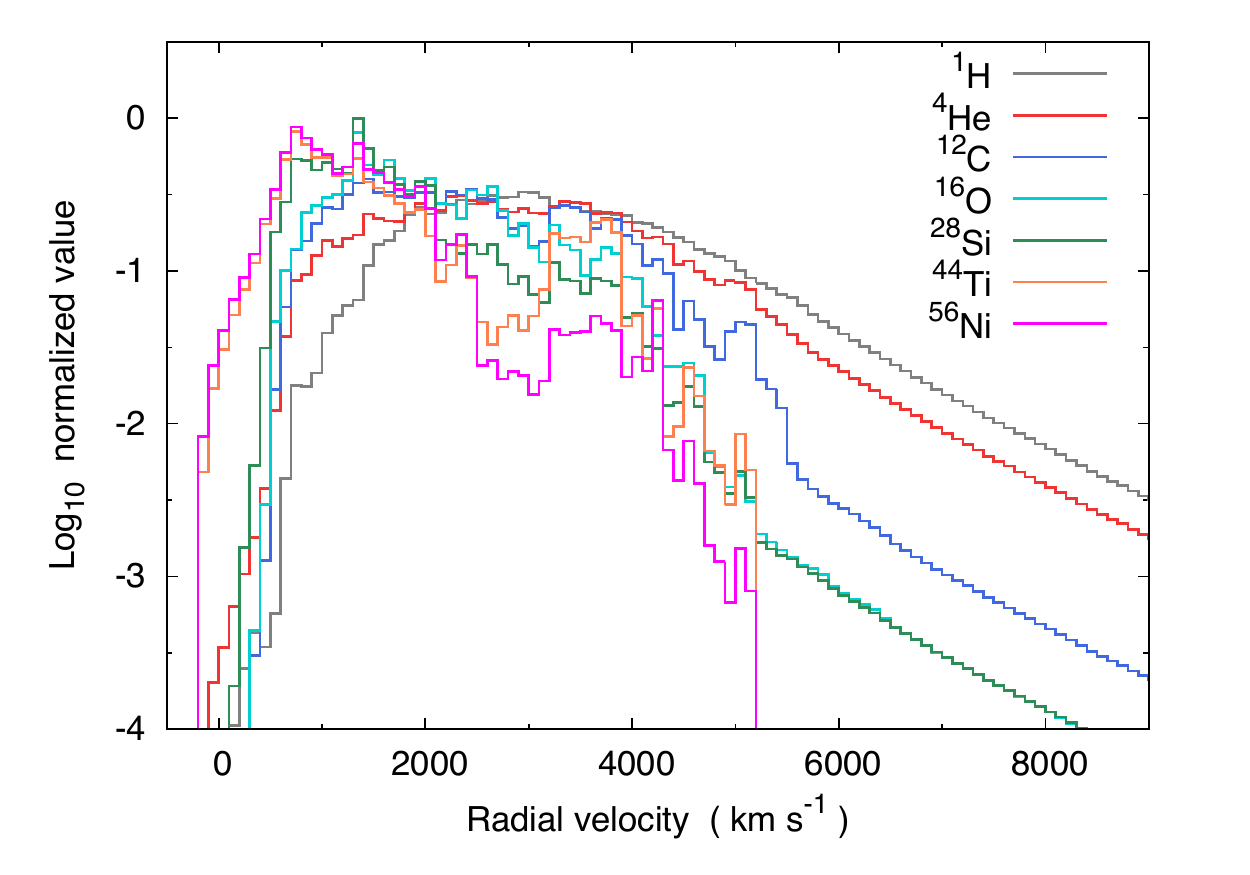}
\includegraphics[scale=0.7]{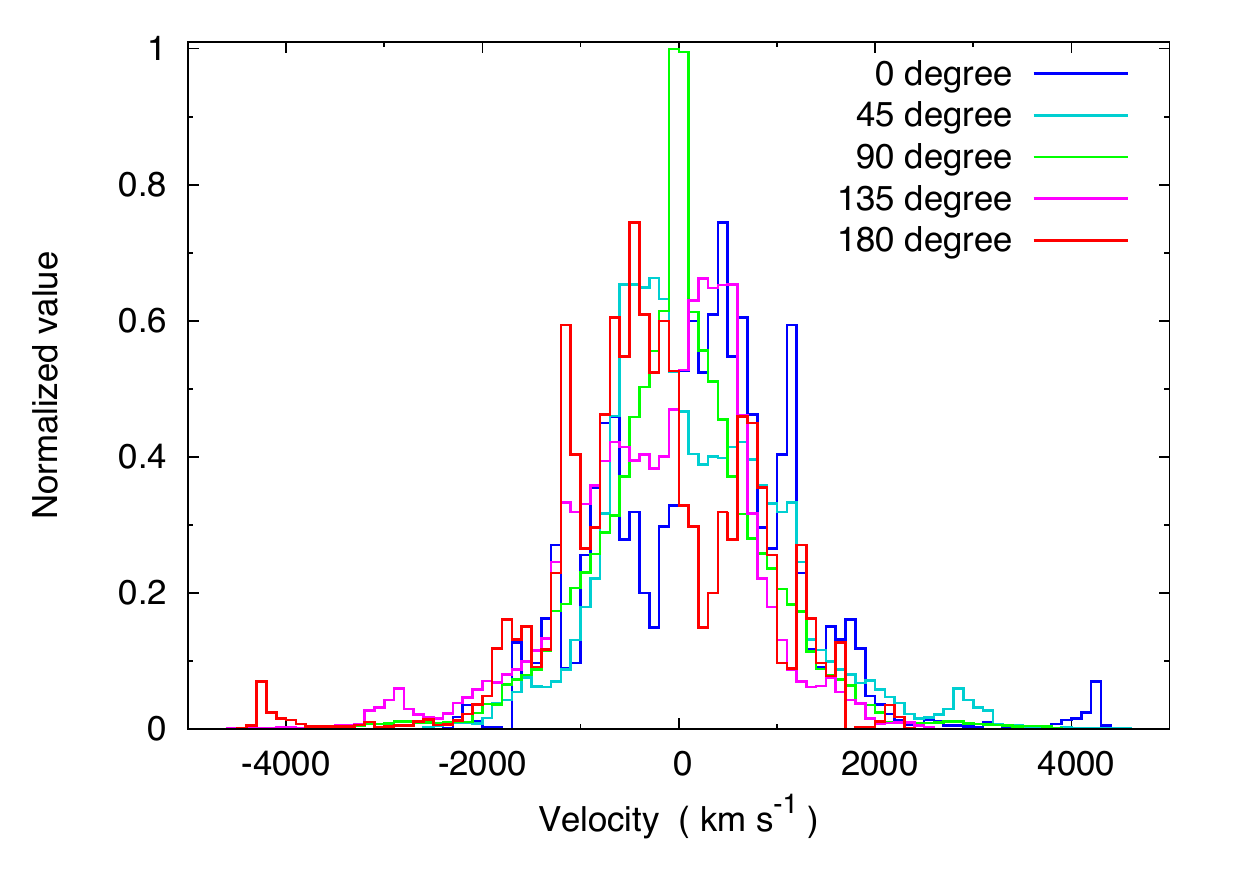}
\caption{Results obtained from the model SC2p50m20f5. Panel descriptions are the same as those in Figure 3.
\label{fig9}}
\end{figure}


\begin{figure}
\includegraphics[scale=0.25]{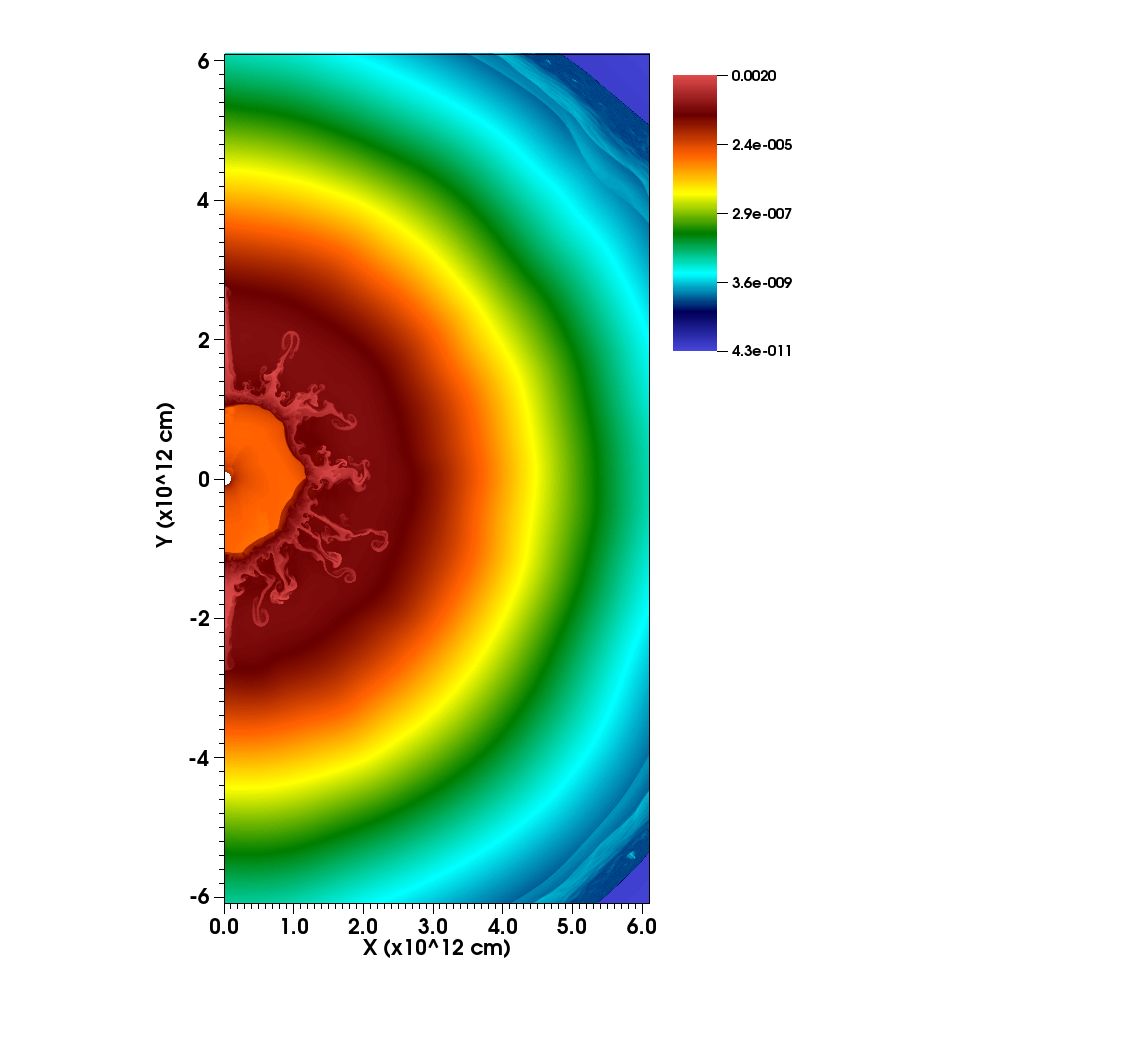}
\includegraphics[scale=0.33]{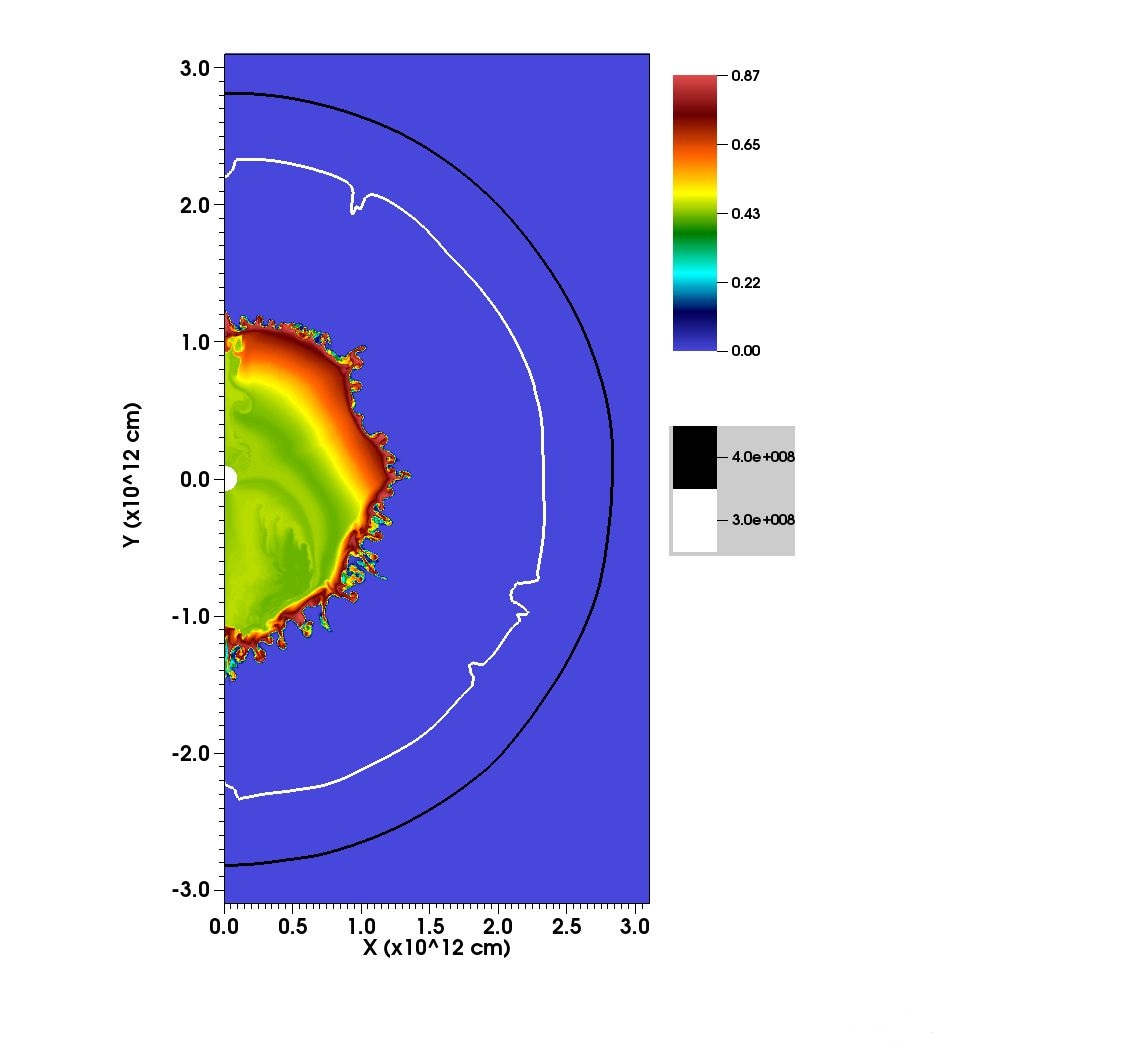}
\includegraphics[scale=0.7]{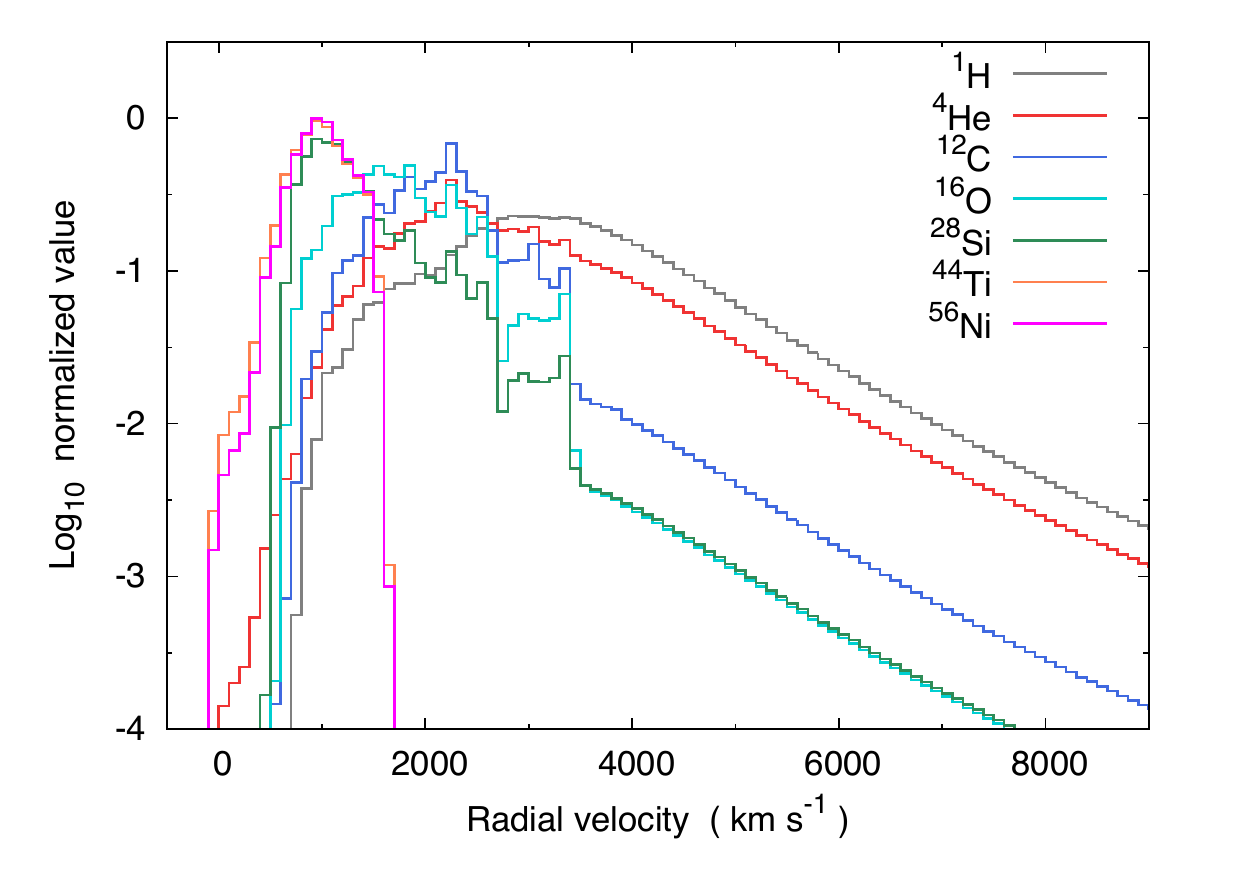}
\includegraphics[scale=0.7]{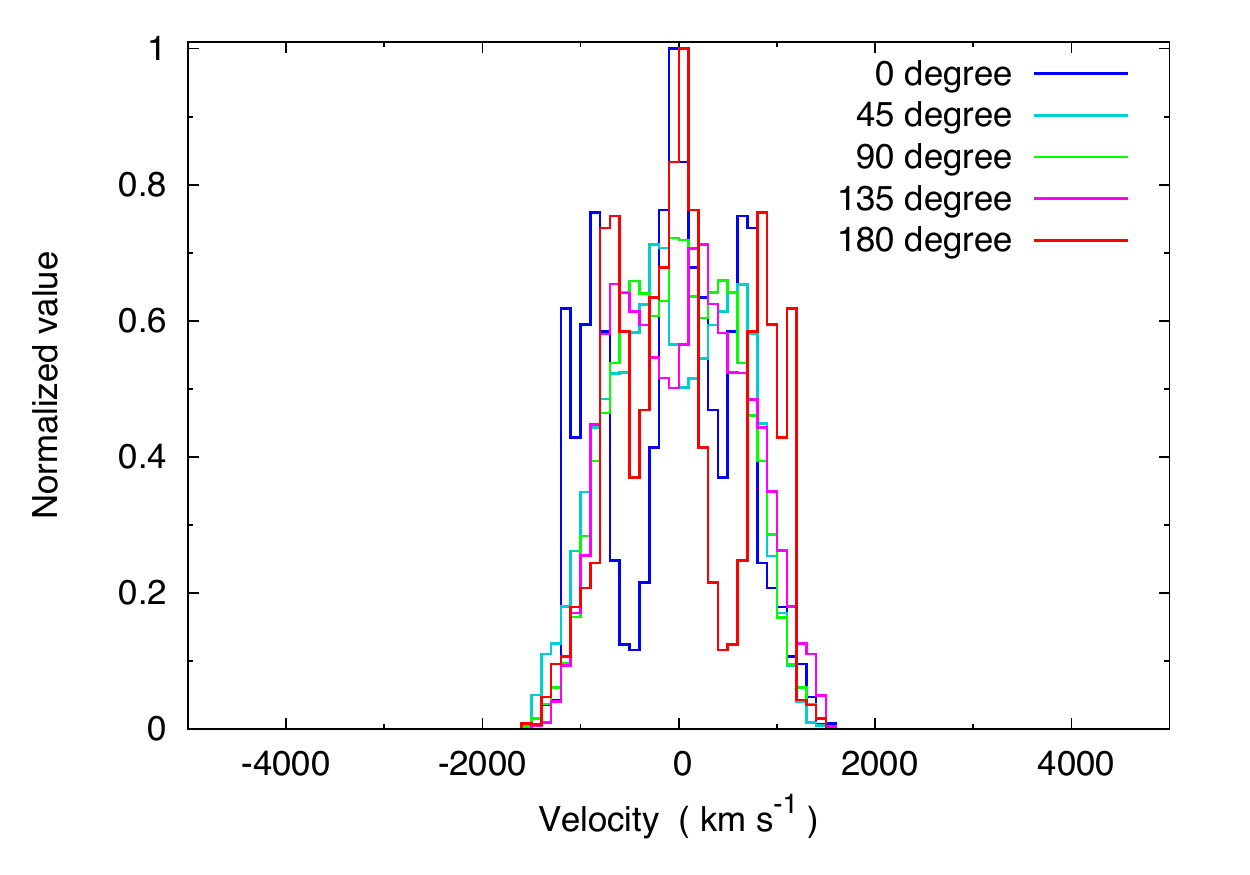}
\caption{Results obtained from the model SC2p10m20f5. Panel descriptions are the same as those in Figure 3.
\label{fig10}}
\end{figure}


\begin{figure}
\includegraphics[scale=0.25]{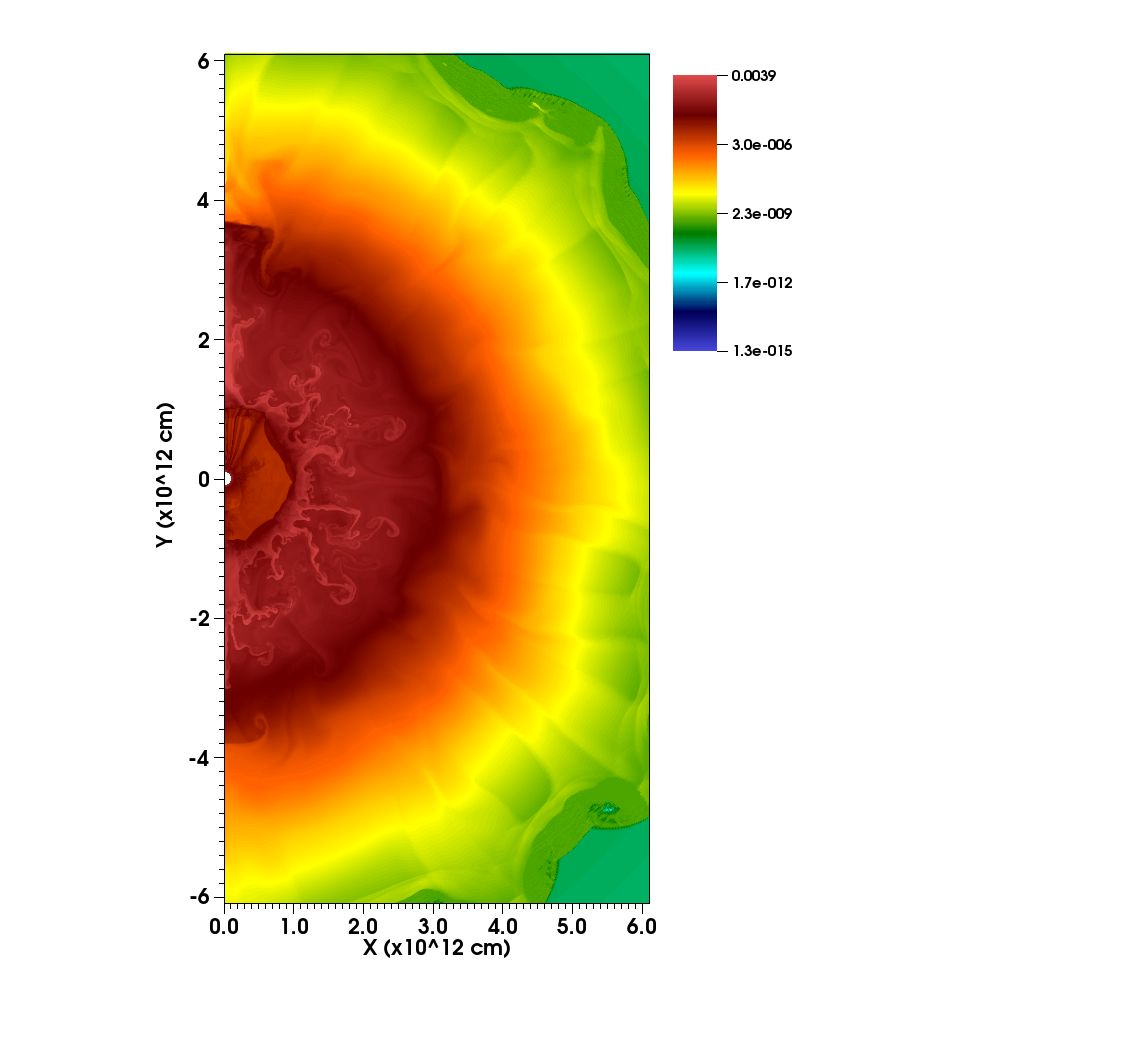}
\includegraphics[scale=0.25]{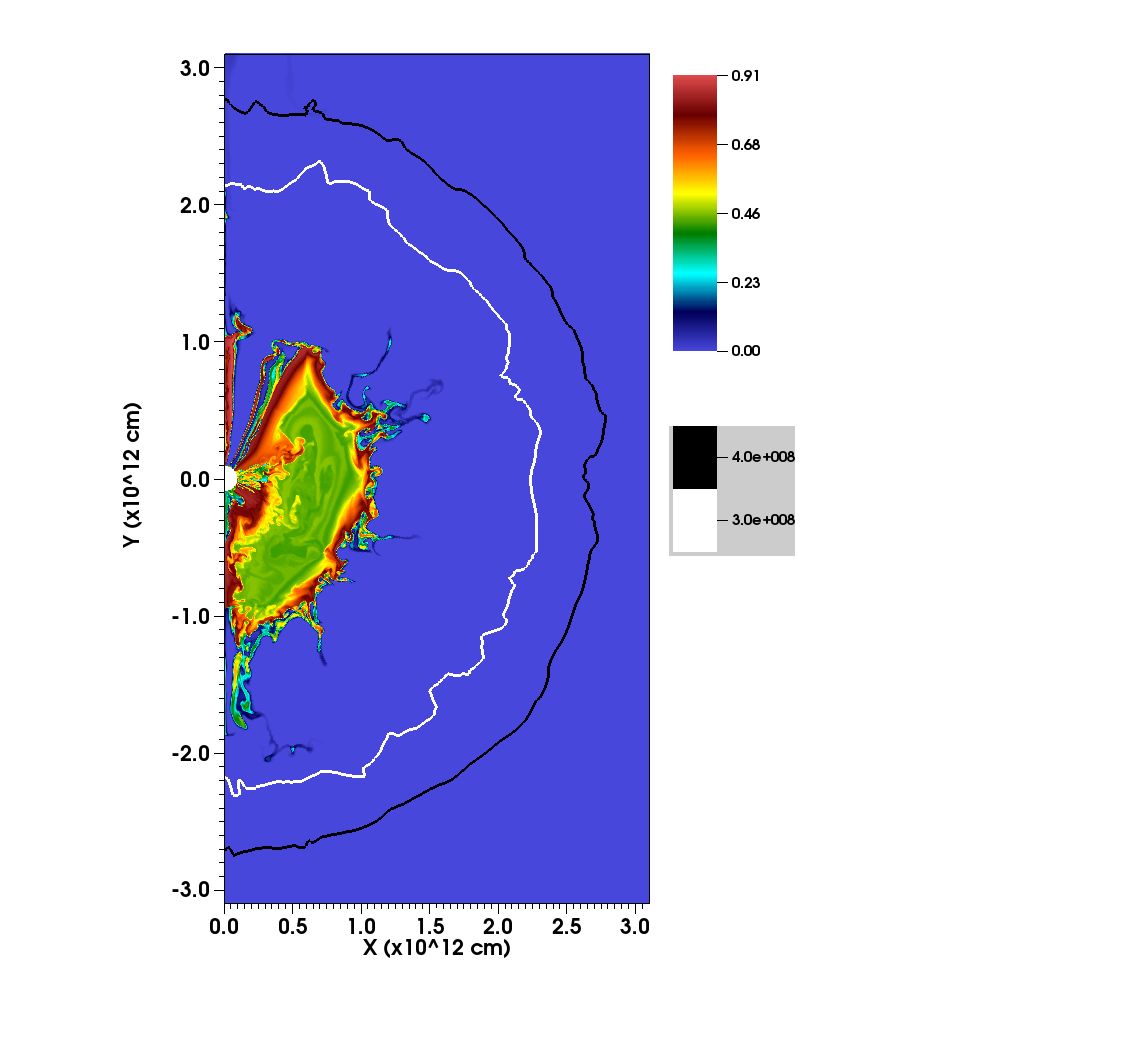}
\includegraphics[scale=0.7]{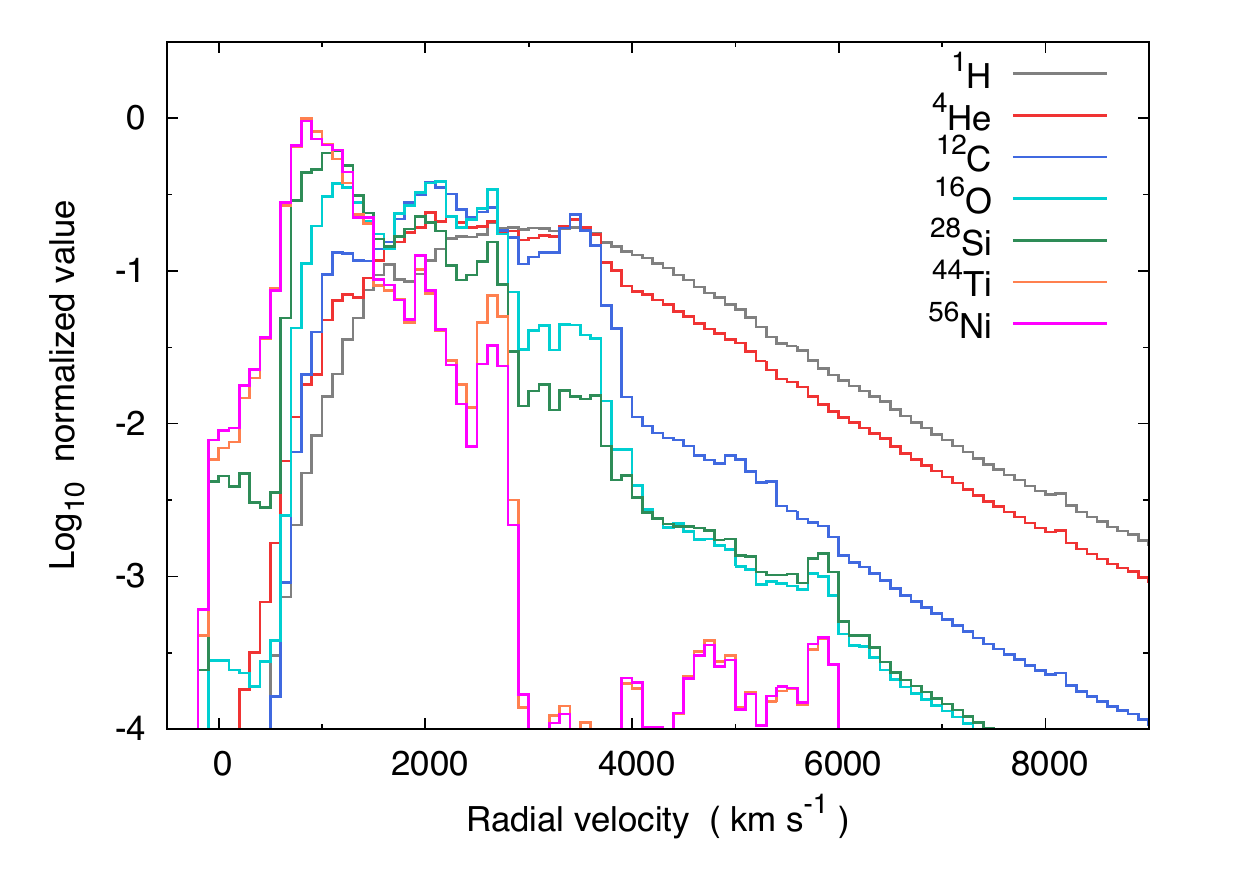}
\includegraphics[scale=0.7]{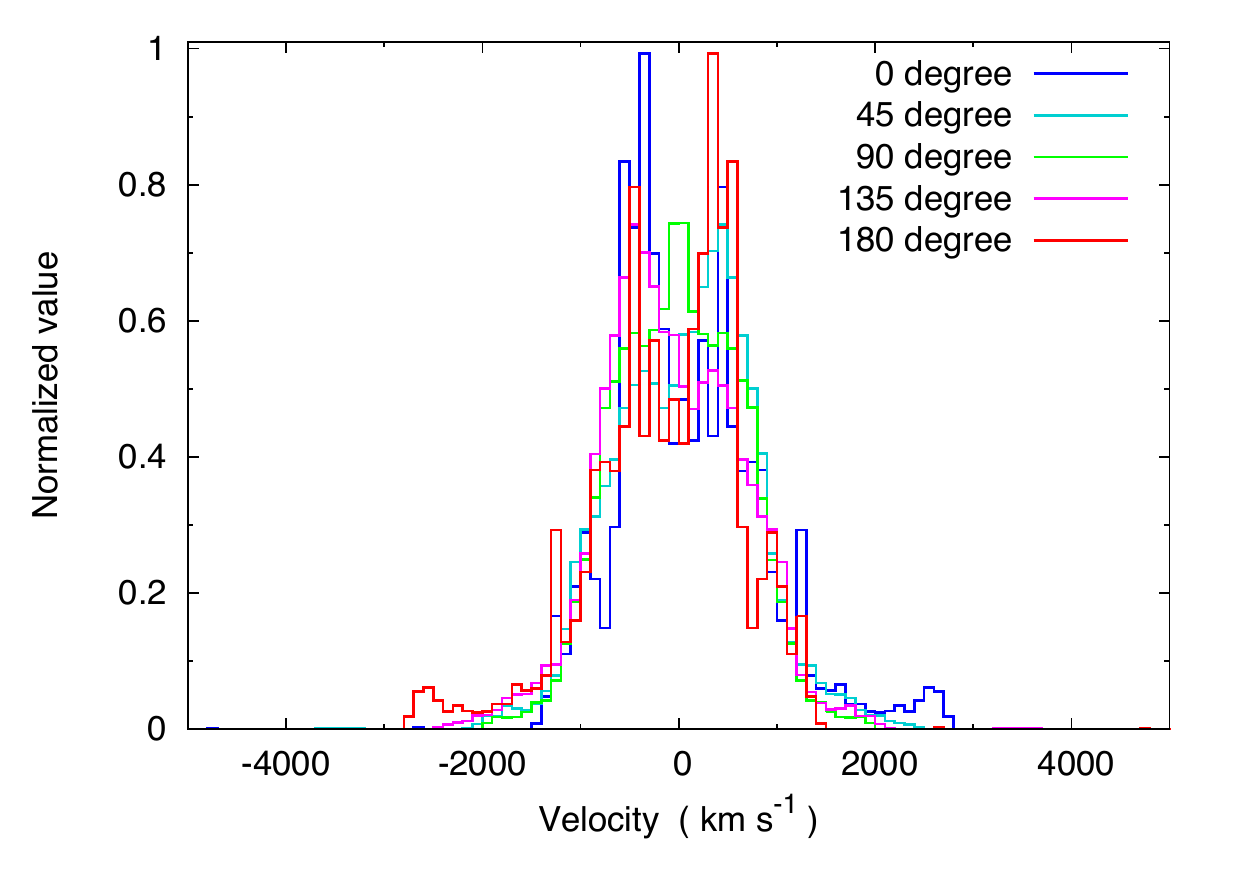}
\caption{Results obtained from the model SC2p50m20f1. Panel descriptions are the same as those in Figure 3.
\label{fig11}}
\end{figure}


\begin{figure}
\includegraphics[scale=0.25]{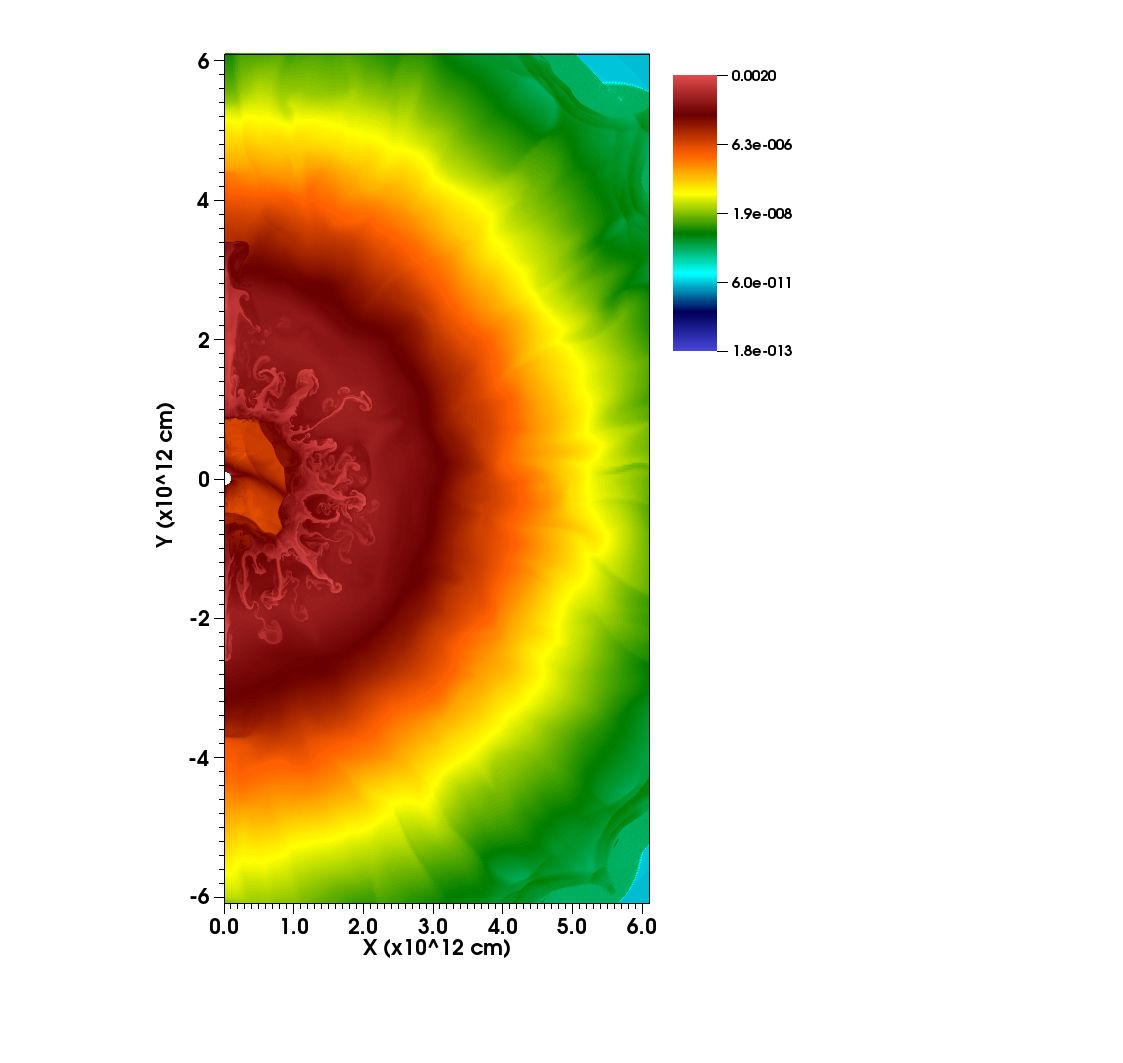}
\includegraphics[scale=0.25]{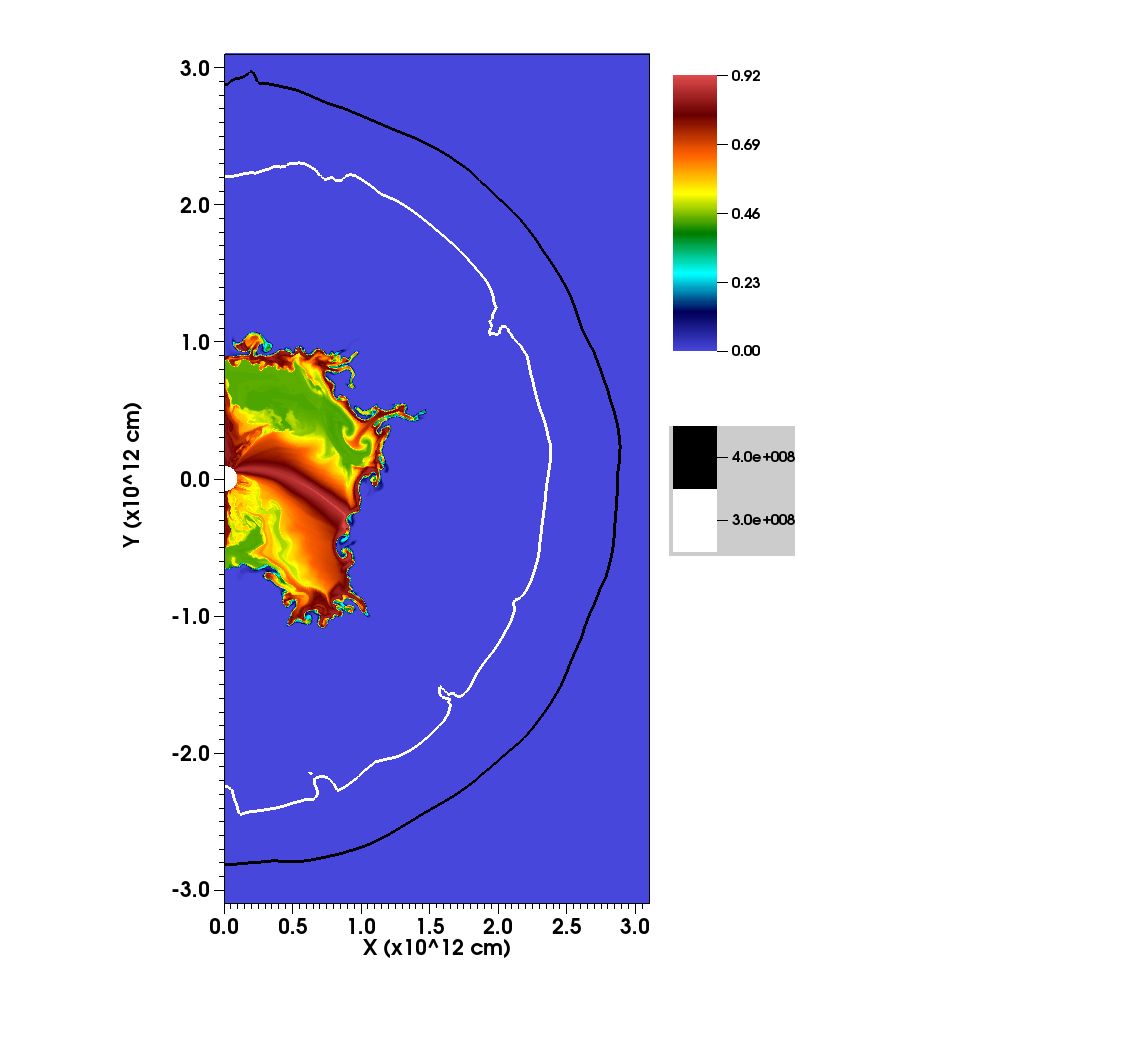}
\includegraphics[scale=0.7]{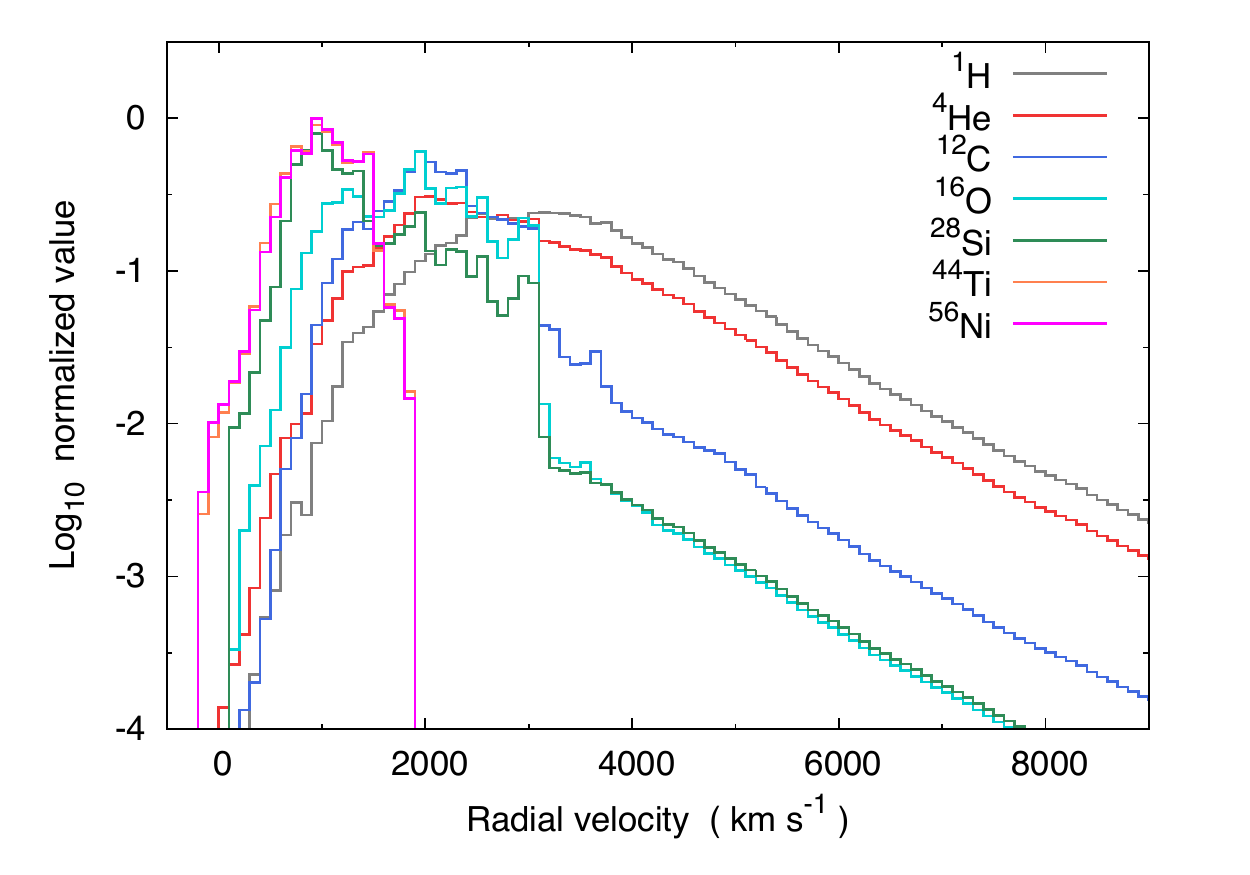}
\includegraphics[scale=0.7]{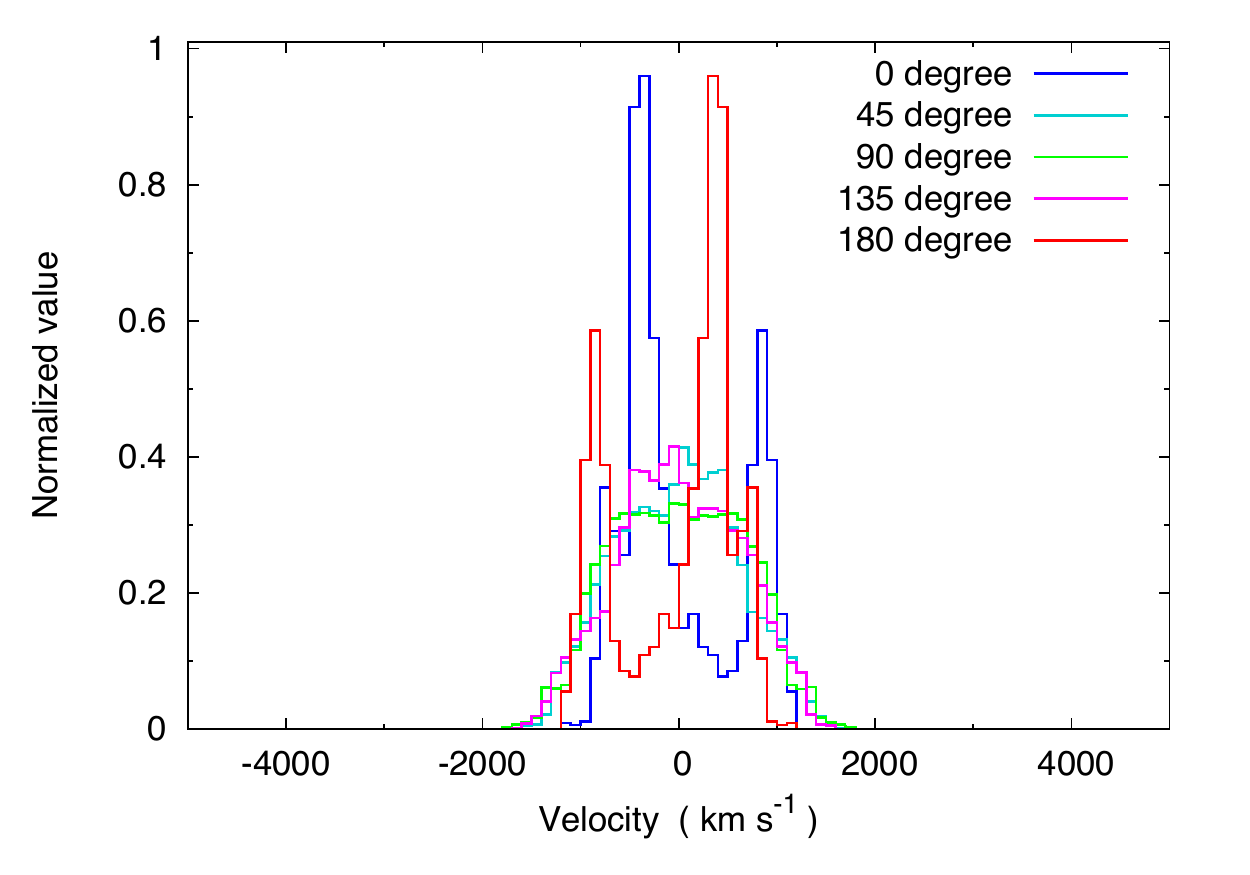}
\caption{Results obtained from the model SC3p50f1. Panel descriptions are the same as those in Figure 3.
\label{fig12}}
\end{figure}


\begin{figure}
\includegraphics[scale=0.25]{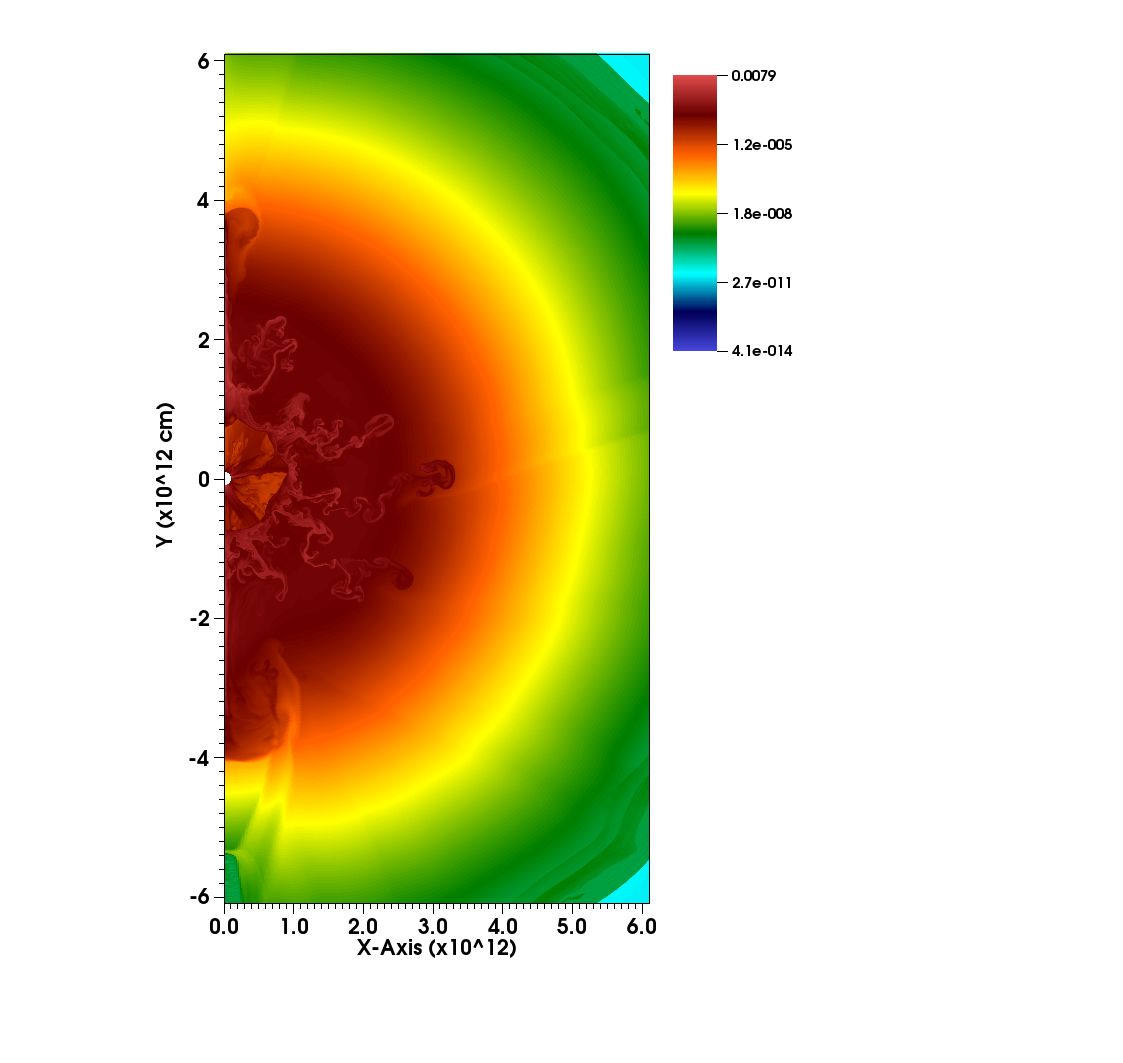}
\includegraphics[scale=0.25]{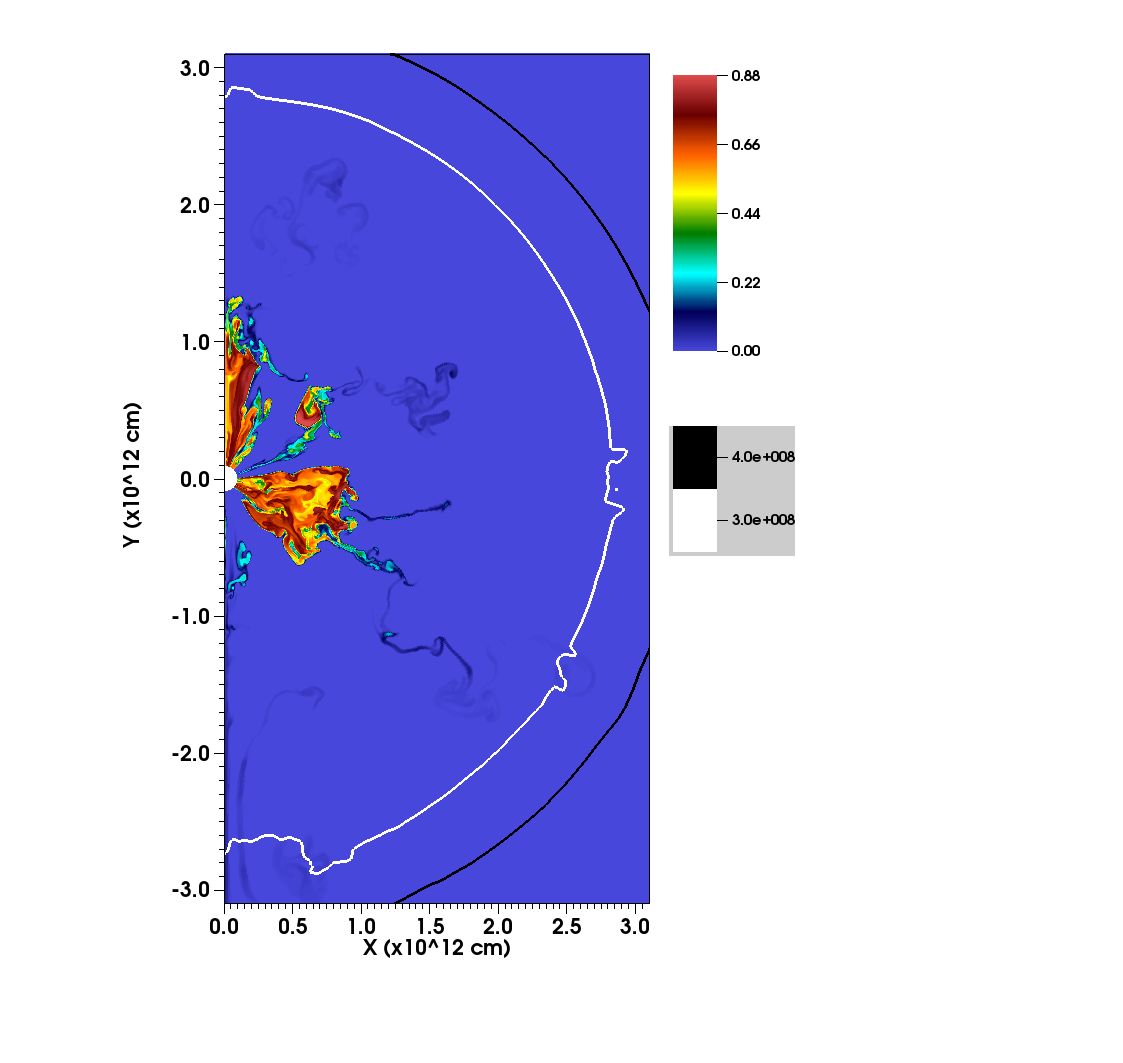}
\includegraphics[scale=0.7]{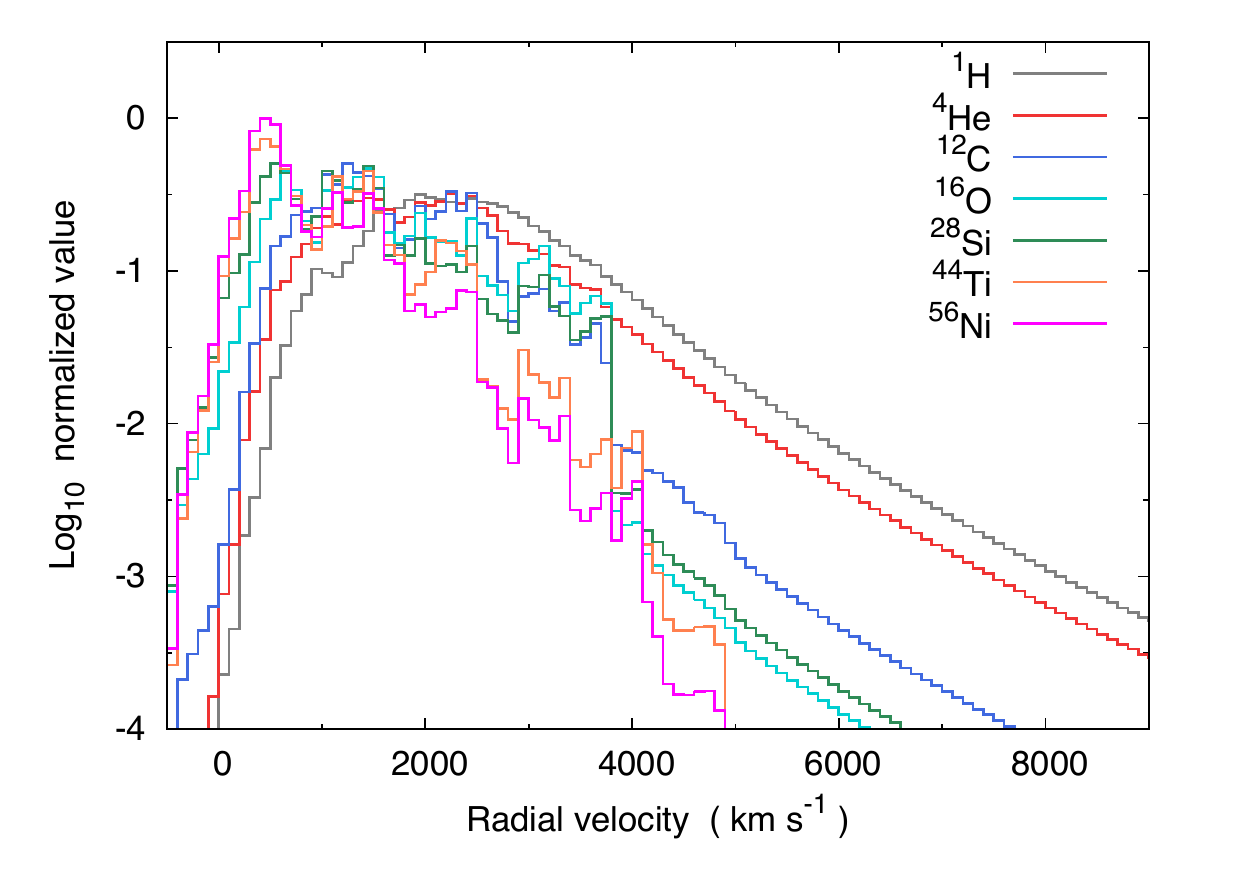}
\includegraphics[scale=0.7]{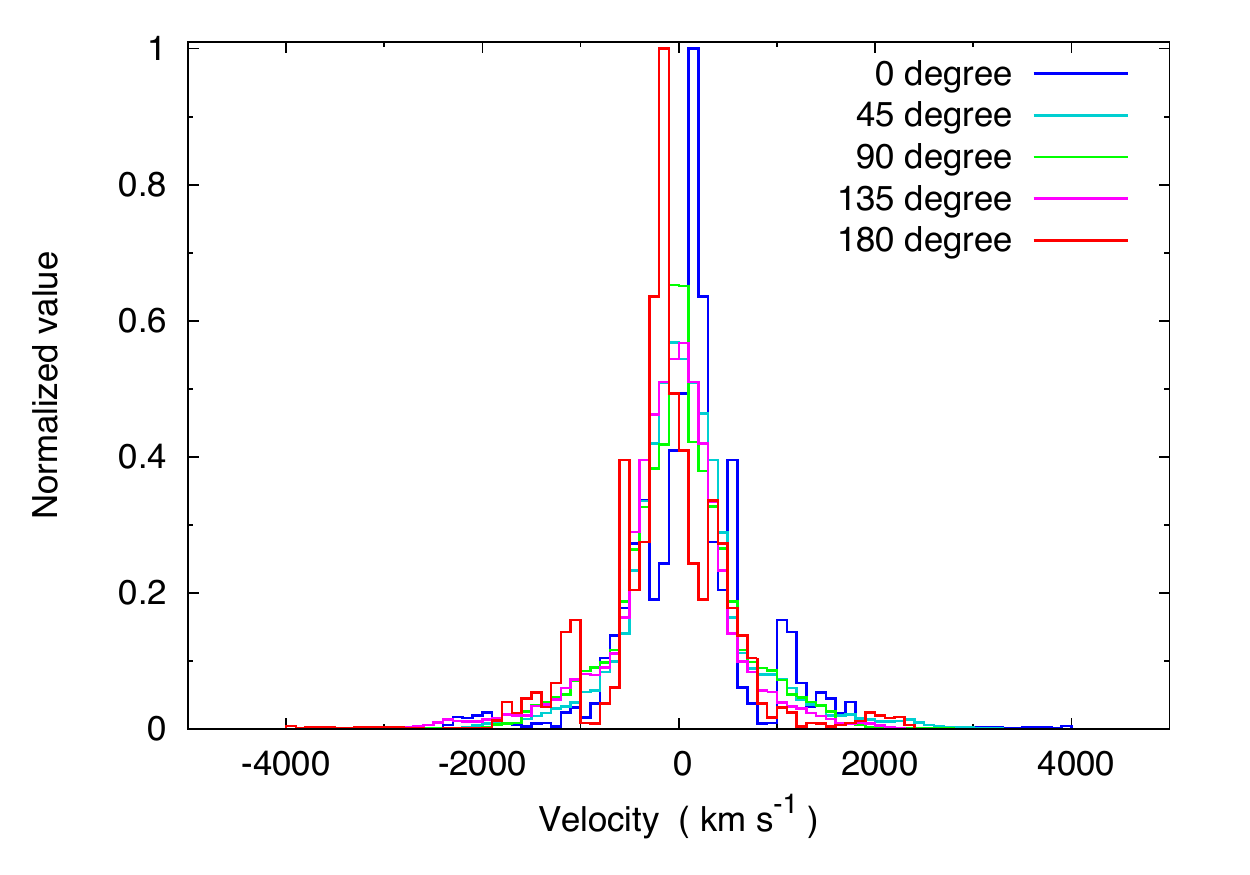}
\caption{Results obtained from the model SC1p50m20($\ast$). Panel descriptions are the same as those in Figure 3.
\label{fig13}}
\end{figure}


\begin{figure}
\includegraphics[scale=0.25]{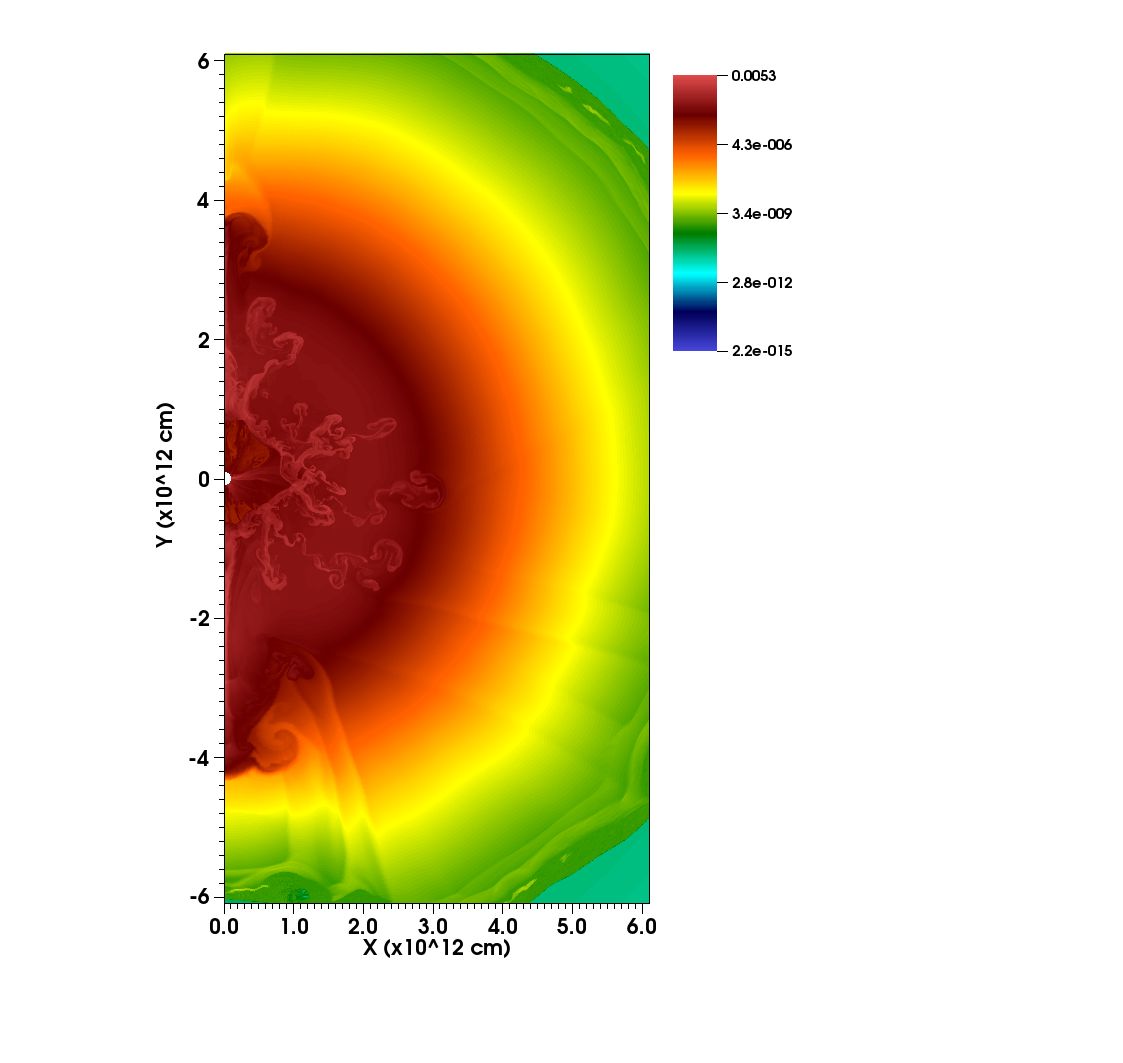}
\includegraphics[scale=0.25]{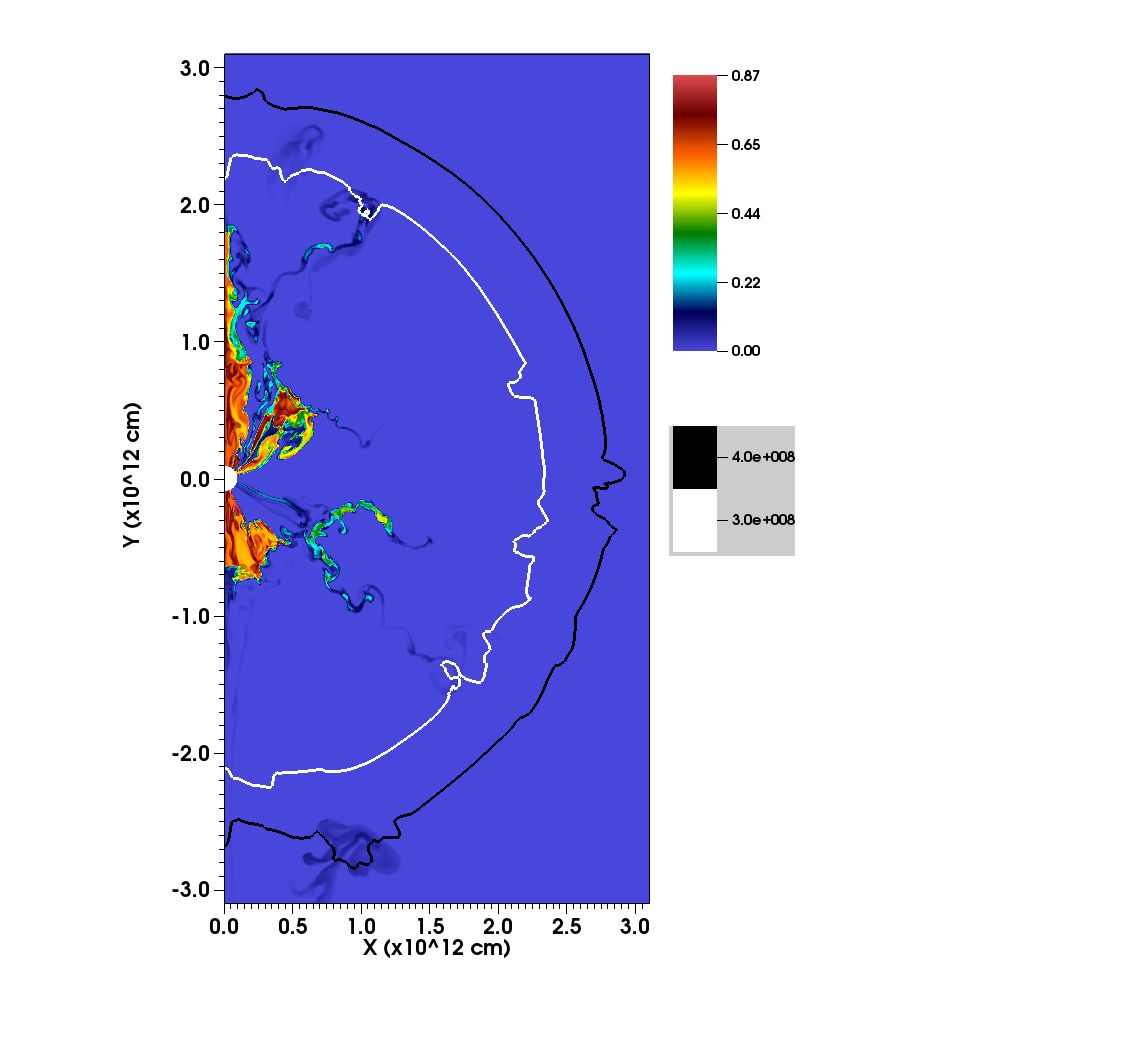}
\includegraphics[scale=0.7]{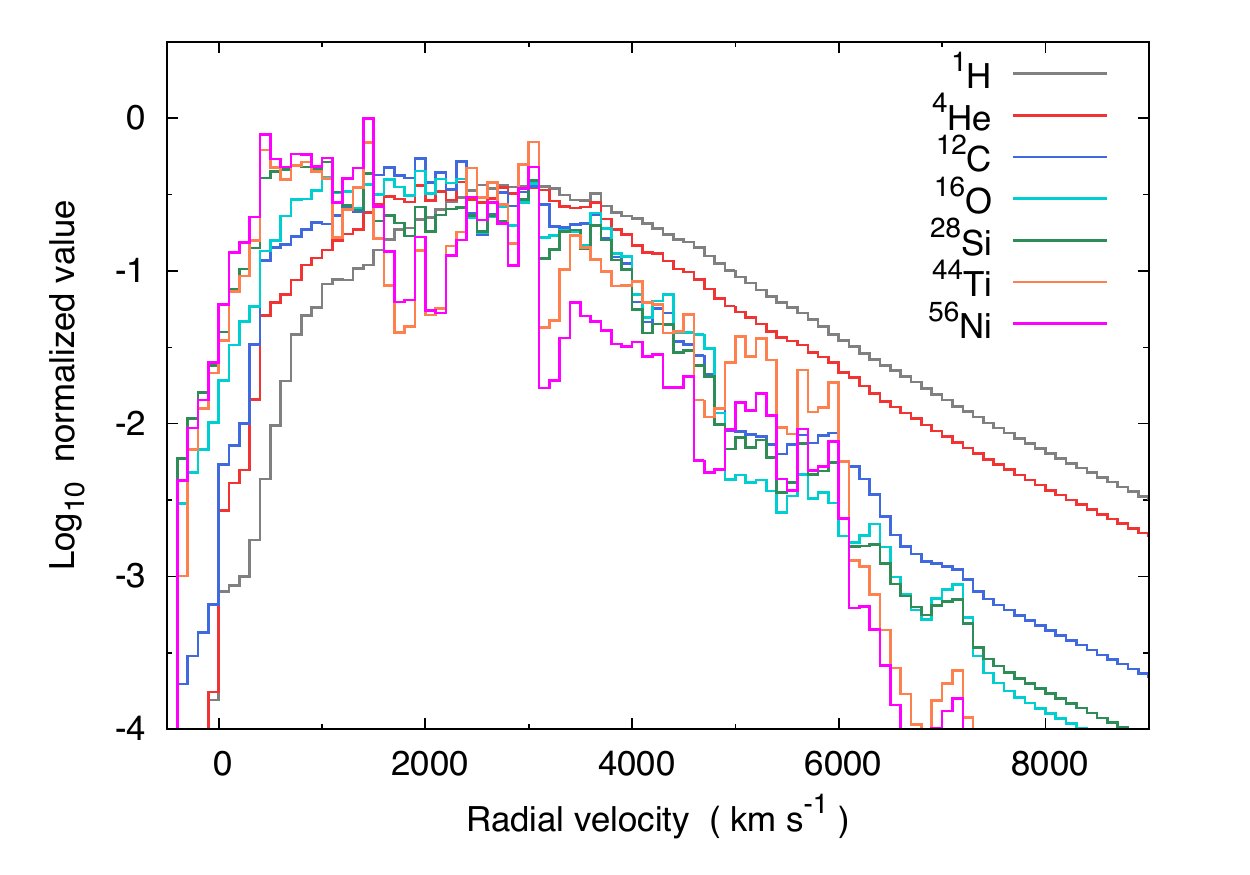}
\includegraphics[scale=0.7]{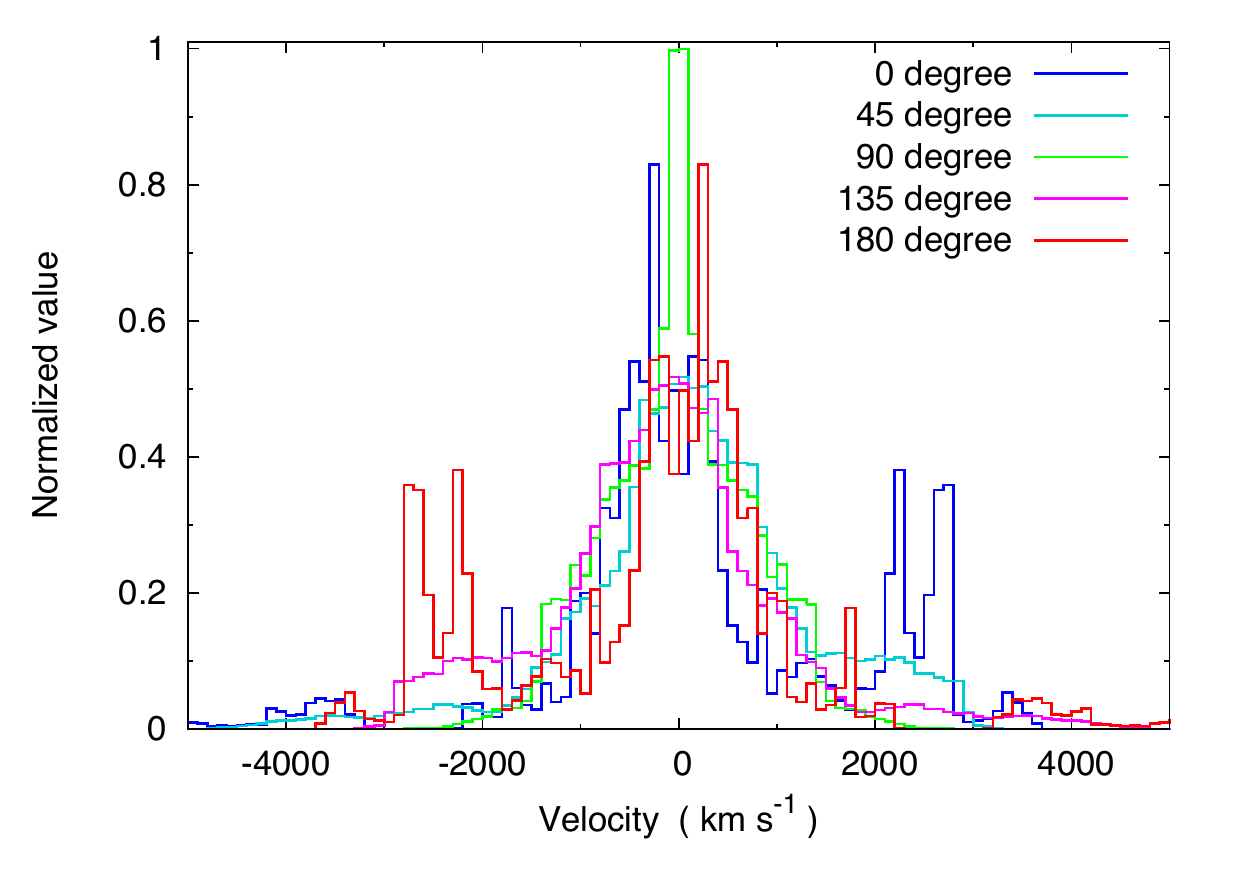}
\caption{Results obtained from the model Bipo+SC1p50m20. Panel descriptions are the same as those in Figure 3.
\label{fig14}}
\end{figure}


\begin{figure}
\includegraphics[scale=0.25]{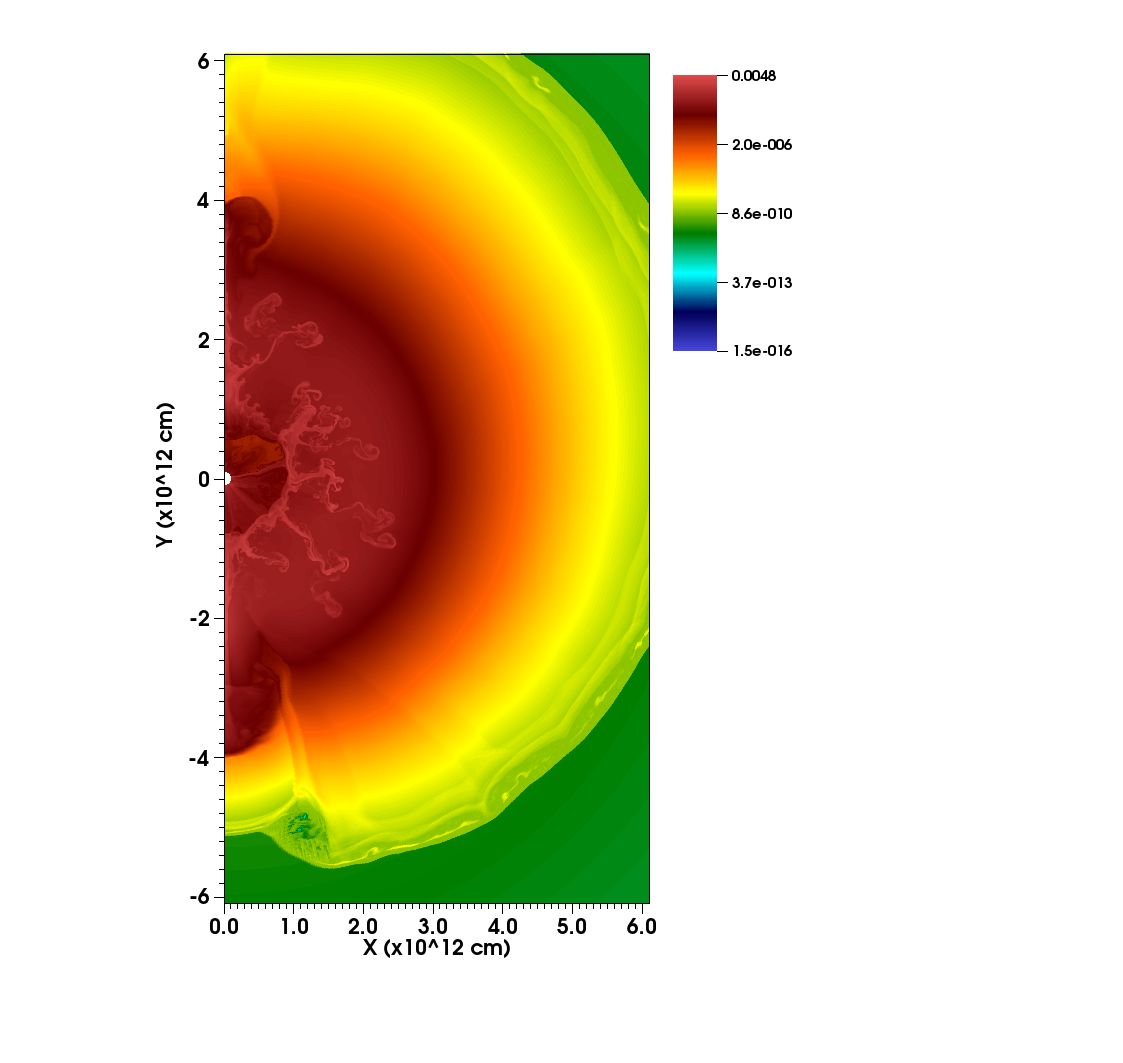}
\includegraphics[scale=0.25]{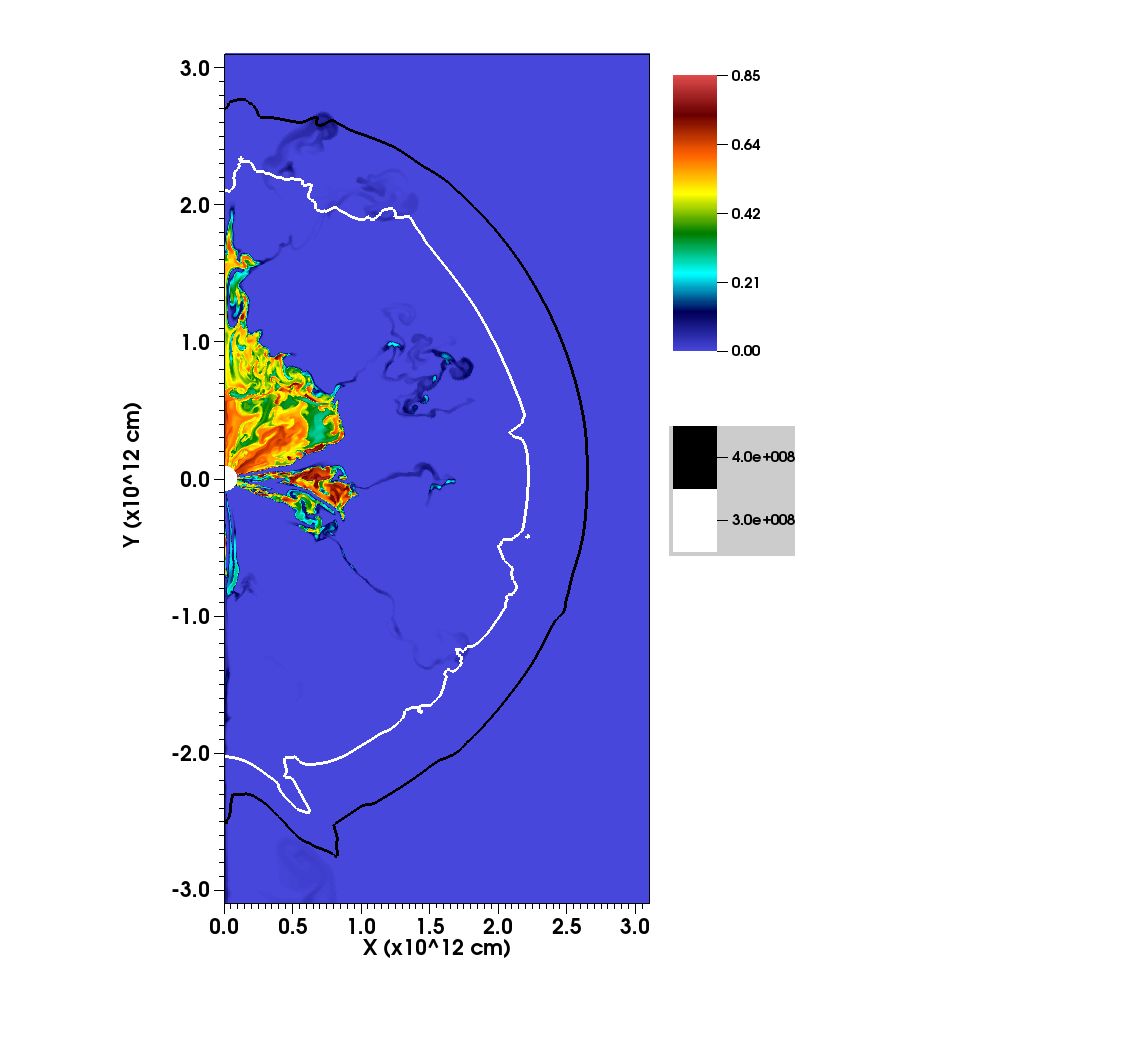}
\includegraphics[scale=0.7]{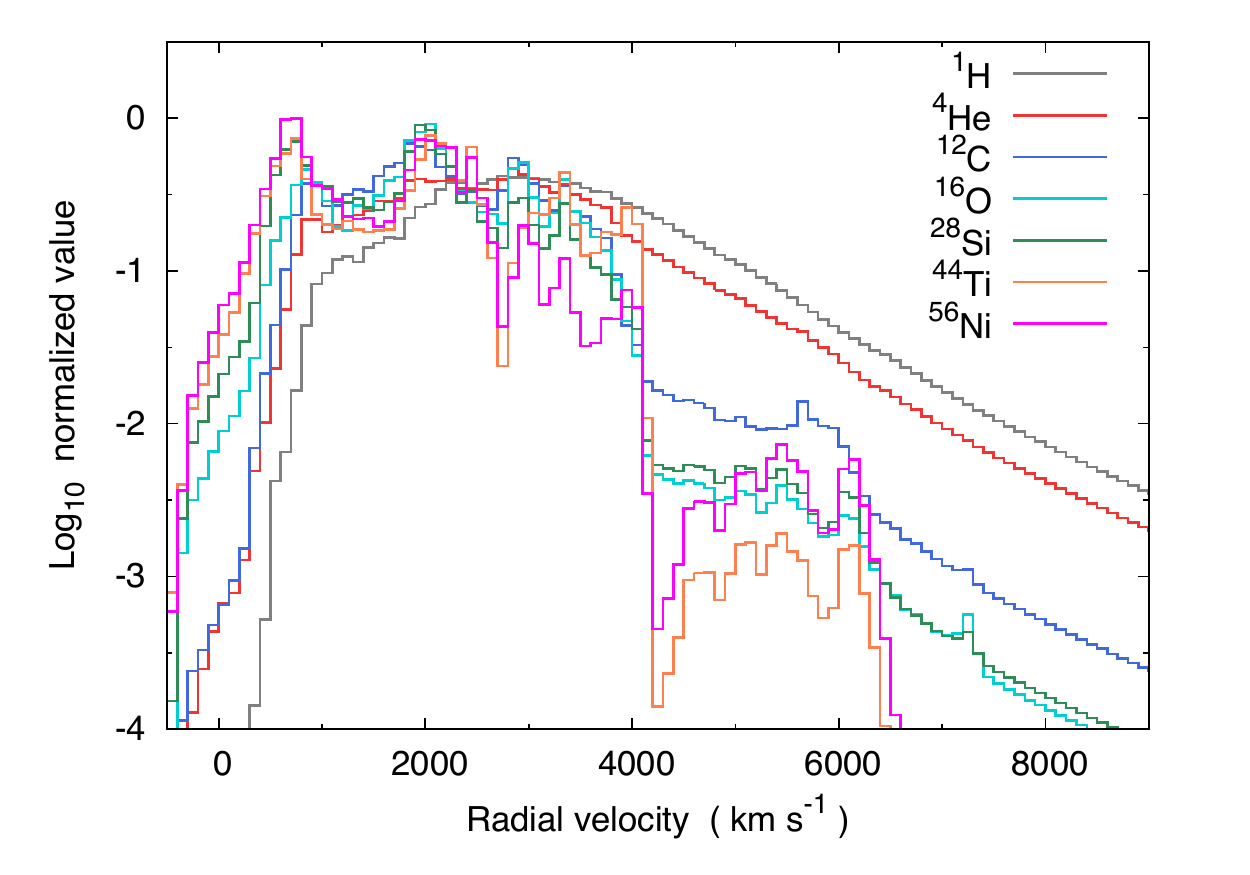}
\includegraphics[scale=0.7]{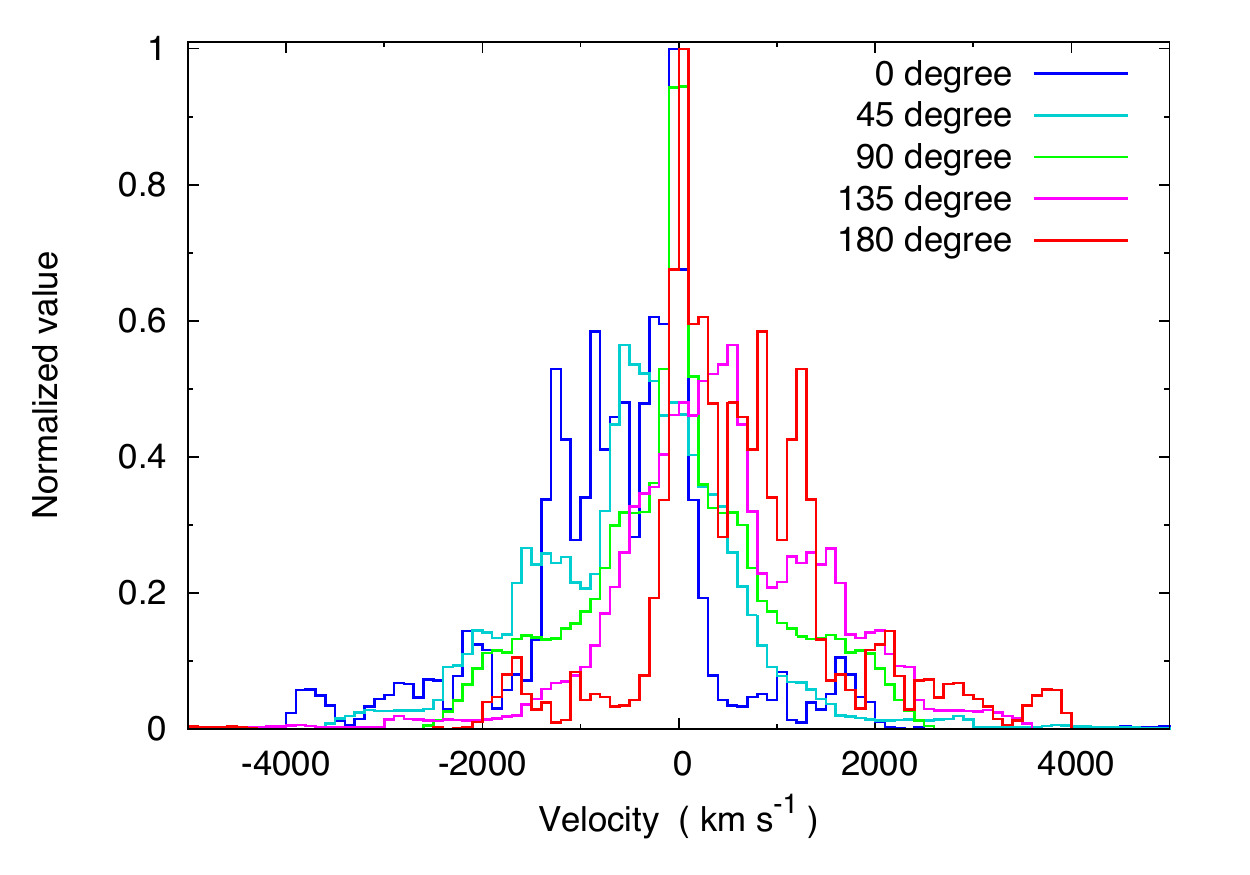}
\caption{Results obtained from the model AsyEqua+SC1p50m20. Panel descriptions are the same as those in Figure 3.
\label{fig15}}
\end{figure}


\begin{figure}
\includegraphics[scale=0.25]{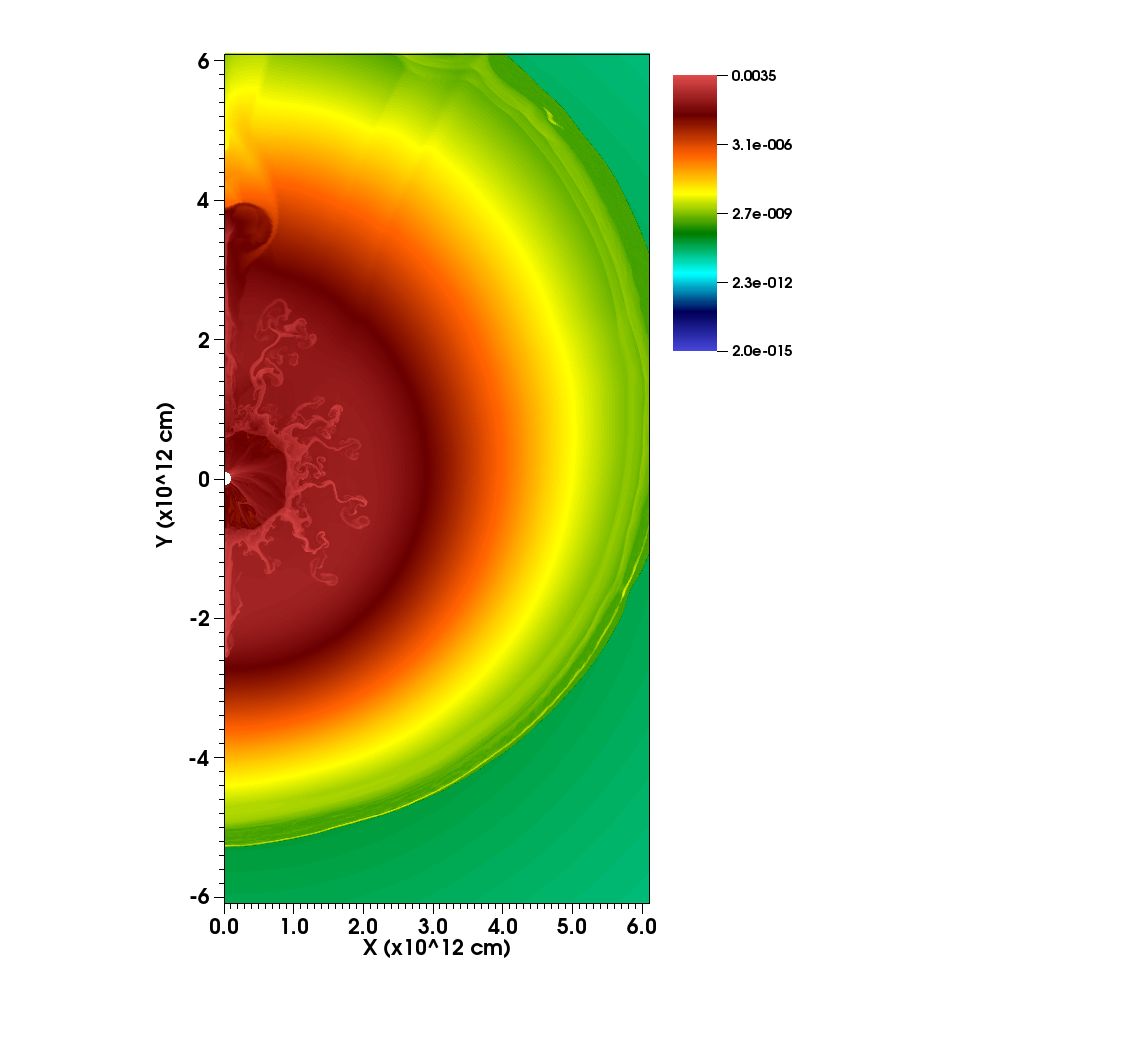}
\includegraphics[scale=0.25]{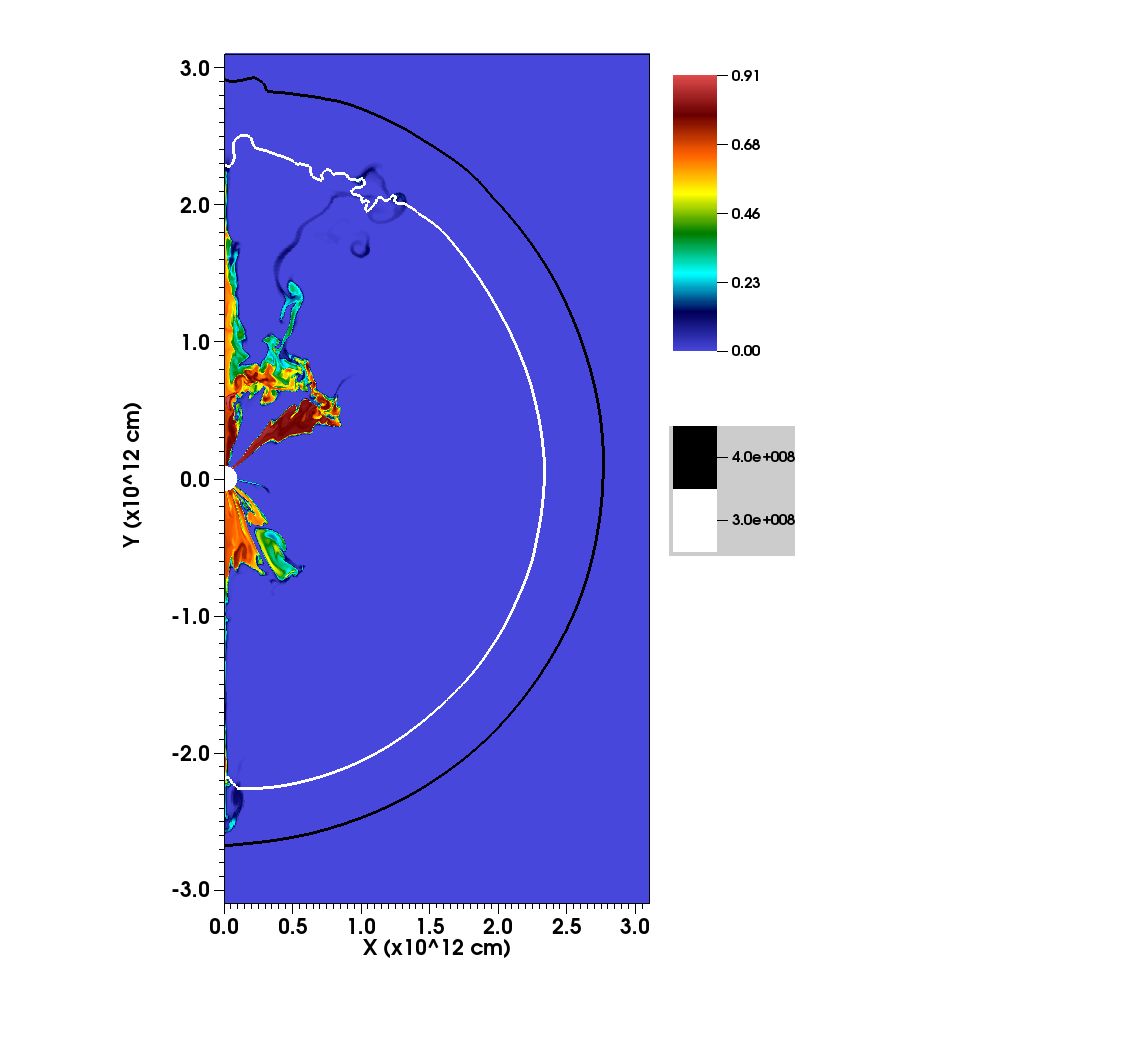}
\includegraphics[scale=0.7]{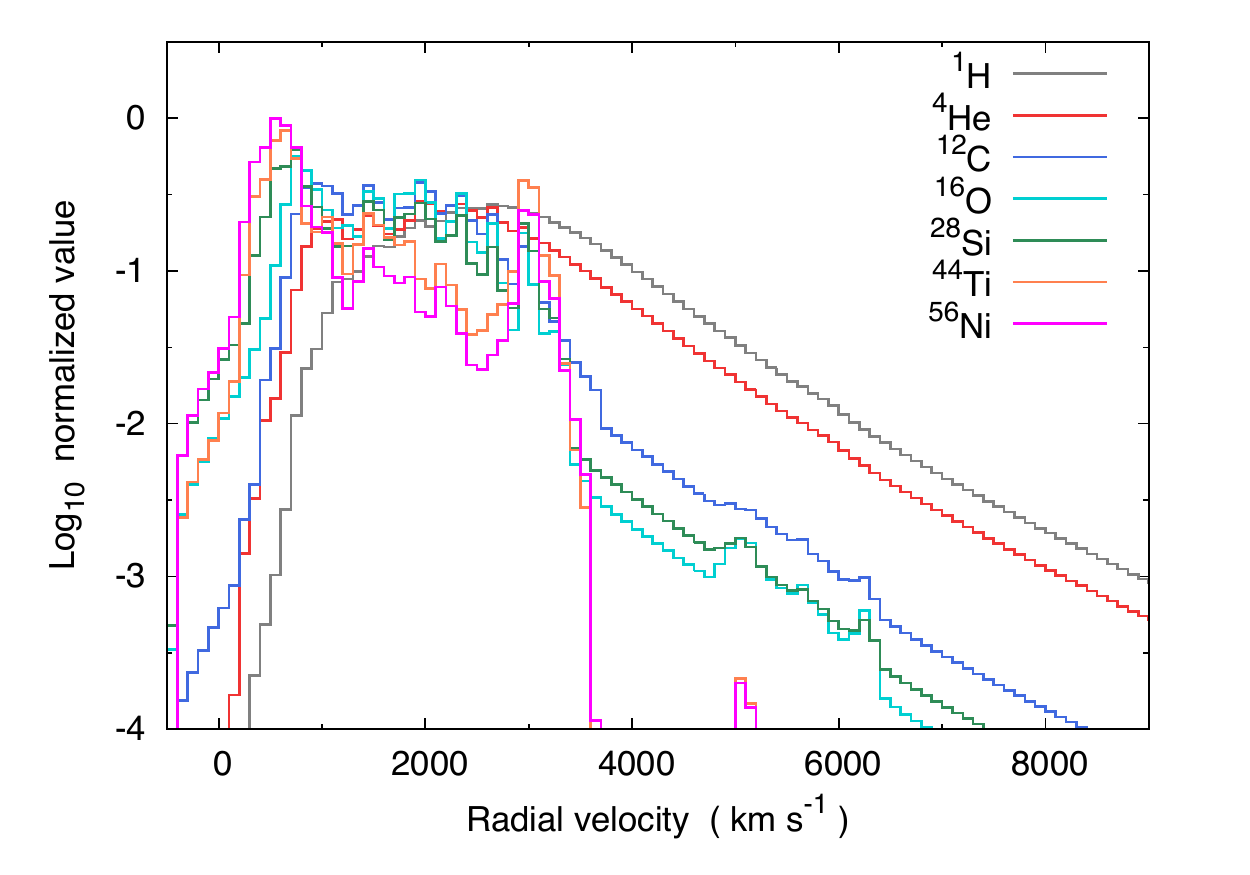}
\includegraphics[scale=0.7]{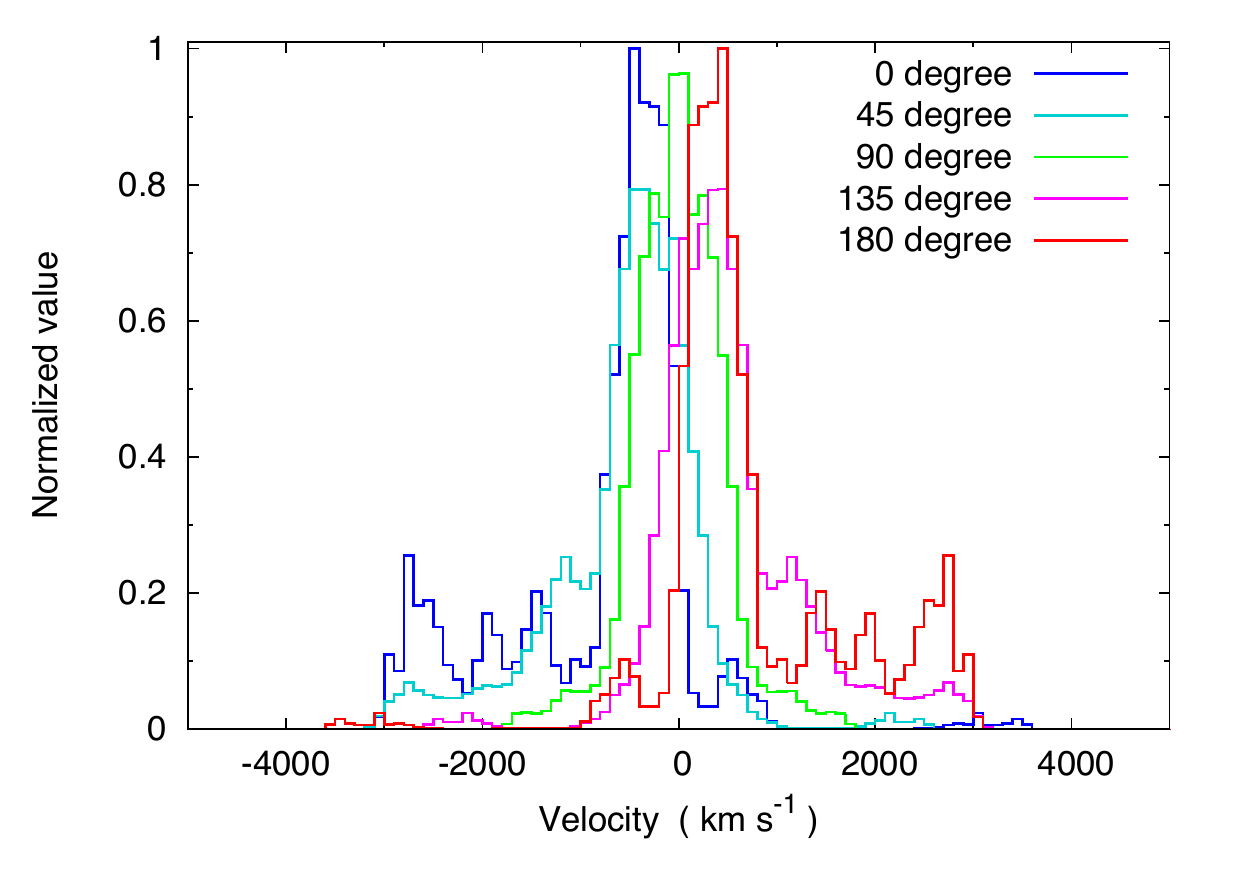}
\caption{Results obtained from the model Asph+SC1p50m20($\ast\ast$). Panel descriptions are the same as those in Figure 3.
\label{fig16}}
\end{figure}


\begin{figure}
\includegraphics[scale=0.7]{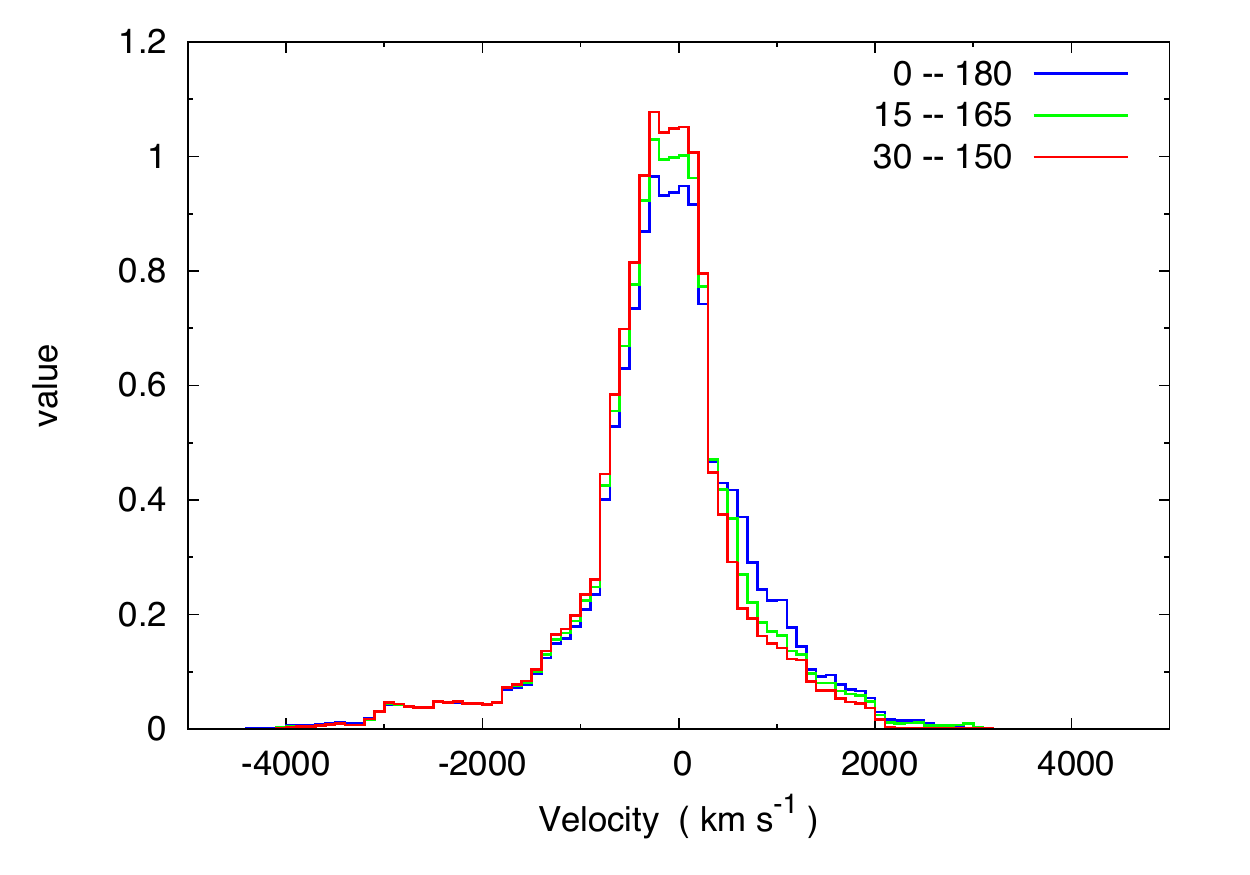}
\includegraphics[scale=0.7]{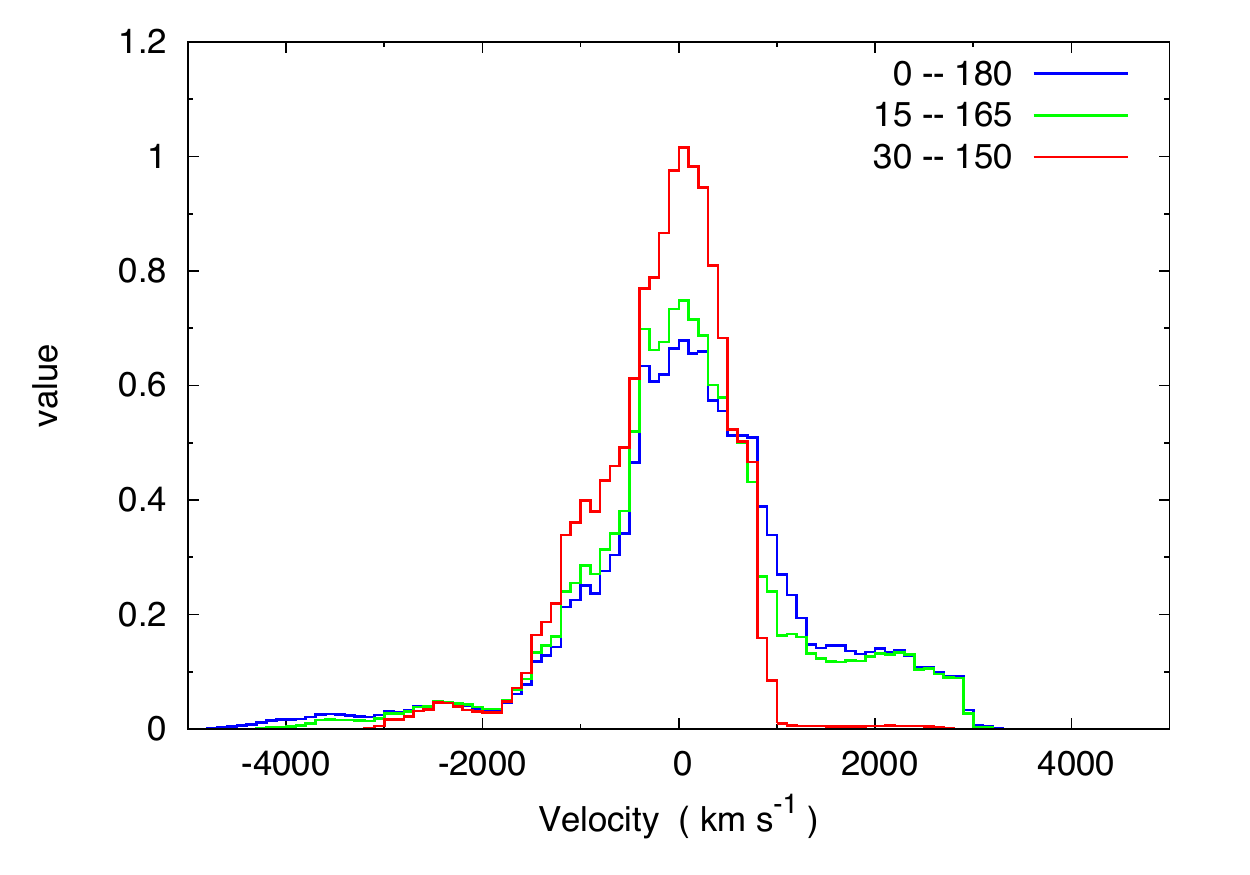}\\
\includegraphics[scale=0.7]{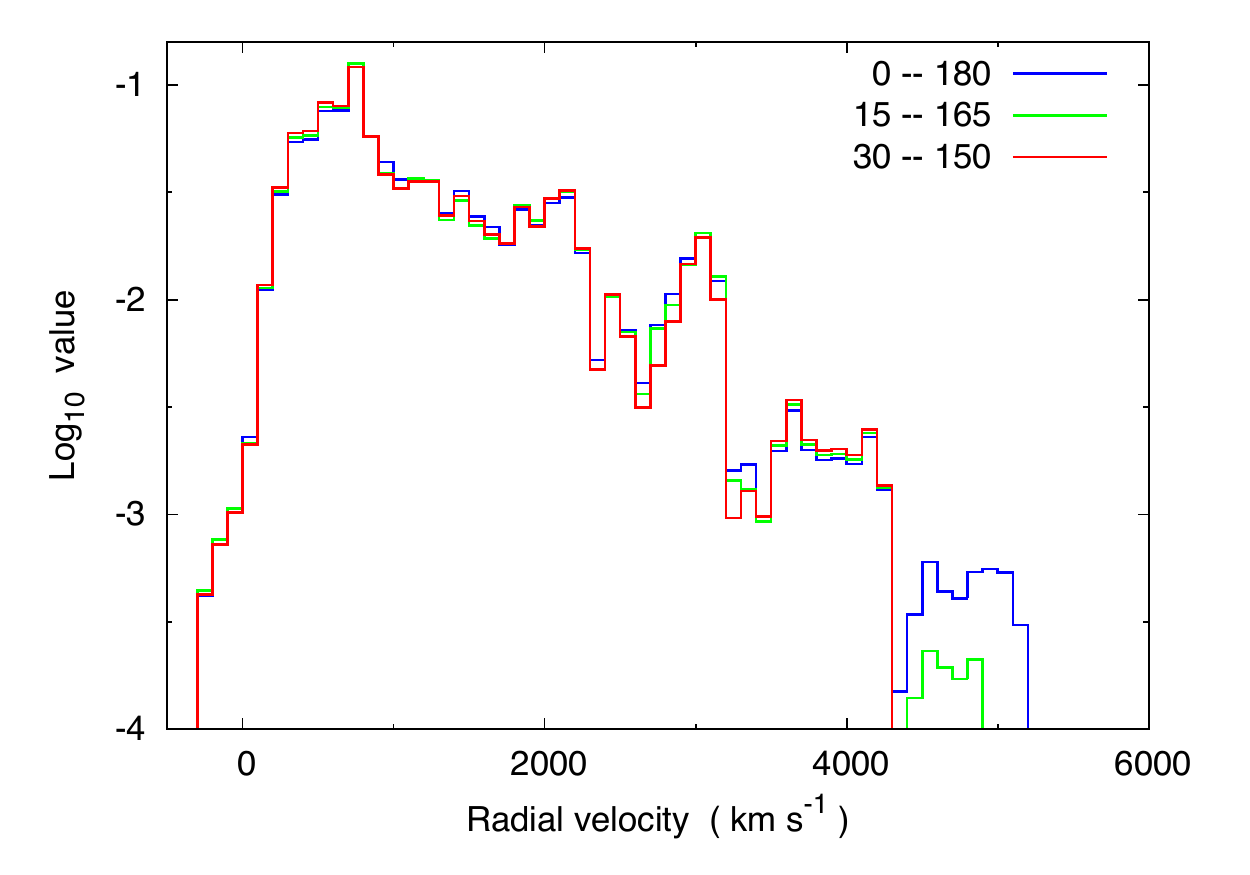}
\includegraphics[scale=0.7]{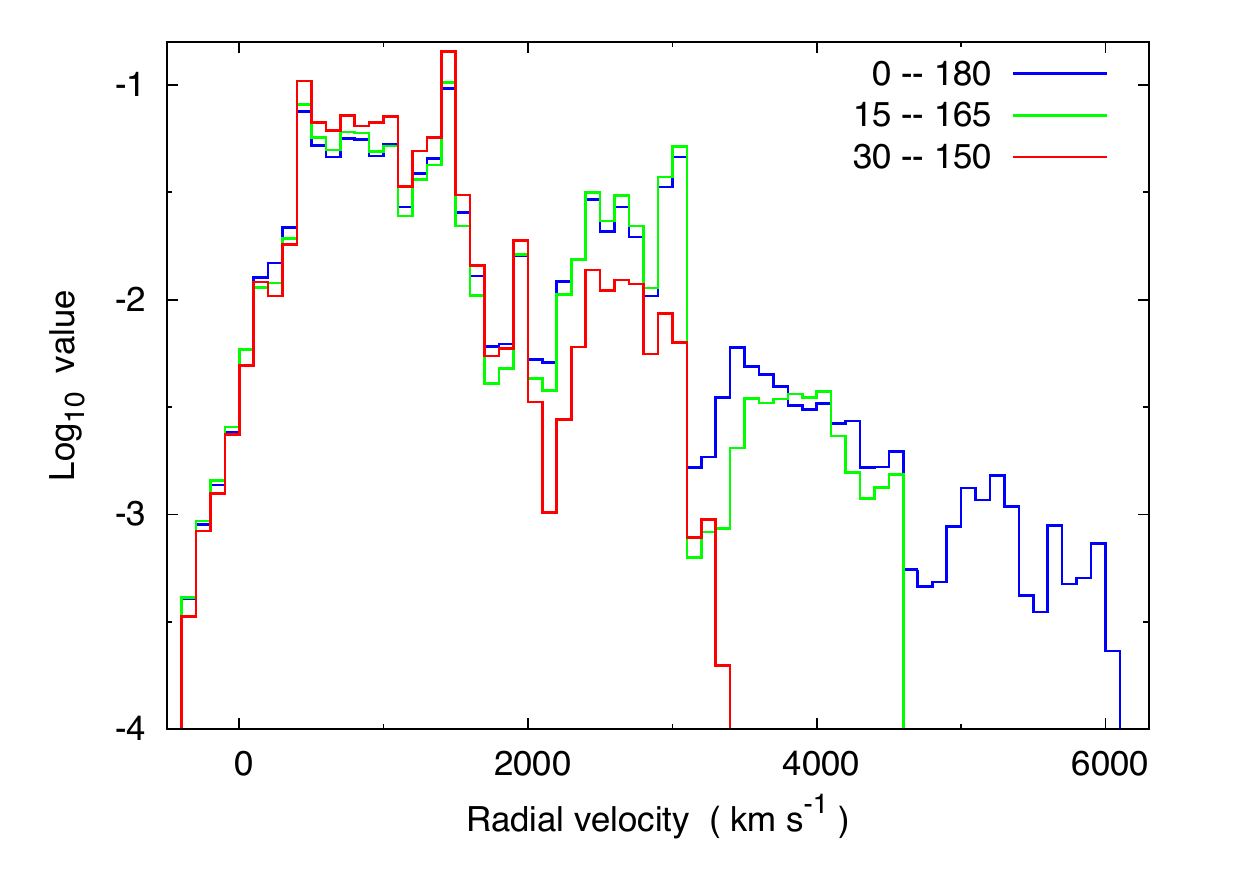}
\caption{Line of sight velocity profiles and radial velocity profiles of $^{56}$Ni for the simple test of the axis-jet effect. The labeled value in the y-axis of each panel is $\Delta M/M$ (without normalization), where $\Delta M$ is the mass in the velocity range of $v\sim v+\Delta v$, and $M$ is the total mass. The view angle is 45 degree and the velocity bin is 100 $\rm{km~s^{-1}}$.
Upper left: Line of sight velocity profiles of the model ``SC1p50m20"; upper right: line of sight velocity profiles of the model ``Bipo+C1p50m20"; lower left: radial velocity profiles of the model ``SC1p50m20"; lower right: radial velocity profiles of the model ``Bipo+C1p50m20". In each panel, ``0$-$180", ``15$-$165", and ``30$-$150" indicate the computational domain within the angular ranges of 0$-$180 degree, 15$-$165 degree, and 30$-$150 degree, respectively.
\label{fig17}}
\end{figure}

\begin{figure}
\includegraphics[scale=0.7]{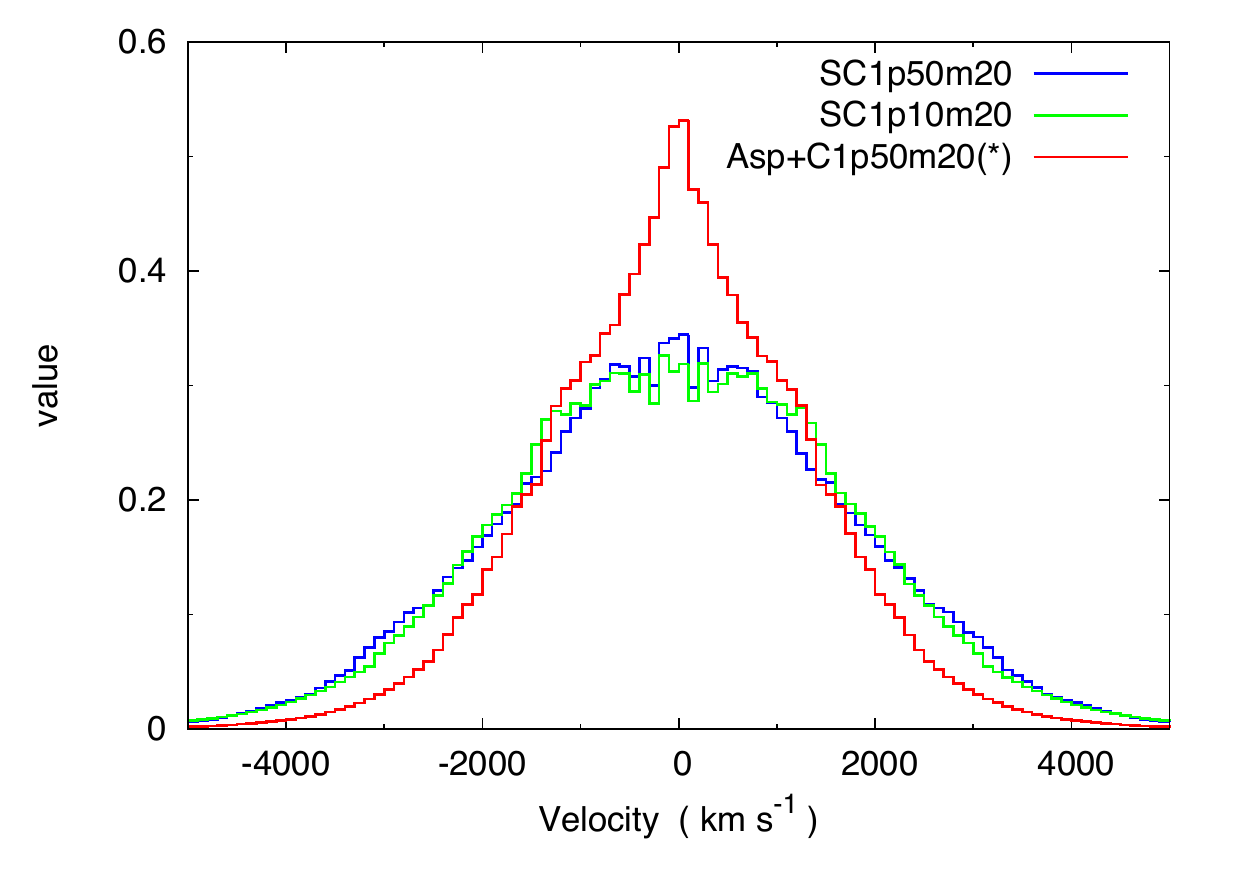}
\includegraphics[scale=0.7]{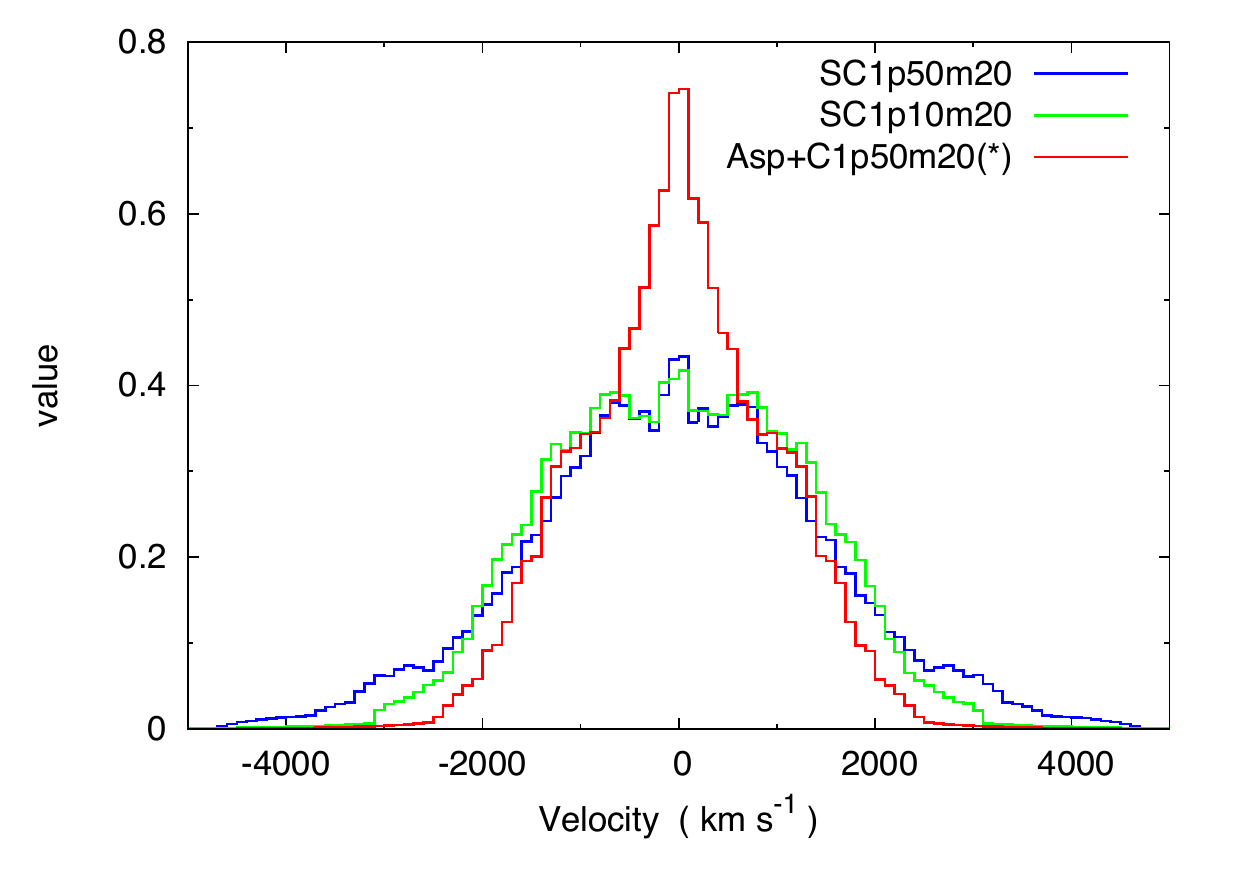}\\
\includegraphics[scale=0.7]{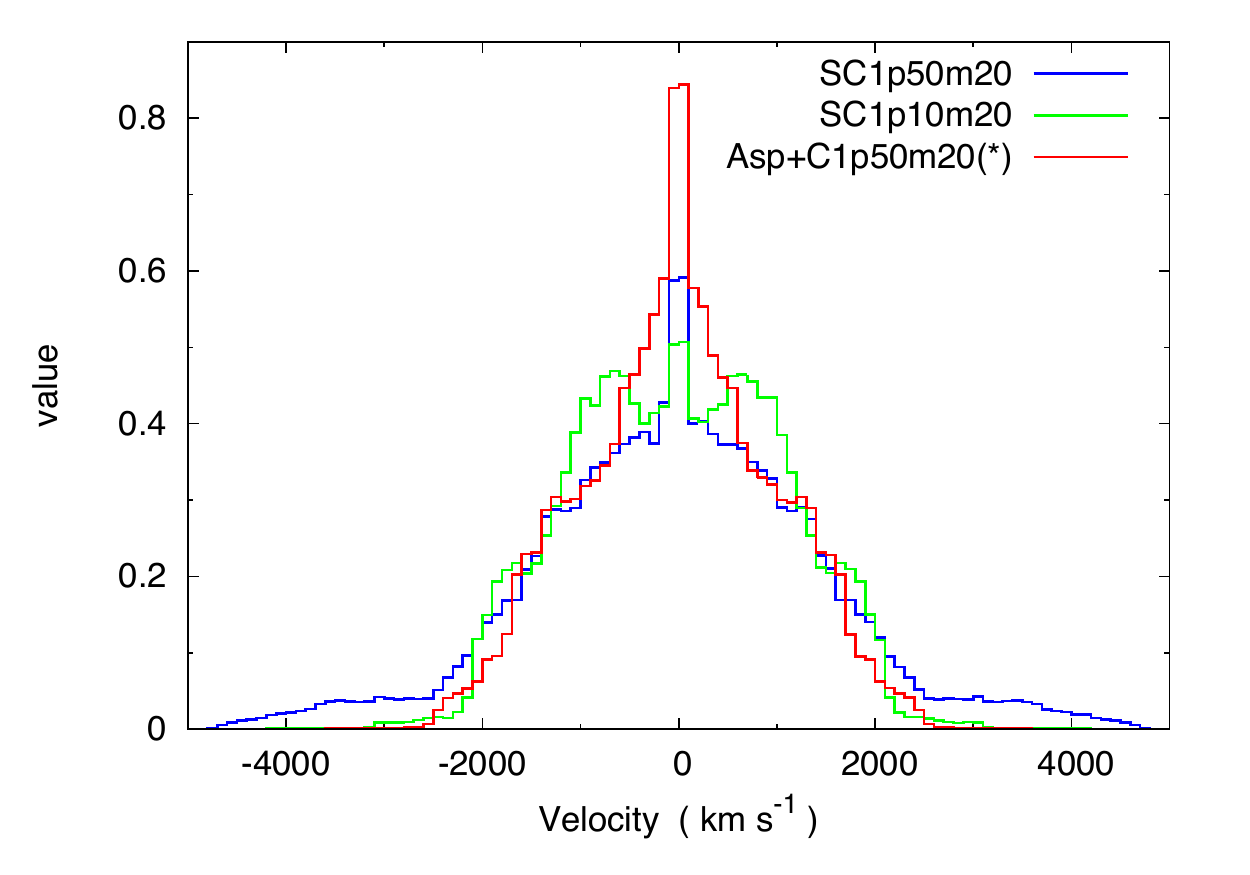}
\includegraphics[scale=0.7]{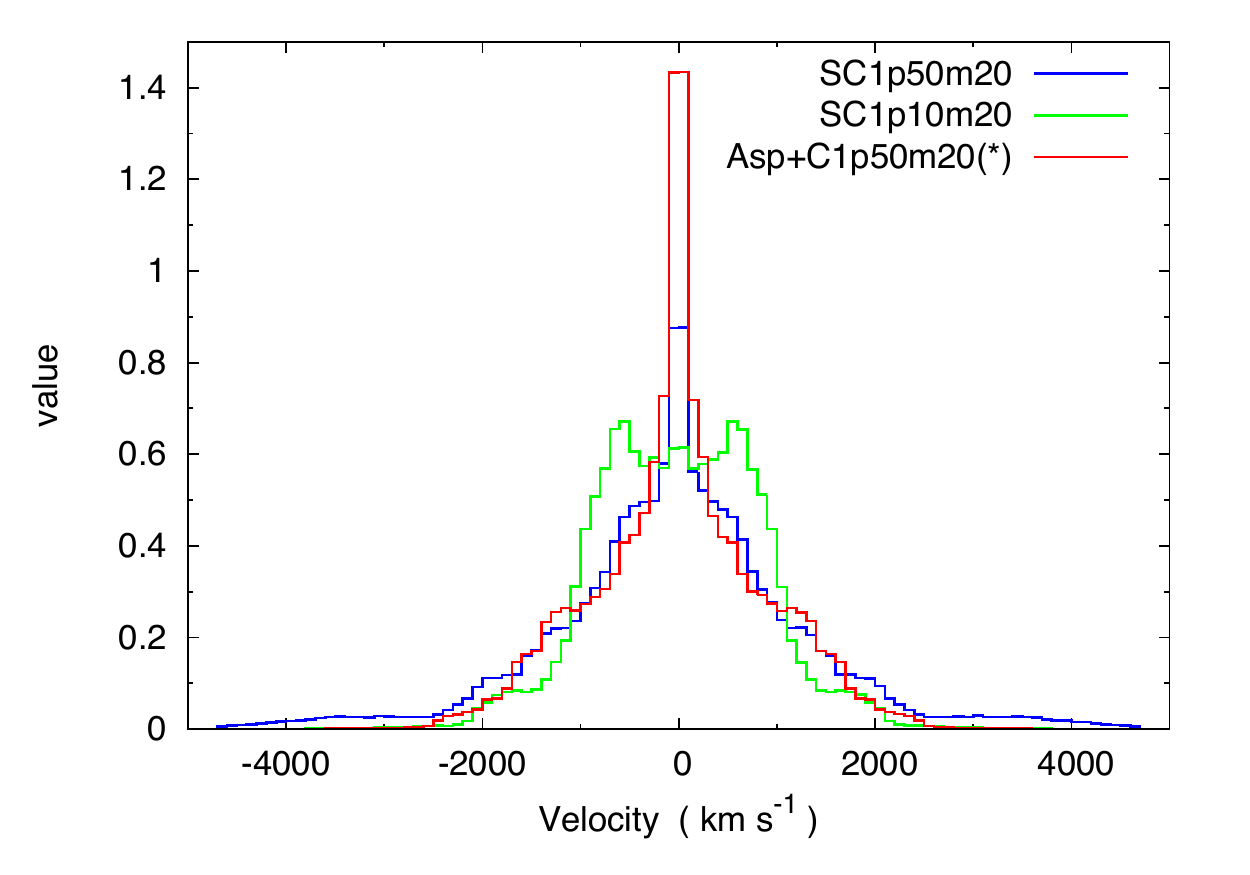}
\caption{Line of sight velocity profiles of four elements. The view angle is 90 degree and the velocity bin is 100 $\rm{km~s^{-1}}$. Upper left: line of sight velocity profiles of $^{4}$He; upper right: line of sight velocity profiles of $^{12}$C; lower left: line of sight velocity profiles of $^{16}$O; lower right: line of sight velocity profiles of $^{28}$Si. The labeled value in the y-axis of each panel is $\Delta M/M$ (without normalization), where $\Delta M$ is the mass in the velocity range of $v\sim v+\Delta v$, and $M$ is the total mass.
Three models are considered in each panel: SC1p50m20 (spherical explosion with 50\% of density perturbation), SC1p10m20 (spherical explosion with 10\% of density perturbation), and Asp+C1p50m20($\ast$) (aspherical explosion with 50\% of density perturbation).
\label{fig18}}
\end{figure}

\begin{figure}
\includegraphics[scale=0.7]{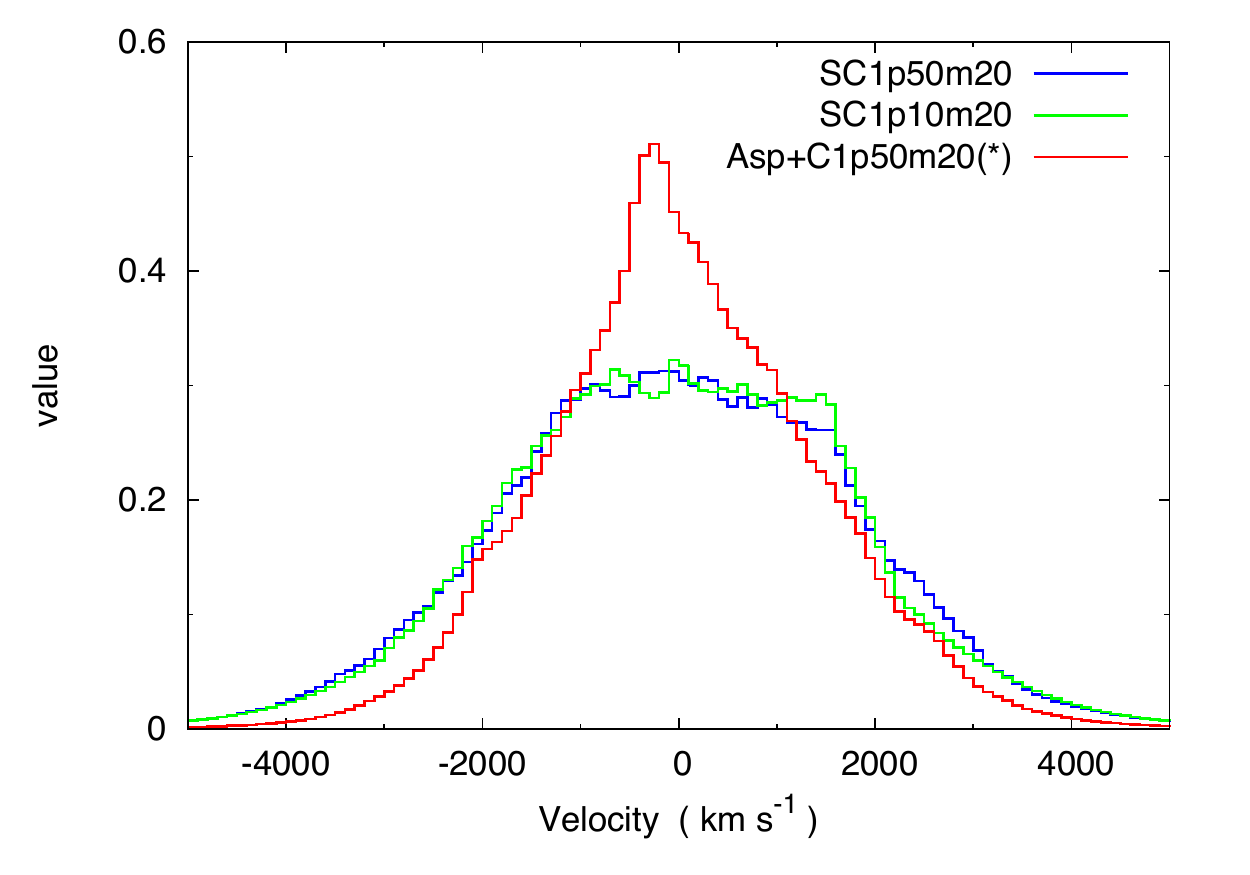}
\includegraphics[scale=0.7]{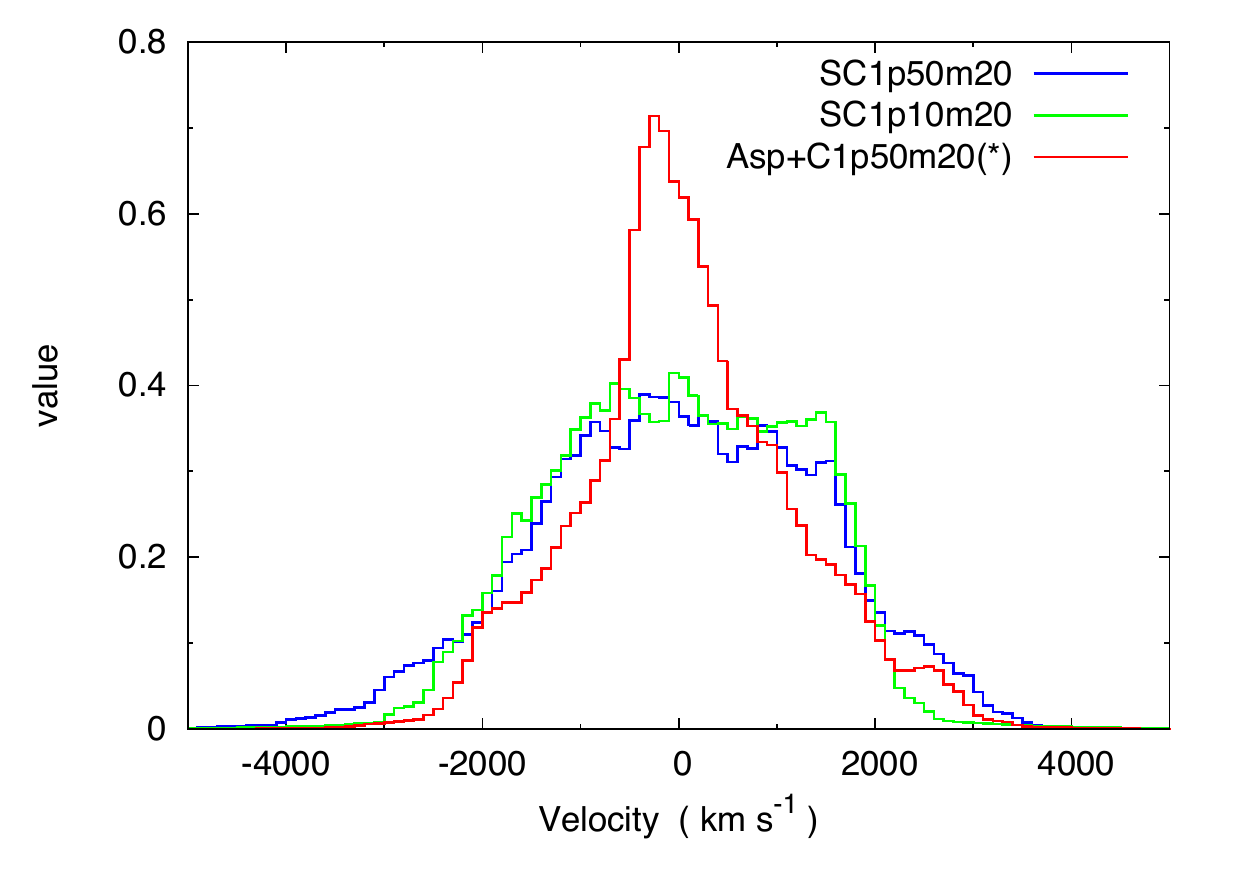}\\
\includegraphics[scale=0.7]{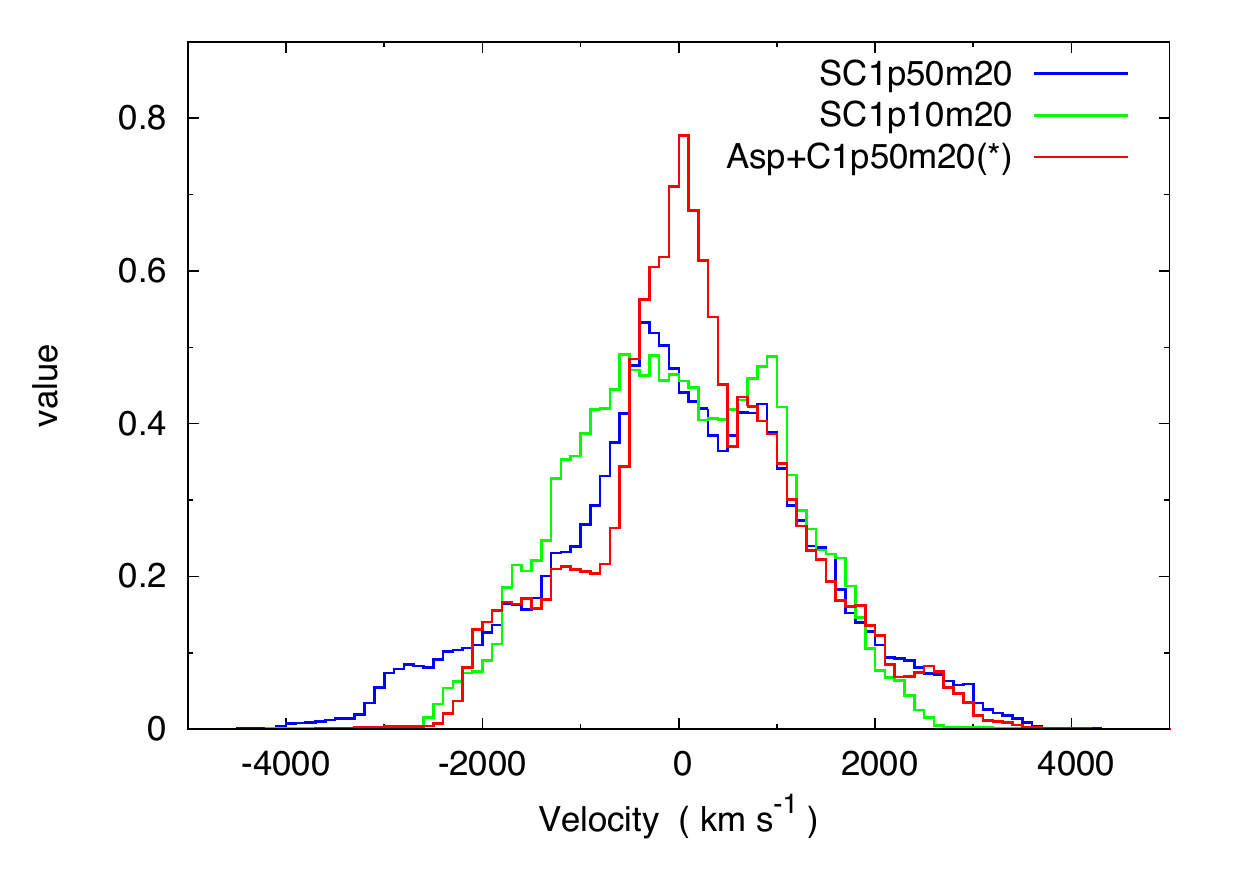}
\includegraphics[scale=0.7]{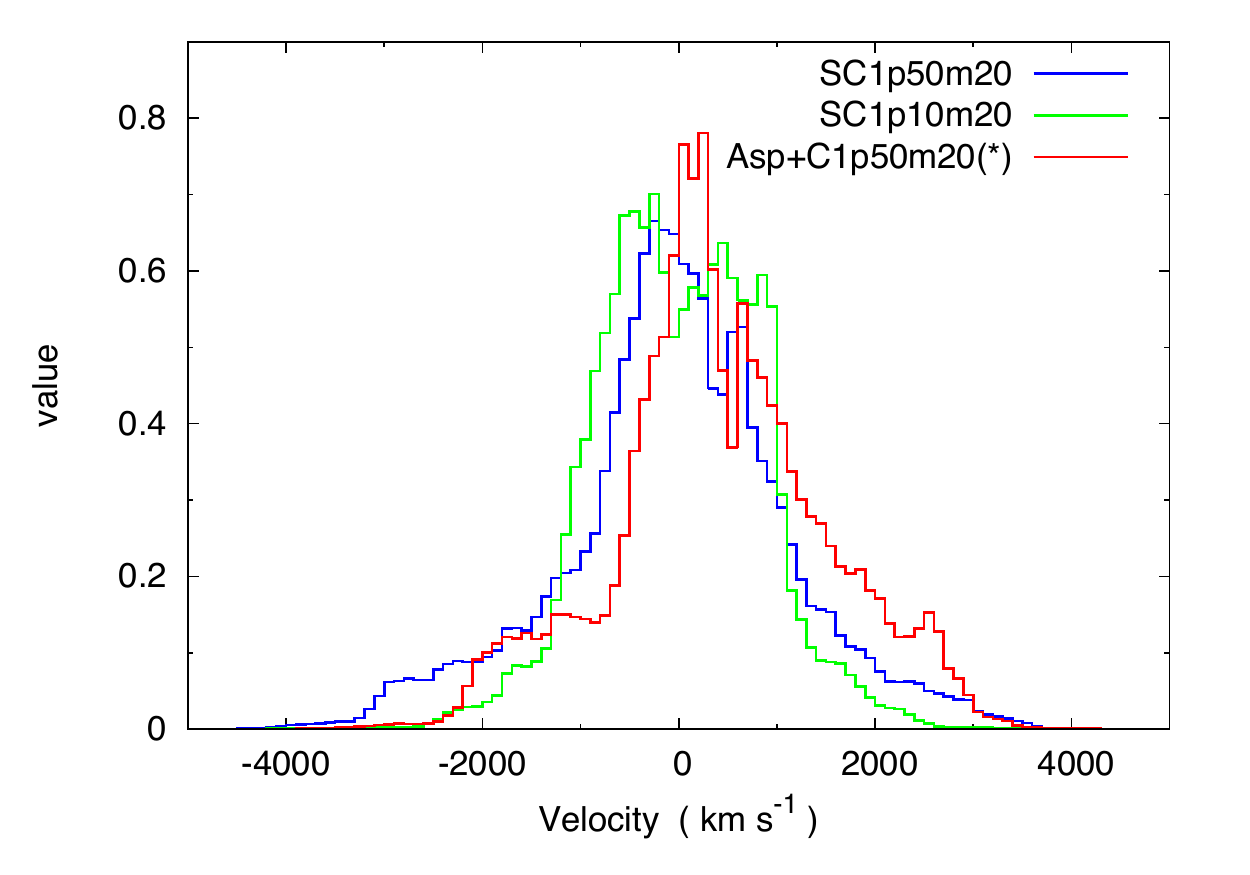}
\caption{Line of sight velocity profiles of four elements. The view angle is 45 degree and the velocity bin is 100 $\rm{km~s^{-1}}$. Upper left: line of sight velocity profiles of $^{4}$He; upper right: line of sight velocity profiles of $^{12}$C; lower left: line of sight velocity profiles of $^{16}$O; lower right: line of sight velocity profiles of $^{28}$Si. The labeled value in the y-axis of each panel is $\Delta M/M$ (without normalization), where $\Delta M$ is the mass in the velocity range of $v\sim v+\Delta v$, and $M$ is the total mass.
Three models are considered in each panel: SC1p50m20 (spherical explosion with 50\% of density perturbation), SC1p10m20 (spherical explosion with 10\% of density perturbation), and Asp+C1p50m20($\ast$) (aspherical explosion with 50\% of density perturbation).
\label{fig19}}
\end{figure}
\clearpage

\begin{table}
\begin{center}
\caption{Parameters in simulation models\label{tab:1}}
\end{center}
\begin{tabular}{lccccccccccccccccc}
\hline \hline
Model/Parameter &$\epsilon_0$& $m$ &  $f$ & $v_{\rm{pol}}/v_{\rm{eq}}$& $v_{\rm{up}}/v_{\rm{down}}$& Perturbation Position \\
\hline
Fiducialmodel   &  0.0       &  0  & -     &       -         &   -              &  no perturbation  \\
\hline
Basicmodel      &  0.5       & 20  &-     &       -         &   -              &  4 critical radii \\
\hline
SC1p50m8        &  0.5       &  8  &  -   &        -        &    -             &  4 critical radii \\
SC1p50m8($\Diamond$)&  0.5       &  8  &  -   &        -        &    -             &  4 critical radii \\
SC1p50m20       &  0.5       &  20 &  -   &        -        &    -             &  4 critical radii   \\
SC1p50m20($\ast$)& 0.5       &  20 &  -   &        -        &    -             &  4 critical radii  \\
SC1p10m8        &  0.1       &  8  &  -   & -               &    -             &  4 critical radii      \\
SC1p10m20       &  0.1       &  20 &  -   & -               &    -             &  4 critical radii      \\
\hline
SC1Si/C+O        &  0.5       &  20 &  -   &        -        &    -             &  Si/C+O interface      \\
SC1C+O/He        &  0.5       &  20 &  -   &        -        &    -             &  C+O/He interface      \\
SC1He/H         &  0.5       &  20 &  -   &        -        &    -             &   He/H interface    \\
SC1C+O/He/H      &  0.5       &  20 &  -   &        -        &    -             &  C+O/He and He/H interfaces     \\
\hline
SC2p50m8f5      &  0.5       &  8  &  5   &        -        &    -             &   radial \& angular modulated  \\
SC2p50m20f5     &  0.5       &  20 &  5   &        -        &    -             &   radial \& angular modulated  \\
SC2p10m8f5      &  0.1       &  8  &  5   &        -        &    -             &   radial \& angular modulated     \\
SC2p10m20f5     &  0.1       &  20 &  5   &        -        &    -             &   radial \& angular modulated     \\
SC2p50m20f1     &  0.5       &  20 &  1   &        -        &    -             &   radial \& angular modulated    \\
\hline
SC3p50f1           &  0.5       &  -  &  1   &        -        &    -             & radial modulated      \\
SC3p50f5           &  0.5       &  -  &  5   &        -        &    -             & radial modulated     \\
\hline
Bipo+C1p50m20  &   0.5       &  20 &  -   &       4.0       &    -             &   4 critical radii  \\
AsyEqua+C1p50m20&  0.5       &  20 &  -   &        -        &    1.8           &   4 critical radii  \\
Asph+C1p50m20   &  0.5       &  20 &  -   &       4.0       &    1.8           &   4 critical radii   \\
Asph+C1p25m20   &  0.25      &  20 &  -   &       4.0       &    1.8           &   4 critical radii   \\
Asph+C1p10m20   &  0.10      &  20 &  -   &       4.0       &    1.8           &   4 critical radii   \\
Asph+C1p50m20($\ast$)& 0.5   &  20 &  -   &       4.0       &    1.8           &   4 critical radii     \\
Asph+C1p25m20($\ast\ast$)&0.25&20&   -    &       4.0       &    1.8           &   4 critical radii     \\
\hline
\tablecomments{\scriptsize{SC1, SC2, and SC3 are three spherical SN explosion cases presented in Section 2.3.1. Bipolar explosion is labeled as ``Bipo" and equatorially asymmetric explosion is labeled as ``AsyEqua". The combination of bipolar and equatorially asymmetric explosions is labeled as ``Asph". The four critical radii in the CCSN progenitor are $3\times 10^8$ cm at Si/C+O interface, $3\times 10^9$ cm at oxygen shell burning position, $6\times 10^9$ cm at C+O/He interface, and $5\times 10^{10}$ cm at He/H interface, respectively. $\epsilon_0$ is the density perturbation amplitude, $m$ is the wave number of perturbations in the angular direction,
and $f$ is the scaling factor of the perturbation length scale.
The ratio of the initial radial velocity along the polar axis to that along the equatorial axis $v_{\rm{pol}}/v_{\rm{eq}}$ is given for bipolar explosion models. The ratio of the initial radial velocity $v_{\rm{up}}/v_{\rm{down}}$ at $\theta=0$ to that at $\theta=\pi$ is
given for equatorially asymmetric models. All models have input energies of $2.5\times 10^{51}$ erg; exceptions are those labeled by ``$\ast$" in which we set the input energy to be $1.7\times 10^{51}$ erg and the one labeled by ``$\ast\ast$" in which we set the input energy to be $2.0\times 10^{51}$ erg.
The model of SC1p50m8($\Diamond$) has a set of random numbers, which are different from the model of sc1p50m8.}
}
\end{tabular}
\end{table}

\clearpage

\begin{table}
\begin{center}
\caption{Main results of simulation models \label{tab:1}}
\end{center}
\begin{tabular}{lccccccccccccccccc}
\hline \hline
Model/Result&  Time &Explosion&  $^{56}\rm{Ni_{tot}}$ & $^{56}\rm{Ni_{ratio}}$ & $^{44}\rm{Ti_{tot}}$&  $v^{\rm{max}}_r$ &FigNo. \\
unit        &     (s)     &  ($10^{51}$erg)  &  ($M_\odot$)& (\%)    &($10^{-4}M_\odot$) &  ($\rm{km~s^{-1}}$)   \\
\hline
Fiducialmodel   &  6959     &   1.91         &  0.21     & -     & 2.8       &1700 &-  \\
\hline
Basicmodel      &  6982     &   1.87         &  0.19     & -     & 3.5       &2000 &-  \\
\hline
SC1p50m8        &  6577     &   1.80         &  0.14     & 23.1  & 2.7       &4800 & Fig. 3  \\
SC1p50m8($\Diamond$)& 6557  &   1.83         &  0.15     & 21.7  & 2.1       &3700 & Fig. 4     \\
SC1p50m20       &  7062     &   1.84         &  0.17     & 5.6   & 7.2       &4200 & Fig. 5  \\
SC1p50m20($\ast$)& 9006     &   1.08         &  0.13     & 0.9   & 3.2       &4100 & Fig. 13 \\
SC1p10m8        &   6978    &   1.89         &  0.19     & -     & 2.1       &1700 & -\\
SC1p10m20       &   7029    &   1.89         &  0.20     & -     & 2.8       &1500 & Fig. 6\\
\hline
SC1Si/C+O         &  7005    &   1.87        &  0.18      &   -  &  5.5      & 1700 & -\\
SC1C+O/He         &  7013    &   1.86        &   0.16     &  0.7 &  2.7      & 3100&  -  \\
SC1He/H         &    7196    &    1.88       &    0.19    &  0.6 &  2.6      & 2100 & -\\
SC1C+O/He/H      &   7087    &  1.83          &  0.15     & 11.5 &  2.3      &4800&  Fig. 7 \\
\hline
SC2p50m8f5       &  5785    &   2.01         &  0.18     & 8.3   & 3.0       &3700 & Fig. 8 \\
SC2p50m20f5      &  6618    &   1.96         &  0.21     & 5.1   & 3.8       &4200 & Fig. 9\\
SC2p10m8f5       &  6762    &   1.94         &  0.21     & -     & 3.0       & 1800& - \\
SC2p10m20f5      &  6892    &   1.93         &  0.21     & -     & 3.0       &1600 & Fig. 10 \\
SC2p50m20f1      &  6636    &   1.89         &  0.21     & 0.1   & 4.4       &2800&  Fig. 11   \\
\hline
SC3p50f1         &  6964    &    1.88         &  0.20    &   -   &  3.2      & 1900 & Fig. 12\\
SC3p50f5         &  6330    &    2.01         &  0.19    &   -   &  2.9      & 2000 &-\\
\hline
Bipo+SC1p50m20  &  6714     &   1.87         &  0.09     & 10.6  & 6.9       &6000& Fig. 14 \\
AsyEqua+SC1p50m20&  6365    &   1.88         &  0.12     & 7.0   & 9.4       &6000& Fig. 15 \\
Asph+SC1p50m20   & 6136     &   1.85         &  0.09     & 19.4  & 7.5       &5300& -  \\
Asph+SC1p25m20   &6101      &   1.91         &  0.09     &  9.2  & 4.0       &4200 & - \\
Asph+SC1p10m20   & 6382     &   1.94         &  0.09     &  1.4  & 2.7       &2100 &- \\
Asph+SC1p50m20($\ast$)& 8133&   1.11         &  0.06     & 10.0  & 4.9      &4500& -  \\
Asph+SC1p25m20($\ast\ast$)& 6984& 1.43         &  0.07     & 6.5  & 2.8      &3400& Fig. 16  \\
\hline
\tablecomments{The physical time at which we stop the simulation is labeled as ``Time" and the explosion energy is labeled as ``Explosion". The total masses of $^{56}$Ni and $^{44}$Ti are written as $^{56}\rm{Ni_{tot}}$ and $^{44}\rm{Ti_{tot}}$, respectively. All of these element mass values are calculated from the 2D FLASH code in which the limited nucleosynthesis network is adopted.
The ratio of the high velocity $^{56}$Ni ($>3000~\rm{km~s^{-1}}$) mass to the total mass of $^{56}$Ni is written
as $^{56}\rm{Ni_{ratio}}$.
The maximum radial velocity of $^{56}$Ni corresponding to the value of $\Delta M/M$ larger than $1.0\times 10^{-3}$ in radial velocity profiles is labeled as $v^{\rm{max}}_r$, where $\Delta M$ is the mass in the velocity range of $v\sim v+\Delta v$, and $M$ is the total mass. ``FigNo." is the figure number corresponding to the related simulation model in this paper.
}
\end{tabular}
\end{table}

\begin{sidewaystable}
\scriptsize
\begin{center}
\caption{Minimum and maximum radial velocities ($v^{\rm{min}}_r$ and $v^{\rm{max}}_r$) of each element
\label{tab:1}}
\end{center}
\begin{tabular}{lcccccccccccccccc}
\hline \hline
Model/Result\footnote{The minimum and maximum radial velocity of each element corresponds to the value of $\Delta M/M$ larger than $1.0\times 10^{-3}$ in radial velocity profiles, where $\Delta M$ is the mass in the velocity range of $v\sim v+\Delta v$, and $M$ is the total mass.}&$v^{\rm{max}}_r$($^{56}$Ni)&$v^{\rm{max}}_r$($^{44}$Ti)&$v^{\rm{min}}_r$\&$v^{\rm{max}}_r$($^{28}$Si)&$v^{\rm{min}}_r$\&$v^{\rm{max}}_r$($^{16}$O)&$v^{\rm{min}}_r$
\&$v^{\rm{max}}_r$($^{12}$C)&$v^{\rm{min}}_r$\&$v^{\rm{max}}_r$($^{4}$He)&$v^{\rm{min}}_r$(H) &FigNo.\footnote{FigNo. is the figure number corresponding to the related simulation model in this paper.} \\
Unit        &($\rm{km~s^{-1}}$)&($\rm{km~s^{-1}}$)&($\rm{km~s^{-1}}$)&($\rm{km~s^{-1}}$)&($\rm{km~s^{-1}}$) &($\rm{km~s^{-1}}$)   \\
\hline
Fiducialmodel         &  1700    & 1700  &1300 \& 1600   & 1300 \& 1800 &1400 \& 4600 &  1500 \& 6700  & 1900 &-       \\
\hline
Basicmodel            &  2000    & 2000  & 300 \& 3000   & 600 \& 3000  & 700 \& 4400   & 700 \& 6700  & 800    &  -  \\
\hline
SC1p50m8              &  4800    & 4800 & 0    \& 4800 & 300 \& 4300 & 700 \& 4900  & 700 \& 6400 & 900 & Fig. 3 \\
SC1p50m8($\Diamond$)  &  3700    & 3700 & 0    \& 3800 & 300 \& 3800  & 500 \& 4500  & 500 \& 6000 & 700 & Fig. 4 \\
SC1p50m20             &  4200    & 4500 & -100 \& 4700 & -200 \& 4700 & 500 \& 5100  & 500 \& 6600 & 600 & Fig. 5 \\
SC1p50m20($\ast$)     &  4100    & 4000 & -300 \& 3700 & -100 \& 3700 & 200 \& 3800  & 200 \& 5300 & 400 & Fig. 13 \\
SC1p10m8              &  1700    & 1700 & 900  \& 2300 & 1000 \& 2500& 1100 \& 4500  & 1100 \& 6700 & 1400& -\\
SC1p10m20             &  1500    & 1500 & 600  \& 3100 & 800  \& 3100 & 900  \& 4500 & 800 \& 6700 & 1000 & Fig. 6\\
\hline
SC1Si/C+O             &  1700     &1700     & 900 \& 2100  & 1100 \& 2200& 1200 \& 4500  & 1200 \& 6700 & 1500  & -\\
SC1C+O/He            &  3100      & 3100  & 400 \& 3200  & 700 \& 3200   & 800 \& 4500    & 800 \& 6700   &1000  &  -  \\
SC1He/H               &  2100     & 2100   & 300 \& 4800  & 300 \& 4800   & 600 \& 5500    & 400 \& 6600   &700  & -\\
SC1C+O/He/H           & 4800      & 4800    & 200\& 4800  & 400 \& 4900  &  500 \& 5300    & 600 \& 6500 & 600  &  Fig. 7 \\
\hline
SC2p50m8f5       &  3700   &  3700& -100 \& 3700  & 300 \& 5600  & 400 \& 5800  & 400 \& 6900  & 700 & Fig. 8\\
SC2p50m20f5      &  4200   &  4600&  400 \& 4600  & 500 \& 4600  & 500 \& 5400  & 500 \& 6800  & 700 & Fig. 9\\
SC2p10m8f5       &  1800     & 1800  & 1000 \& 2300   &1100 \& 2800   & 1200 \& 4600  &1100 \& 6800   & 1400& - \\
SC2p10m20f5      &  1600     & 1500    & 500 \& 3300    & 600 \& 3400   & 800 \& 4500   & 700 \& 6800   & 900 & Fig. 10 \\
SC2p50m20f1      &  2800     & 2700 & 600 \& 3700    & 700 \& 3900   & 700 \& 5000   & 600 \& 6700   & 900 &  Fig. 11   \\
\hline
SC3p50f1         &  1900     & 1800 & 100 \& 3100    & 400 \& 3100   & 700 \& 4700   & 600 \& 6800   &900 & Fig. 12\\
SC3p50f5         &  2000     & 2000 & 800 \& 2200    & 1000 \& 2200  & 1200 \& 4800  &1300 \& 7400   &1900&-\\
\hline
Bipo+SC1p50m20  &  6000    & 6000 & -400 \& 6400 & -400 \& 6400 & 0 \& 7100 & 0 \& 9800 & 300 & Fig. 14 \\
AsyEqua+SC1p50m20& 6000    & 5600 & -400 \& 6300 & -400 \& 6300 &200 \& 7200 &200 \& 9900 & 500 & Fig. 15 \\
Asph+SC1p50m20   & 5300    & 5300  & 0 \& 5300     &    0 \& 5300 & 200 \& 5300&300 \& 6600 & 600 &  - \\
Asph+SC1p25m20   & 4200    & 4300  & -200 \& 3900 & 200 \& 3900 &500 \& 4400&500 \& 6800& 800 & - \\
Asph+SC1p10m20   & 2100      & 3500  & 100 \& 3800 & 400 \& 3800   & 800 \& 4500 &800 \& 6800& 1100 & - \\
Asph+SC1p50m20($\ast$)&4500  & 3900   & -400 \& 3900& -400 \& 3800  & -100 \& 4100 & 0 \& 5400  & 400 & -  \\
Asph+SC1p25m20($\ast\ast$)&3400  &3400& -300 \& 3400& -100 \& 3300  & 400 \& 4000 & 400 \& 6000  & 700 & Fig. 16  \\
\hline
\end{tabular}
\end{sidewaystable}


\begin{thebibliography}{}
\bibitem[Ahmad et al.(2006)]{ahmad} Ahmad, I., Greene, J. P., Moore, E. F., Ghelberg, S., Ofan, A., Paul, M., \& Kutschera, W. 2006, Phys. Rev. C., 74, 065803
\bibitem[Arnett(1994)]{arnett94} Arnett, W. D. 1994, \apj, 427, 932
\bibitem[Arnett et al.(1989)]{arnett89} Arnett, W. D., Fryxell, B., \& M\"{u}ller, E. 1989, \apj, 341, L63
\bibitem[Arnett et al.(2009)]{arnett09} Arnett, W. D., Meakin, C., \& Young, P. A. 2009, \apj, 690, 1715
\bibitem[Arnett \& Meakin(2011a)]{arnett11a} Arnett, W. D., \& Meakin, C. 2011a, \apj, 733, 78
\bibitem[Arnett \& Meakin(2011b)]{arnett11b} Arnett, W. D., \& Meakin, C. 2011b, \apj, 741, 33
\bibitem[Bazan \& Arnett(1994)]{bazan94} Bazan, G., \& Arnett, W. D. 1994, \apj, 433, L41
\bibitem[Bazan \& Arnett(1998)]{bazan98} Bazan, G., \& Arnett, W. D. 1998, \apj, 496, L316
\bibitem[Benz \& Thielemann(1990)]{benz90} Benz, W., \& Thielemann, F.-K. 1990, \apj, 348, L17
\bibitem[SASI, Blondin et al.(2003)]{blondin03} Blondin, J. M., Mezzacappa, A., \& DeMarino, C. 2003, \apj, 584, 971
\bibitem[Bruenn et al.(2013)]{bruenn13} Bruenn, S. W., Mezzacappa, A., Hix, W. R. et al. 2013, \apj, 767, L6
\bibitem[Burrows et al.(1995)]{burrows95} Burrows, A., Hayes, J., \& Fryxell, B. A. 1995, \apj, 450, 830
\bibitem[Burrows et al.(2007)]{burrows07} Burrows, A., Dessart, L., Livne, E., Ott, C. D., \& Murphy, J. 2007, \apj, 664, 416
\bibitem[Burrows(2013)]{burrows13} Burrows, A. 2013, RvMP, 85, 245
\bibitem[Canuto \& Dubovikov(1998)]{canuto98} Canuto, V. M., \& Dubovikov, M. 1998, \apj, 493, 834
\bibitem[Cawthorne \& Cobb(1990)]{cawthorne90} Cawthorne, T. V., \& Cobb, W. K. 1990, \apj, 350, 536
\bibitem[Chevalier(1976)]{chevalier76} Chevalier, R. A. 1976, \apj, 207, 872
\bibitem[Coburn \& Boggs(2003)]{coburn03} Coburn, W., \& Boggs, S. E. 2003, Nature, 423, 415
\bibitem[Couch \& Ott(2013)]{couch13} Couch, S. M., \& Ott, C. D. 2013, \apj, 778, L7
\bibitem[Couch \& Ott(2014)]{couch14} Couch, S. M., \& Ott, C. D. 2014, \apj, 799, 5
\bibitem[Decin et al.(2012)]{decin12} Decin, L., Cox, N. L. J., Royer, P., et al. 2012, \aap, 548, A113
\bibitem[Deng et al.(2006)]{deng06} Deng, L., Xiong, D. R., \& Chan, K. L. 2006, \apj, 643, 426
\bibitem[Dotani et al.(1987)]{dotani87} Dotani, T., Hayashida, K., Inoue, H., et al. 1987, Nature, 330, 230
\bibitem[Dwek et al.(2010)]{dwek10} Dwek, E., Arendt, R. G., Bouchet, P., et al. 2010, \apj, 722, 425
\bibitem[Ebisuzaki et al.(1989)]{ebi89} Ebisuzaki, T., Shigeyama, T., \& Nomoto, K. 1989, \apj, 344, L65
\bibitem[Ellinger et al.(2012)]{ellinger12} Ellinger, C. I., Young, P. A., Fryer, C. L., \& Rockefeller, G. 2012, \apj, 755, 160
\bibitem[Fransson \& Kozma(1993)]{fransson93} Fransson, C., \& Kozma, C. 1993, \apj, 408, L25
\bibitem[Fryxell et al.(2000)]{fryxell00} Fryxell, B., Olson, K., Picker, P., et al. 2000, ApJS, 131, 273
\bibitem[Gawryszczak et al.(2010)]{gaw10} Gawryszczak, A., Guzman, J., Plewa, T. \& Kifonidis, K. 2010, \aap, 521, A38
\bibitem[Grebenev et al.(2012)]{grebenev12} Grebenev, S. A., Lutovinov, A. A., Tsygankov, S. S.\& Winkler, C. 2012, \apj, 490, 373
\bibitem[Grefenstette et al.(2014)]{gre14} Grefenstette, B. W., Harrison, F. A., Boggs, S. E., et al. 2014, Nature, 506, 339
\bibitem[Gr\"{o}ningsson et al.(2008)]{gro08} Gr\"{o}ningsson, P., Fransson, C., Leibundgut, B., Lundqvist, P., Challis, P.,
Chevalier, R. A., \& Spyromilio, J. \aap, 492, 481
\bibitem[Grossman et al.(1993)]{grossman93} Grossman, S. A., Narayan, R., \& Arnett, D. 1993, \apj, 407, 284
\bibitem[Haas et al.(1990)]{haas90} Haas, M. R., Erickson, F. A., Lord, S. D., et al. 1990, \apj, 360, 257
\bibitem[Hammer et al.(2010)]{hammer10} Hammer, N. J., Plewa, T., \& M\"{u}ller, E. 2010, \apj, 714, 1371
\bibitem[Handy et al.(2014)]{handy14} Handy, T., Janka, H.-T., \& Odrzywolek, A. 2014, \apj, 783, 125
\bibitem[Hanke et al.(2013)]{hanke13} Hanke, F., M\"{u}ller, B., Wongwathanarat, A., Marek, A., \& Janka, H.-T. 2013, \apj, 770, 66
\bibitem[Hashimoto et al.(1989)]{hashimoto89} Hashimoto, M., Nomoto, K., \& Shigeyama, T. 1989, \aap, 210, L8
\bibitem[Hashimoto(1995)]{hashimoto95} Hashimoto, M. 1995, Progress of Theoretical Phyics, 94, 663
\bibitem[Hanuschik et al.(1988)]{hanuschik88} Hanuschik, R., Thimm, G., \& Dachs, J. 1988, \mnras, 234, 41
\bibitem[Herant \& Benz(1991)]{herant91} Herant, M., \& Benz, W. 1991, \apj, 370, L81
\bibitem[Herant \& Benz(1992)]{herant92} Herant, M., \& Benz, W. 1992, \apj, 387, 294
\bibitem[Herant et al.(1994)]{herant94} Herant, M., Benz, W., Hix, W. R., Fryer, C. L., \& Colgate, S. A. 1994, \apj, 435, 339
\bibitem[Hotta et al.(2014)]{hotta14} Hotta, H., Rempel, M., \& Yokoyama, T. 2014, \apj, 786, 24
\bibitem[Hungerford et al.(2003)]{hung03} Hungerford, A. L., Fryer, C. L., \& Warren, M. S. 2003, \apj, 594, 390
\bibitem[Hungerford et al.(2005)]{hung05} Hungerford, A. L., Fryer, C. L., \& Rockefeller, G. 2005, \apj, 635, 487
\bibitem[Janka \& M\"{u}ller (1995)]{janka95} Janka, H.-T., \& M\"{u}ller, E. 1995, \apj, 448, L109
\bibitem[Janka(2012)]{janka12} Janka, H.-T. 2012, Annual Review of Nuclear and Particle Science, 62, 407
\bibitem[Jerkstrand et al.(2011)]{jerkstrand11} Jerkstrand, A., Fransson, C., \& Kozma, C. 2011, \aap, 530, A45
\bibitem[Joggerst et al.(2009)]{jogg09} Joggerst, C. C., Woosley, H., \& Heger, A. 2009, \apj, 693, 1780
\bibitem[Joggerst et al.(2010a)]{jogg10a} Joggerst, C. C., Alngren, A., Bell, J., Heger, A., Whalen, D., \& Woosley, S. E., 2010a,
\apj, 709, 11
\bibitem[Joggerst et al.(2010b)]{jogg10b} Joggerst, C. C., Alngren, A., \& Woosley, S. E., 2010b, \apj, 723, 353
\bibitem[Kane et al.(2000)]{kane00} Kane J., Arnett, D., Remington, B. A., Glendinning, S. G., Baz\'{a}n, G., M\"{u}ller, E., Fryxell, B. A., \& Teyssier, R. 2000, \apj, 528, 989
\bibitem[Kifonidis et al.(2003)]{kif03} Kifonidis, K., Plewa, T., Janka, H.-T., \& M\"{u}ller, E. 2003, \aap, 408, 621
\bibitem[Kifonidis et al.(2006)]{kif06} Kifonidis, K., Plewa, T., Scheck, L., Janka, H.-T., \& M\"{u}ller, E. 2006, \aap, 453, 661
\bibitem[Kj{\ae}r et al.(2010)]{kjar10} Kj{\ae}r, K., Leibundgut, B., Fransson, C., Jerkstrand, A., \& Spyromilio, J. 2010, \aap, 517, A51
\bibitem[Kotake et al.(2004)]{kotake04} Kotake K., Sawai, H., Yamada, S., \& Sato, K. 2004, \apj, 608, 391
\bibitem[Kotake et al.(2006)]{kotake06} Kotake K., Sato, K., \& Takahashi, K. 2006, Rep. Prog. Phys. 69, 971
\bibitem[Kotake et al.(2012)]{kotake12} Kotake K., Takiwaki, T., Suwa, Y., Iwakami, N. W., Kawagoe, S., Masada, Y., \& Fujimoto, S. 2012, Adv. Astron. 428757
\bibitem[Kuroda et al.(2012)]{kuroda12} Kuroda, T., Kotake, K., \& Takiwaki, T. 2012, \apj, 755, 11
\bibitem[Larsson et al.(2013)]{larsson13} Larsson, J., Fransson, C., Kjaer, K., et al. 2013, \apj, 768, 89
\bibitem[Leising \& Share(1990)]{leising90} Leising, M. D., \& Share, G. H. 1990, \apj, 357, 638
\bibitem[Li \& Yang(2007)]{li07} Li, Y., \& Yang, J. Y. 2007, \mnras, 375, 388
\bibitem[Li(2012)]{li12} Li, Y. 2012, \apj, 756, 37
\bibitem[Kippenhahn \& Weigert(1990)]{kipp90} Kippenhahn, R., \& Weigert, A. 1990, Stellar Sturcture and Evolution (Berlin: Springer)
\bibitem[Magkotsios et al.(2010)]{mag10} Magkotsios, G., Timmes, F. X., Hungerford, A. L., Fryer, C. L., Young, P. A., \& Wiescher, M. 2010, ApJS, 191, 66
\bibitem[Matz et al.(1988)]{matz88} Matz, S. M., Share, G. H., Leising, M. D., Chupp, E. L., Vestrand, W. T., Purcell, W. R., Strickman, M. S., \& Reppin, C. 1988, Nature, 331, 416
\bibitem[Meakin \& Arnett(2006)]{meakin06} Meakin, C. A. \& Arnett, W. D. 2006, \apj, 637, L53
\bibitem[Meakin \& Arnett(2007a)]{meakin07a} Meakin, C. A. \& Arnett, W. D. 2007a, \apj, 667, 448
\bibitem[Meakin \& Arnett(2007b)]{meakin07b} Meakin, C. A. \& Arnett, W. D. 2007b, \apj, 665, 690
\bibitem[Montarges et al.(2014)]{montarges14} Montarges, M., Kervella, P., Perrin, G., Ohnaka, K., Chiavassa, A., Ridgway, S. T., \& Lacour, S. 2014, \aap, 572, A17
\bibitem[M\"{u}ller et al.(1991)]{muller91} M\"{u}ller, E., Fryxell, B., \& Arnett, W. D. 1991, \aap, 251, 505
\bibitem[M\"{u}ller et al.(2012)]{muller12} M\"{u}ller, B., Janka, H.-T., \& Heger, A. 2012, \apj, 761, 72
\bibitem[Murphy \& Meakin(2011)]{murphy11} Murphy, J. W., \& Meakin, C. 2011, \apj, 742, 74
\bibitem[Murphy et al.(2013)]{murphy13} Murphy, J. W., Dolence, J. C., \& Burrows, A. 2013, \apj, 771, 52
\bibitem[Nagataki et al.(1997)]{nagataki97} Nagataki, S., Hashimoto, M., Sato, K., \& Yamada, S. 1997, \apj, 486, 1026
\bibitem[Nagataki et al.(1998)]{nagataki98} Nagataki, S., Shimizu, T. M., \& Sato, K. 1998, \apj, 495, 413
\bibitem[Nagataki(2000)]{nagataki00} Nagataki, S. 2000, ApJS, 127, 141
\bibitem[Nakamura et al.(2014)]{nakamura14} Nakamura, K., Kuroda, T., Takiwaki, T., \& Kotake, K. 2014, \apj, 793, 45
\bibitem[Nomoto \& Hashimoto(1988)]{nomoto88} Nomoto, K., \& Hashimoto, M. 1988, Phys. Rep., 163, 13
\bibitem[Nordhaus et al. (2010)]{nordhaus10} Nordhaus, J., Burrows, A., Almgren, A., \& Bell, J. 2010, \apj, 720, 694
\bibitem[Ono et al.(2009)]{ono09} Ono, M., Hashimoto, M, Fujimoto, S., Kotake, K.,  \& Yamada, S. 2009, Prog. Theor. Phys. 122, 755
\bibitem[Ono et al.(2013)]{ono13} Ono, M., Nagataki, S., Ito, H., Lee, S.-H., Mao, J., Hashimoto, M., \& Tolstov, A. 2013, \apj, 773, 161
\bibitem[Ott et al.(2013)]{ott13} Ott, C. D., Abdikamalov, E., Mosta, P., et al. 2013, \apj, 768, 115
\bibitem[Sawai et al.(2005)]{sawai05} Sawai, H., Kotaki, K., \& Yamada, S. 2005, \apj, 631, 446
\bibitem[Sawai et al.(2013)]{sawai13} Sawai, H., Yamada, S., \& Suzuki, H. 2013, \apj, 770, L19
\bibitem[Shigeyama et al.(1987)]{shigeyama87} Shigeyama, T., Nomoto, K., Hashimoto, M., \& Sugimoto, D. 1987, Nature, 328, 320
\bibitem[Shigeyama et al.(1988)]{shigeyama88} Shigeyama, T., Nomoto, K., \& Hashimoto, M. 1988, \aap, 196, 141
\bibitem[Shigeyama \& Nomoto(1990)]{shigeyama90} Shigeyama, T., \& Nomoto, K. 1990, \apj, 360, 242
\bibitem[Sinnott et al.(2013)]{sinnott13} Sinnott, B., Welch, D. L., Rest, A., Sutherland, P. G., \& Bergmann, M. 2013, \apj, 767, 45
\bibitem[Smith(2007)]{smith07} Smith, N. 2007, \aj, 133, 1034
\bibitem[Smith \& Arnett(2014)]{smith14} Smith, N., \& Arnett, W. D. 2014, \apj, 785, 82
\bibitem[Spyromilio et al.(1990)]{spyromilio90} Spyromilio, J., Meikle, W. P. S., \& Allen, D. A. 1990, \mnras, 242, 669
\bibitem[Steffen(1990)]{steffen90} Steffen, M. 1990, \aap, 239, 443
\bibitem[Sunyaev et al.(1987)]{sunyaev87} Sunyaev, R. A., Kaniovsky, A., Efremov, V., et al. 1987, Nature, 330, 227
\bibitem[Suwa et al.(2010)]{suwa10} Suwa, Y., Kotake, K., Takiwaki, T., Whitehouse, S. C., Liebend\"{o}rfer, M., \& Sato, K. PASJ, 62, L49
\bibitem[Suwa et al.(2011)]{suwa11} Suwa, Y., Kotake, K., Takiwaki, T., Liebend\"{o}rfer, M., \& Sato, K. 2011, \apj, 738, 165
\bibitem[Takiwaki et al.(2009)]{takiwaki09} Takiwaki, T., Kotaki, K., \& Sato, K. 2009, \apj, 691, 1360
\bibitem[Takiwaki et al.(2014a)]{takiwaki14} Takiwaki, T., Kotaki, K., \& Sato, K. 2014, \apj, 786, 83
\bibitem[Tanaka et al.(2012)]{tanaka12} Tanaka, M., Kawabata, K. S., Hattori, T., et al. 2012, \apj, 754, 63
\bibitem[Tian et al.(2009)]{tian09} Tian, C.-L., Deng, L.-C., \& Chan, K.-L. 2009, \mnras, 398, 1011
\bibitem[Tziamtzis et al.(2011)]{tziamtzis11} Tziamtzis, A., Lundqvist, P., Gr\"{o}ningsson, P., \& Nasoudi-Shoar, S. 2011, \aap, 527, A35
\bibitem[Utrobin et al.(1995)]{utrobin95} Utrobin, V. P., Chugai, N. N., \& Andronova, A. A. 1995, \aap, 295, 129
\bibitem[Wongwathanarat et al.(2013)]{wong13} Wongwathanarat, A., Janka, H.-T., \& M\"{u}ller, E. 2013, \aap, 552, A126
\bibitem[Viallet et al.(2013)]{viallet13} Viallet, M., Meakin, C., Arnett, D., \& Moc\'{a}k, M. 2013, \apj, 769, 1
\bibitem[Wongwathanarat et al.(2015)]{wong14} Wongwathanarat, A., \& M\"{u}ller, E., \& Janka, H.-T. 2015, \aap, 577, A48
\bibitem[Woosley \& Arnett(1973)]{woosley73} Woosley, S. E., \&  Arnett, W. D. 1973, ApJS, 26, 231
\bibitem[Woosley \& Hoffman(1991)]{woosley91} Woosley, S. E., \&  Hoffman, R. D. 1991, \apj, 368, L31
\bibitem[Xiong(1977)]{xiong77} Xiong, D. R. 1977, Acta Astron. Sinica, 18, 86
\bibitem[Xiong et al.(1997)]{xiong97} Xiong, D. R., Cheng, Q. L., \& Deng, L. 1997, ApJS, 108, 529
\bibitem[Yamada \& Sato(1991)]{yamada91} Yamada, S., \& Sato, K. 1991, \apj, 382, 594
\bibitem[Zhang \& Li(2012a)]{zhang12a} Zhang, Q.-S., \& Li, Y. 2012a, \apj, 746, 50
\bibitem[Zhang \& Li(2012b)]{zhang12b} Zhang, Q.-S., \& Li, Y. 2012b, \apj, 750, 11
\end{thebibliography}
\end{document}